\documentclass[11pt]{article}

\usepackage{putex}
\usepackage{cite}
\usepackage[utf8]{inputenc}
\usepackage{amsmath,amsfonts,amssymb}
\usepackage{physics}
\usepackage{cancel}
\usepackage{bbm}
\usepackage{appendix}
\usepackage{caption}
\usepackage{subcaption}
\usepackage{color}
\usepackage{datetime}
\usepackage{float}
\usepackage{graphicx}
\usepackage{feynmp-auto}
\usepackage[dvipsnames]{xcolor}
\usepackage{bm}
\usepackage{makeidx}
\usepackage{epsf}
\usepackage{slashed}
\usepackage{mathtools}
\usepackage{xcolor}
\usepackage{dsfont}
\usepackage{braket}
\usepackage{hyperref}

\newcommand{\mL}{\mathcal{L}}
\newcommand{\mO}{\mathcal{O}}

\newcommand{\pd}{\partial}
\renewcommand{\c}{\chi}
\newcommand{\cdag}{c^{\dagger}}

\newcommand{\normord}[1]{\vcentcolon\mathrel{#1}\vcentcolon}
\providecommand{\vcentcolon}{\mathrel{\mathop{:}}}

\numberwithin{equation}{section}

\begin{document}             

\preprint{PUPT-2662\\MIT-CTP/6036}

\institution{a}{SISSA, Via Bonomea 265, 34136 Trieste, Italy}
\institution{b}{INFN, Sezione di Trieste, Via Valerio 2, 34127 Trieste, Italy}
\institution{c}{Center for Theoretical Physics - a Leinweber Institute, Massachusetts Institute of Technology, Cambridge, MA 02139, USA}
\institution{d}{Joseph Henry Laboratories, Princeton University, Princeton, NJ  08544, USA}
\institution{e}{Princeton Center for Theoretical Science, Princeton University, Princeton, NJ 08544, USA}

\title{The two-flavor  
Schwinger model at 50:\\ Solving Coleman's puzzles 
}

\authors{Gabriel Cuomo,\worksat{\a,\b} Ross Dempsey,\worksat{\c} Andrei Katsevich,\worksat{\d} Igor R.~Klebanov,\worksat{\d,\e}\\[.5em]Ilia V.~Kochergin,\worksat{\d} Silviu S.~Pufu,\worksat{\d,\e} Benjamin T.~S{\o}gaard\worksat{\d}}

\abstract{
In his 1976 paper ``More about the massive Schwinger model" \cite{Coleman:1976uz}, Coleman introduced $1+1$-dimensional Quantum Electrodynamics coupled to two charged massive fermions.  
By applying Abelian bosonization, he elucidated much of the physics of this two-flavor Schwinger model, but he listed three puzzles at the end of his paper. We present new analytical and numerical calculations to solve Coleman's three puzzles and thereby deepen our understanding of this model. These puzzles pertain to the theory with equal fermion masses at $\theta = 0$ and at $\theta = \pi$, as well as the size of isospin-breaking effects when the fermion masses are unequal. For the puzzle at $\theta = \pi$, the solution is related to the structure of the zero-temperature phase diagram \cite{Dempsey:2023gib}: for equal fermion masses $m$, the model exhibits spontaneous breaking of charge conjugation symmetry and absence of confinement for any value of the gauge coupling $g$, so that there is a smooth interpolation from weak to strong coupling. Using two-loop Renormalization Group and integrability methods, we show that the mass gap behaves as $\sim m e^{-0.111 g^2/m^2}$ in the strong coupling regime $m\ll g$. Our numerical results using the lattice Hamiltonian are in good agreement with this behavior. For the puzzle at $\theta = 0$, the solution is related to a level crossing between two isosinglet particles with different discrete quantum numbers; we demonstrate the necessity of such a crossing by comparing integrability and weak coupling calculations, and we also exhibit the crossing numerically. Finally, we provide a new estimate for the size of isospin-breaking effects caused by different fermion masses at strong coupling.
}

\date{September 2026}

\maketitle

\tableofcontents

\section{Introduction and summary}

An important problem in theoretical physics is proving the existence of the ``mass gap" in $3+1$-dimensional quantum Yang-Mills theory.  The mass gap is the mass of the lightest glueball, a color-singlet bound state of gluons. Due to the asymptotic freedom of Yang-Mills theory \cite{Gross:1973id,Politzer:1973fx}, this mass is related by an exponentially small factor to the ultraviolet (UV) cutoff, which in the context of lattice gauge theory \cite{Wilson:1974sk, Kogut:1974ag} is the inverse lattice spacing. There is extensive numerical evidence for the non-vanishing mass gap (see, for example, \cite{Chen:2005mg}) and the exponential hierarchy of scales, sometimes called ``dimensional transmutation" \cite{Coleman:1973jx}. However, finding an analytical proof has been a formidable challenge for more than 50 years. 

Due to the difficulty of the mass gap problem in Yang-Mills theory, it is useful to consider analogous problems in $1+1$-dimensional models. For instance, the O$(N)$ sigma models, which are asymptotically free \cite{Polyakov:1975rr}, develop an exponentially small gap, as evidenced by integrability methods \cite{Polyakov:1983tt, Wiegmann:1984pw, Hasenfratz:1990zz}. Another model that exhibits dimensional transmutation, and with which we will make connections in this paper, is the SU$(2)$ Thirring model \cite{Banks:1975xs, Amit:1979ab}, which is equivalent to the chiral Gross-Neveu model \cite{Gross:1974jv}. 

A class of $1+1$-dimensional gauge theories that have been used as models for Quantum Chromodynamics (QCD) is the Schwinger model \cite{Schwinger:1962tp, Lowenstein:1971fc} and its generalization that include a fermion mass term and the topological angle $\theta$ \cite{Coleman:1975pw}.
The Schwinger model has been used as a simple model of quark confinement 
\cite{Casher:1974vf, Coleman:1975pw, Coleman:1976uz}
(in the papers \cite{Coleman:1975pw, Coleman:1976uz}, it is sometimes called ``quark trapping.").  For $\theta=\pi$, Coleman \cite{Coleman:1976uz} discovered a zero-temperature confinement/deconfinement phase transition: when the fermion mass $m$ is kept fixed, it occurs at a critical value of the gauge coupling, $g=g_*$.  This quantum phase transition is in the 2D Ising universality class \cite{Byrnes:2002nv}, and $m/g_*$ is very close to $\frac{1}{3}$ but deviates from it slightly \cite{ArguelloCruz:2024xzi, Fujii:2024reh}. 
For $g>g_*$, there is a unique vacuum whose lowest excitation is a massive fermion-antifermion bound state, while for $g< g_*$ there are two vacua. The excitations above them contain no bound states, only a continuum due to decay into two or more charged ``half-asymptotic particles," signaling the loss of confinement \cite{Coleman:1976uz, Byrnes:2002nv}. The half-asymptotic particles are the solitons\footnote{Additional names that have been used for these objects are kinks, lumps, and domain walls. } interpolating between the two vacua. The absence of bound states on the weak coupling side of the phase transition has been confirmed numerically \cite{Byrnes:2002nv, Dempsey:2025wia}. 

It is of further interest to generalize the Schwinger model to include $N_f$ massive fermions. Work in this direction began with Coleman's seminal study of the $N_f=2$ case \cite{Coleman:1976uz}, where the Abelian bosonization \cite{Coleman:1974bu, Mandelstam:1975hb} was used to access its strong coupling behavior. 
For equal flavor masses, $m_1=m_2$, the model has $\text{SU}(2)/\mathbb{Z}_2\simeq  \text{SO}(3)$ symmetry, which Coleman called ``isospin" by analogy with QCD\@. The work on the two-flavor Schwinger model continued in many subsequent papers, including \cite{Harada:1993va, Hetrick:1995wq, Smilga:1996pi, Hosotani:1998za, Hosotani:1998kd, Berruto:1999ga, Misumi:2019dwq, Georgi:2020jik, Georgi:2022sdu}. In this paper, we use a mix of analytical and numerical techniques to carry out a detailed investigation of this model and present solutions of the three puzzles about its strong coupling behavior, which Coleman listed at the end of his paper \cite{Coleman:1976uz}.

Our paper builds on \cite{Dempsey:2023gib}, where the two-flavor massive Schwinger model was shown to exhibit quantum phase transitions and dimensional transmutation when the topological angle $\theta$ is set exactly to $\pi$. Using the effective field theory (EFT) obtained via bosonization \cite{Coleman:1976uz}, it was shown \cite{Dempsey:2023gib} that, for equal fermion masses $m_1=m_2=m\ll g$, 
the theory undergoes the asymptotically free Berezinski-Kosterlitz-Thouless (BKT) 
Renormalization Group (RG) flow \cite{Berezinskii:1970pzv, Kosterlitz:1973xp, Kosterlitz:1974nba, Jose:1977gm, Amit:1979ab}. The BKT RG flow analysis suggests that the lightest states have masses suppressed by the factor $e^{-0.111 g^2/m^2}$ \cite{Dempsey:2023gib}. Numerical support for the exponential smallness of the mass gap was presented in \cite{Dempsey:2023gib} and in further work \cite{Itou:2023img,Itou:2024psm,Kanno:2024elz}. In this paper, we provide considerable further evidence for this claim. Furthermore, using a mix of numerical and analytical methods, including integrability, we describe the dependence of the mass spectrum on the three dimensionless parameters of the model:
$m_1/g$, $m_2/g$, and $\theta$, in various regimes. When $m_1$ and/or $m_2$ vanishes, there is no dependence on $\theta$: due to the chiral anomaly, $\theta$ can be rotated away. For 
$m_1=m_2=0$, the mass gap vanishes. Using non-Abelian bosonization \cite{Witten:1983ar}, the low-energy limit of this model is described \cite{Gepner:1984au,Affleck:1985wa} by the $\text{SU}(2)_1$ Wess-Zumino-Witten (WZW) model, which is a conformal field theory of a compact scalar field.

For $m_1=m_2=m\ll g$ and $\theta\neq \pi$, the lightest bound states are an $\text{SO}(3)$ triplet of pseudoscalar mesons (the pions $\pi^0$, $\pi^+$, $\pi^-$) followed by a singlet $\sigma$ particle of positive parity~\cite{Coleman:1976uz}. This spectrum is reminiscent of QCD with two light flavors, although in the strongly-coupled Schwinger model, $\sigma$ is a bound state rather than a resonance. For $\theta=\pi$, we find that the isosinglet becomes exactly degenerate with the isotriplet (for earlier numerical results suggesting this degeneracy, see \cite{Itou:2023img,Itou:2024psm}). We will show that this degeneracy signals the beginning of a continuum of energy levels. As in the $N_f=1$ model, this continuum arises due to the half-asymptotic particles. The continuous spectrum of gauge-invariant states implies that confinement is lost for $\theta=\pi$ due to the presence of the half-asymptotic particles.

The zero-temperature phase diagram of the $N_f=2$ massive Schwinger model at $\theta=\pi$ as a function of the two masses was proposed in \cite{Dempsey:2023gib} (for the generalization to two flavors with different charges, see \cite{Mouland:2025dct}). It is redrawn in the $\theta=0$ convention in Figure~\ref{fig:Phase_Diagram} 
(the shift of $\theta$ by $\pi$ is equivalent to the transformation $(m_1, m_2)\rightarrow (- m_1, m_2)$ \cite{Georgi:2022sdu}). 
This phase diagram has a qualitative similarity to the phase diagram of QCD with two light quark flavors of masses $m_u$ and $m_d$ \cite{Creutz:2018vgl}. 
The latter is well-known to contain a region where the $CP$ symmetry is spontaneously broken \cite{Dashen:1970et, Creutz:1995wf,  Smilga:1998dh, Creutz:2010ts, Gaiotto:2017yup}. This symmetry breaking is analogous to the symmetry breaking observed in the two-flavor Schwinger model \cite{Dempsey:2023gib}, where along the $\text{SO}(3)$-symmetric line the theory has two degenerate vacua.  There is no phase transition along the $\text{SO}(3)$-symmetric line: for any coupling $g$, we find a continuous spectrum and no confinement of charges.
As shown in Section~\ref{sec_theta_Pi}, at weak coupling the mass of a half-asymptotic particle may be expanded as $m(1- g^2/2\pi m^2 + \ldots )$. At strong coupling, we use the two-loop Renormalization Group to show that this mass approaches $\sim m e^{-0.111 g^2/m^2}$, and we provide support for this behavior using the Bethe ansatz solution \cite{Belavin:1979pq,Andrei:1979sq,Forgacs:1991nk} of the SU$(2)$ Thirring model. Our numerical results, which are shown in Figures~\ref{fig:kink_mass} and \ref{fig:kink_massB}, agree with these limiting behaviors and demonstrate a smooth interpolation between them.

In Section \ref{confret}, we discuss the effects of isospin-breaking fermion masses at $\theta=\pi$. For any non-vanishing masses $m_1\neq m_2$, the theory exhibits a quantum phase transition at some critical coupling $g_*(m_1, m_2)$.  This phase transition is in the Ising universality class. For $g<g_*$, the theory is in the non-confining phase with broken $C$, while for $g> g_*$ it is in the confining phase with unbroken $C$ (see the phase diagram in Figure~\ref{fig:Phase_Diagram}).  As discussed in~\cite{Dempsey:2023gib}, in the $N_f=2$ model with $m_2\gg m_1$, the critical coupling $g_*$ is $\approx 3 m_1$, as in the $N_f=1$ model. The behavior of the critical coupling $g_*$ in the limit of small isospin breaking is given in \eqref{critcoupl} below. 

The remainder of this paper is organized as follows. In Section~\ref{threepuzzles} we copy the three puzzles from \cite{Coleman:1976uz} and briefly state our solutions. 
In Section~\ref{sec:setup}, we explain the setup of the two-flavor Schwinger model and its symmetries.  We then study the case $\theta \neq \pi$ in Sections~\ref{sec_equal_m_theta_Not_Pi} and~\ref{isobreak}, and then turn to the theory at $\theta = \pi$ in Sections~\ref{sec_theta_Pi} and~\ref{confret}. Many technical details are relegated to several appendices.

\section{The three puzzles}
\label{threepuzzles}

The seminal paper \cite{Coleman:1976uz} has stimulated both analytic and numerical research for 50 years, establishing the two-flavor Schwinger model as a highly non-trivial yet tractable QFT\@. At the end of \cite{Coleman:1976uz}, Coleman listed three strong coupling phenomena that he found puzzling:

\begin{enumerate}

\item{``Why are the lightest particles in the theory a degenerate isotriplet, even if one quark is $10$ times heavier than the other?"}

\item{``Why does the next lightest particle have $I^{PG}= 0^{++}$, rather than $0^{--}$?"}

\item{``For $|\theta|=\pi$, how can an isodoublet quark and an isodoublet antiquark, carrying opposite electric charges, make an isodoublet bound state of electric charge zero?"  }

\end{enumerate}

Our solutions to these puzzles can be briefly stated as follows:
\begin{enumerate}
    \item  When $m_1 = m_2$, the lightest particles form an isotriplet of degenerate pions.  The near-degeneracy of the lightest three particles even when $m_1\neq m_2$ is explained by an accidental SO$(3)$ symmetry of the low energy EFT of the two-flavor Schwinger model at strong coupling. This is explained in Section~\ref{isobreak}. We find that, for $\theta=0$ and strong coupling, the leading term in $\left (M_{\pi^\pm}^2 - {M_{\pi^0}^2}\right )/{M_{\pi^0}^2} $ is positive and proportional to $(m_1-m_2)^2/g^2$. (This puzzle was discussed in past work including \cite{Delphenich:1997ex, Georgi:2020jik, Albandea:2025dhu}, but the quantitative details of our results are different.)
    
    \item  At weak coupling (large $m/g$), the lightest isosinglet has\footnote{The definition of $G$-parity is given around \eqref{GDef}.} $I^{PG} = 0^{--}$ and the second lightest has $I^{PG} = 0^{++}$.
    As we explain in Section~\ref{sec_equal_m_theta_Not_Pi}, this is  consistent with the lightest isosinglet having $I^{PG} = 0^{++}$ at strong coupling (small $m/g$) because there is a level crossing between the two states at $m/g \approx 0.15$ (see Figure~~\ref{fig:mass_triplet_singlet_theta0}). At strong coupling, the $0^{++}$ state can be interpreted as a bound state of two pions  \cite{Harada:1993va, Delphenich:1997ex}. 
    
    \item The phase diagram of the theory at $\theta = \pi$, which was proposed in \cite{Dempsey:2023gib}, implies that there is no phase transition in the $m_1 = m_2$ theory when $g$ is increased. In this paper, we show that there is no confinement for any value of $g$, so that the lightest excitations are the charged half-asymptotic particles (solitons). Therefore, contrary to Coleman's expectation, the charge-neutral spectrum contains no bound states: the lowest energy states are a two-body continuum formed by a pair of charged half-asymptotic particles. In Section~\ref{sec_theta_Pi}, we show that the mass and electric charge of the half-asymptotic particles decrease as $g$ increases, but both are nonzero for any finite value of $g$, and hence there is no massive charge-zero isodoublet hypothesized by Coleman.
\end{enumerate}

\section{The setup}\label{sec:setup}

In this paper, we are primarily interested in the two-flavor massive Schwinger model, which is the $1+1$-dimensional QED theory with two unit-charge Dirac fermions. The Lagrangian density reads \cite{Dempsey:2023gib}
\begin{equation}\label{eq_Lag}
 \mL=\sum_{\alpha=1}^{2}
    \left(i\bar{\psi}^\alpha\slashed{D}\psi_\alpha-m_\alpha\bar{\psi}^\alpha\psi_\alpha\right)
   -\frac{1}{4g^2}F_{\mu\nu}^2-\frac{\theta}{4\pi}\varepsilon^{\mu\nu}F_{\mu\nu}\,,
\end{equation}
where the $\theta$ angle is $2\pi$-periodic, $D_\mu \psi_\alpha = (\partial_\mu - i A_\mu) \psi_\alpha$, $F_{\mu\nu}$ is the field strength of the gauge field $A_\mu$, $g$ is the gauge coupling, and $m_1$ and $m_2$ are the two mass parameters. We work with a diagonal metric $\eta_{00}=- \eta_{11}= 1$ and Levi-Civita tensor such that $\varepsilon^{01}=1$, and we use the gamma matrix convention $\gamma^0 = \sigma_2$, $\gamma^1 = i \sigma_1$, with $\sigma_A$ being the Pauli matrices.  The chirality matrix is $\gamma^5 = \gamma^0 \gamma^1 = \sigma_3$. 

A case of particular interest is the equal-mass theory, $m_1=m_2$, for which the theory enjoys an $\text{SU}(2)/\mathbb{Z}_2 \simeq \text{SO}(3)$ flavor symmetry, referred to as isospin symmetry.\footnote{Note that under the action of the center of $\text{SU}(2)$, we have $\psi_\alpha\rightarrow-\psi_\alpha$, which is equivalent to a gauge transformation.  Thus, the gauge-invariant states stay invariant under the center of $\text{SU}(2)$.} 
The isospin generators are 
\begin{equation}
   Q_A=\frac12\int dx\,\bar{\psi}^\alpha\gamma^0(\sigma_A)_\alpha^{\;\beta}\psi_\beta\,.
\end{equation}
The internal symmetry is further enhanced to $\left[ \text{SU}(2)\times \text{SU}(2) \right]/\mathbb{Z}_2 \cong \text{SO}(4)$ at the massless point. We also recall that due to the Schwinger anomaly, axial transformations shift the $\theta$ angle, implying in particular the equivalence of couplings $(g, m_1, m_2,\theta)$, $(g, m_1, -m_2,\theta + \pi)$, and $(g, -m_1, m_2,\theta + \pi)$. For real masses, the model is invariant under parity ($P$) and charge conjugation ($C$) at $\theta = 0, \pi$, where these symmetries act on the fields as
\begin{align}
   (A_\mu(t,x) ,\psi_\alpha(t,x))&\stackrel{P}{\longrightarrow}(\eta_{\mu\mu}A_\mu(t,-x) ,\gamma^0\psi_\alpha(t,-x))\,,\\
   (A_\mu(t,x) ,\psi_\alpha(t,x))&\stackrel{C}{\longrightarrow}(-A_\mu (t,x),\gamma^5\psi_\alpha^*(t,x))\,.
\end{align}
Note that the model is $CP$-invariant for all values of $\theta$.

At $\theta = 0$, it is also convenient to introduce $G$-parity \cite{Lee:1956sw,Itou:2023img} as a combination of charge conjugation and an $\text{SO}(3)$ isospin transformation generated by $Q_2$:\footnote{
We note that $e^{i\pi Q_2}$ can be taken to act as 
$ 
   \left(\begin{array}{c}
       \psi_1  \\
         \psi_2    \end{array}\right)
{\longrightarrow}
\left(\begin{array}{c}
       \psi_2  \\
         -\psi_1    \end{array}\right)
$.} 
\begin{equation} \label{GDef}
G = Ce^{i\pi Q_2}\ .
\end{equation}
$G$-parity commutes with the isospin symmetry and has a $\pm 1$ eigenvalue. Therefore, particles may be labeled by their $G$-quantum number rather than $C$, via the $I^{PG}$ notation, that stands for isospin $I$, parity $P$, and $G$-parity $G$.
This was the notation used in \cite{Coleman:1976uz}.

\begin{figure}[t]
    \centering
\includegraphics[width=0.7\linewidth]{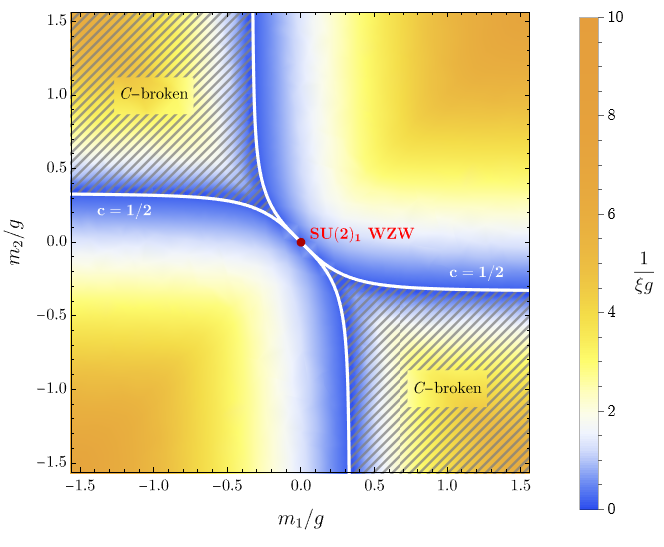}
    \caption{The phase diagram of the two-flavor Schwinger model with $\theta=0$. The top-left and bottom-right regions exhibit charge liberation and spontaneous breaking of charge conjugation $C$. They are separated from the $C$-preserving, confining phase by curves described by the $c = \frac{1}{2}$ Ising CFT. These curves meet only at the origin, where they merge into a $c = 1$ CFT, the $\text{SU}(2)_1$ WZW model. The heat map in the background is the inverse of the correlation length $\xi$ (in coupling units), calculated using infinite matrix product states on a Hamiltonian lattice with fixed small lattice spacing, $a = \frac{0.35}{g}$.
    }
    \label{fig:Phase_Diagram}
\end{figure}

The equal mass two-flavor Schwinger model also exhibits an anomaly in coupling space between $\theta$ and the $\text{SO}(3)$ symmetry~\cite{Cordova:2019jnf,Itou:2023img}, as well as a mixed anomaly between charge conjugation and the flavor symmetry at $\theta=\pi$~\cite{Komargodski:2017dmc,Sulejmanpasic:2020lyq}.
The mixed anomaly implies that we cannot have a non-degenerate gapped vacuum at $\theta=\pi$ for arbitrary values of $m$. The phase diagram proposed in~\cite{Dempsey:2023gib}, shown in Figure~\ref{fig:Phase_Diagram}, is compatible with this requirement. We will discuss further consequences of these anomalies in Section~\ref{sec_theta_Pi}.

The model~\eqref{eq_Lag} is super-renormalizable, meaning that it is UV-finite up to a cosmological constant renormalization contribution. The UV-finiteness makes this model a particularly interesting toy model of QCD with two quark flavors in which it is possible to obtain exact results for the spectrum in certain regimes, including in the strong coupling region $g\gg m_1,m_2$.

\section{The equal mass theory at \texorpdfstring{$\theta \neq \pi$}{θ≠π}}\label{sec_equal_m_theta_Not_Pi}

In this section, we analyze the mass spectrum of the $m_1 = m_2 \equiv m$ theory, with a particular emphasis on the strong coupling regime. This problem has been extensively studied in the literature~\cite{ Harada:1993va, Hetrick:1995wq, Delphenich:1997ex, Hosotani:1998za, Georgi:2020jik, Georgi:2022sdu,Affleck:1985wa,Itou:2023img,Itou:2024psm,Kanno:2024elz}. Nevertheless, we present a self-contained treatment that includes several original results. In particular, we demonstrate that the solution of the second Coleman puzzle, concerning the parity assignment of the next-to-lightest state at strong coupling, arises from a level-crossing phenomenon between a mesonic particle and a tetraquark state.

At $\theta=0$, the system has a unique $C$-preserving vacuum where the electric field vanishes in expectation value,  $\langle F_{01}\rangle=0$. The vacuum remains unique also for $\theta\neq 0$ (as long as $\theta \neq \pi$), where charge conjugation is explicitly broken. For this reason, it is convenient to discuss all values $\theta \neq \pi$ together. In two dimensions, the gauge field gives rise to a linearly rising Coulomb potential already at the classical level. As a result, the theory confines for arbitrary values of the coupling. Below we discuss the low-lying states of the theory.

\subsection{Analytical results at weak coupling}

\begin{figure}[t]
    \centering
    \includegraphics[width=0.6\linewidth]{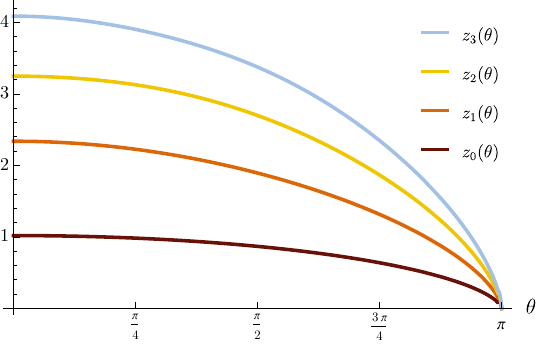}
    \caption{The functions $z_n(\theta)$, which appear in the non-relativistic energy eigenvalues in~\eqref{eq_E_bound}. These functions are defined implicitly in terms of the Airy function in \eqref{eq_app_cond_z}.}
    \label{fig:app_z_NR_QM}
\end{figure}

Let us first review the weak coupling regime $g\ll m$.  In this regime, the low-mass particles (referred to as ``mesons'') are non-relativistic electron-positron bound states of mass $M=2m+ E_{\text{NR}}$, where $E_{\text{NR}}$ is the non-relativistic interaction energy. At leading order in the weak coupling expansion,  $E_{\text{NR}}$ is determined by the Schr\"odinger equation associated with the non-relativistic Hamiltonian (see for instance~\cite{Kenway:1977dk,Ziyatdinov:2010vg,Shifman:2012zz}): 
\begin{equation}\label{eq_H_NR}
    H_{\text{NR}}=\frac{p^2}{m}+\frac{g^2}{2}\left(|x|-\frac{\theta}{\pi}x\right)\,,
\end{equation}
where in the kinetic term we used the reduced mass $m/2$ of the electron-positron system. At this order, the Hamiltonian is the same for particles in the singlet or triplet representations of the $\text{SO}(3)$ isospin symmetry, and so these particles have equal masses at leading order. We review the solution of the Schr\"odinger equation in the linear potential~\eqref{eq_H_NR} in Appendix~\ref{app_weak_coupling_NR}. The energy levels are given by\footnote{Note that the scaling $E_{\text{NR}}\sim g^{4/3}/m^{1/3}$ as well as the power of the corrections (due to contact interactions and the relativistic kinetic term) can be seen by combining the virial theorem $p^2/m\sim g^2|x|$ and the uncertainty principle $p\sim 1/x$.}
\begin{equation}\label{eq_E_bound}
    E_{\text{NR}}^{(n)} =\frac{g^\frac{4}{3}}{(4m)^\frac{1}{3}}z_n(\theta)+\mathcal{O}\left(\frac{g^\frac{8}{3}}{m^\frac{5}{3}}\right)\,,
\end{equation}
where the values $z_n(\theta)$ are related to the matching conditions for Airy functions; for instance, $z_0(0)\simeq  1.01879$ is the smallest positive number such that $\text{Ai}'(-z_0(0))=0$. In Figure~\ref{fig:app_z_NR_QM}, we plot the coefficients $z_n(\theta)$ for the lowest four states; we will remark on the vanishing of these coefficients at $\theta = \pi$ shortly.

Note that at $\theta=0$, the ground state of the non-relativistic Hamiltonian \eqref{eq_H_NR} has an even wavefunction, and since the elementary particles in the QFT are fermions, the lightest meson therefore corresponds to a parity-odd state. Since the Hamiltonian \eqref{eq_H_NR} is flavor-independent at leading order, it follows that both the lightest isotriplet and isosinglet mesons are parity-odd. The next lightest mesons are instead parity-even, and so on. The same is true for the action of charge conjugation on the $Q_3$-neutral state of the representation, and hence we say that all mesons are $CP$-even.  The $G$-parity of the meson is related to charge conjugation by $G=(-1)^I C$, where $I$ is the isospin and the eigenvalue of $C$ is computed for the $Q_3$-neutral state in this representation.\footnote{To show this, assume that $\ket{\psi_0}$ is the $Q_3$-neutral state in an isospin $I$ irreducible representation: $Q_3 \ket{\psi_0} = 0$; note that it is unique. From the commutation relations $C Q_A = (-1)^A Q_A C$ and uniqueness it follows that $\ket{\psi_0}$ is an eigenstate of $C$. Then we construct the $Q_2$-eigenbasis $\{\ket{J}\}$: $J = -I,\dots,I$; $Q_2\ket{J} = J\ket{J}$. We also define the ladder operators $Q_{\pm} = Q_3 \pm iQ_1$, then $Q_3 = \frac{Q_+ + Q_-}{2}$. From $\bra{I}Q_3\ket{\psi_0} = 0$ it follows that $\braket{I-1|\psi_0} = 0$. Repeating this for $\bra{I-2k}Q_3\ket{\psi_0}$ we recursively show that $\braket{I-2k+1|\psi_0} = 0$, so $\ket{\psi_0} = \sum_{k=0}^I c_k \ket{-I+2k}$. Therefore, $G \ket{\psi_0} = C e^{i \pi Q_2} \ket{\psi_0} = (-1)^I C \ket{\psi_0}$. Since $G$ commutes with the isospin symmetry, it has the same eigenvalue for all states in the representation.}

\begin{figure}[t]
    \centering
    \includegraphics[width=0.6\linewidth]{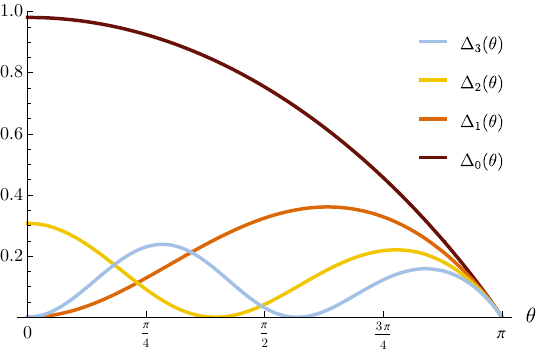}
    \caption{The functions $\Delta_n(\theta)$, which appear in the singlet-triplet splittings given in~\eqref{eq_stsplitting}. They are defined in \eqref{eq_app_Delta}.}
    \label{fig:app_Delta_NR_QM}
\end{figure}

To the leading nontrivial order, the energy difference between the singlet and triplet mesons corresponding to the same eigenstate of $H_\text{NR}$ arises from the contact interaction induced by a photon exchange in the $s$-channel~\cite{Coleman:1976uz}:
\begin{equation}\label{eq_DeltaH_NR}
    \Delta H_{\text{singlet}}=\frac{ g^2}{2m^2}\delta(x)\,,
\end{equation}
which implies (see Appendix~\ref{App:SingletTriplet} for details)
\begin{equation} 
\label{eq_stsplitting}
    M_{\text{singlet}}-M_{\text{triplet}}\simeq \frac{g^\frac{8}{3}}{2^\frac{7}{3}  m^\frac{5}{3}}\Delta_n(\theta)+\mathcal{O}\left(\frac{g^{4}}{m^{3}}\right)\,,
\end{equation}
where the expressions for $\Delta_n(\theta)$ are implicitly given in terms of Airy functions in \eqref{eq_app_Delta}.  
See Figure~\ref{fig:app_Delta_NR_QM} for a plot of the functions $\Delta_n(\theta)$ for the four lightest singlet-triplet pairs, respectively.  We find $\Delta_0(0)=1/z_0(0)$ and $\Delta_1(0) = 0$, as per \eqref{GotDeltan}.  (The vanishing of $\Delta_n(0)$ for odd $n$ is due to the $n$th excited eigenfunction of $H_\text{NR}$ being odd.)  Note from Figure~\ref{fig:app_Delta_NR_QM} that $\Delta_1(\theta) \leq \Delta_0(\theta)$ for all $\theta$, and thus the singlet-triplet mass splitting for the second-lightest singlet-triplet pair is uniformly smaller than that for the lightest singlet-triplet pair. 

At weak coupling there are therefore many mesons with non-relativistic energies $E_\text{NR}\sim g^{\frac{4}{3}}/m^{\frac{1}{3}}$, lying above the two lowest states and below the two-particle threshold at $\sim 4m$. Additionally, contact interactions may induce meson-meson bound states slightly below the two-particle threshold, but we do not study those here.

As we vary $\theta$ from $0$ to $\pi$ at weak coupling, both the binding energy and the singlet-triplet mass difference decrease and eventually approach zero continuously, as we see in Figure~\ref{fig:app_Delta_NR_QM}. This is because the confining potential in~\eqref{eq_H_NR} vanishes for $x>0$ at $\theta=\pi$. We refer to the theory as \emph{not confining} whenever charge can escape to infinity. Then the Schwinger model with $\theta\neq \pi$, as well as QCD with dynamical quarks, are called \emph{confining}. We emphasize, however, that this terminology is unrelated to the behavior of Wilson loops,\footnote{We may say that a theory is \emph{flux-confining} when Wilson loops for probe charges obey an area law. By contrast, the notion of confinement we are using in this paper is sometimes called color confinement. The two-flavor Schwinger model is color-confining except for in its $C$-broken phase. For integer-charge Wilson loops, it is never flux-confining; for fractional-charge Wilson loops, it is flux-confining except when one of the masses vanishes.} which always obey a perimeter law (for integer charges) due to the absence of one-form symmetries under which the unit charge Wilson loop is charged. We will analyze the theory at $\theta=\pi$ in greater detail in the next section.

\subsection{Analytical results at strong coupling}
\label{sec:analytical}

Let us now move on to the strong coupling regime $g\gg m$. It is convenient to rely on bosonization techniques. In particular, below we will discuss the theory both via Abelian~\cite{Coleman:1974bu,Mandelstam:1975hb} and non-Abelian~\cite{Witten:1983ar} bosonization. The non-Abelian bosonization makes the symmetries of the model manifest, and is thus particularly convenient to analyze its qualitative features. The Abelian bosonization is technically simpler, and we will rely on it (and on integrability) to obtain precise quantitative results.

Let us first use non-Abelian bosonization.  Following~\cite{Affleck:1985wa}, the model~\eqref{eq_Lag} can be recast as a\footnote{In terms of the fields 
$U$ and $\Phi$, the $\mathbb{Z}_2$ factor implies an identification relating the center of $\text{SU}(2)$ to a half-shift of the compact boson.} 
\begin{equation}
    \text{U}(2)_1 \simeq\frac{\text{SU}(2)_{1}\times \text{U}(1)}{\mathbb{Z}_2}
\end{equation}
WZW CFT deformed by a mass term and coupled to a $\text{U}(1)$ gauge field with a standard kinetic term. The action can be written explicitly in terms of a periodic scalar $\Phi\sim\Phi+2\pi$, and an electrically neutral ``pion'' matrix $U\in \text{SU}(2)$. Explicitly,~\eqref{eq_Lag} is equivalent to
\begin{equation}\label{eq_non_Abelian_exact}
    S=S_\text{WZW}+\int d^2x\left(\frac{1}{4\pi}\pd_\mu \Phi \pd^\mu \Phi-\frac{F_{01}}{2\pi}\left(\theta+2\Phi\right)+\frac{1}{2g^2}F_{01}^2+\tilde{c}\,m\sqrt{\tilde{\mu}}\Re
    \,N_{\tilde{\mu}}e^{i\Phi}\text{Tr}\left[:U: \right]\right)\,,
\end{equation}
where $\tilde{c}$ is a $\mO(1)$ coefficient, whose precise value is irrelevant for the moment, $N_{\tilde{\mu}}e^{i \Phi}$ denotes the dimension-$\frac{1}{2}$ vertex operator normal-ordered  with respect to an arbitrary mass scale $\tilde\mu$, and the first term is the well-known $k=1$ WZW action,
\begin{equation} \label{SWZW}
    S_\text{WZW}=\frac{1}{8\pi} \int d^2x\,  \mathrm{Tr}\left( \partial_\mu U^\dagger  \partial^\mu U \right)+\frac{1}{12\pi} \int_{B} d^3y \, \epsilon^{ijk}
  \mathrm{Tr}\left(
  U^\dagger \partial_i U 
  U^\dagger \partial_j U 
  U^\dagger \partial_k U
  \right)\,.
\end{equation}
Here, $B$ is a three-dimensional manifold whose boundary $\pd B=\mathbb{R}^2$ is the physical space. In ~\eqref{eq_non_Abelian_exact}, $:U_\alpha{}^{\beta}:$ denotes the operator with dimension $1/2$ in the WZW model normalized in order to ensure that its time-ordered two-point function obeys
\begin{equation}
\left\langle T \normord{U_\alpha{}^{\beta}(x)} \,\normord{(U^\dagger)_{\beta'}{}^{\alpha'}(0)}\right\rangle\stackrel{\sqrt{-x^2}\rightarrow 0}{=}\frac{\delta_{\alpha}^{\alpha'}\delta^{\beta}_{\beta'}}{\sqrt{-x^2+i\epsilon}}\,.
\end{equation}
That the mass $\tilde\mu$ mentioned above is arbitrary can be seen using the re-normal ordering identity~\cite{Coleman:1974bu}
\begin{equation}
    \sqrt{\mu_1}N_{\mu_1}e^{i \Phi}=\sqrt{\mu_2}N_{\mu_2}e^{i \Phi}\,.
\end{equation}

It is convenient to integrate out the electric field and canonically normalize $\Phi$ as
\begin{equation}
\Phi= \sqrt{2\pi}\,\phi_+\,,
\end{equation}
so that the action becomes 
\begin{equation}\label{eq_bosonized_NonAbelian}
\begin{split}
S&=S_{\text{WZW}}\\
&+\int d^2x\left(\frac12 \pd_\mu \phi_+ \pd^\mu \phi_+ -\frac{g^2}{\pi }\min_{n\in\mathds{Z}}\left(\phi_++\frac{\theta +2\pi n}{2\sqrt{2\pi }}\right)^2
+\tilde{c}\,m\sqrt{\tilde{\mu}}\Re  \,N_{\tilde{\mu}}e^{i\sqrt{2\pi}\phi_+}\text{Tr}\left[:U: \right]\right)
\,.
\end{split}
\end{equation}
Note that we have been careful about the global quantization of the electric field when we wrote down the potential for $\phi_+$ that results from integrating out $A_\mu$ (see \cite{Komargodski:2020mxz}). 

For $m=0$, the bosonized action~\eqref{eq_bosonized_NonAbelian} provides a concrete illustration of the general result that a massless two-dimensional gauge theory with fermionic matter factorizes into a decoupled massive sector and a CFT~\cite{Kutasov:1994xq}. The massless sector coincides with the $k=1$ WZW model, which possesses an $\text{SO}(4) \cong \left[ \text{SU}(2)\times \text{SU}(2) \right]/\mathbb{Z}_2$ symmetry acting as $U\rightarrow g_L U g_R^\dagger$,\footnote{We mod out by $\mathbb{Z}_2$ because the diagonal $\mathbb{Z}_2$ center element $g_L = g_R = -\mathds{1}$ leaves $U$ invariant.} matching the flavor symmetry of the UV theory.  Here, $g_L$ and $g_R$ are $\text{SU}(2)$ unitary matrices in the fundamental representation.  In the massive sector, the excitation of the field $\phi_+$ consists of a massive particle, commonly denoted by $\eta$, with mass 
\begin{equation}\label{eq_eta_mass}
    \mu=\sqrt{\frac2\pi}g\,,
\end{equation}
which is a pseudoscalar meson.

To analyze the limit of small but nonzero mass, it is convenient to integrate out the
massive scalar $\phi_+$ and work within an EFT for the light pion field matrix $U$. To the leading non-trivial order in $m$ we simply set the massive field at its minimum, $\phi_+=-\theta/\sqrt{8\pi}$, and use re-normal ordering to compute the expectation value of the vertex operator:
\begin{equation}
   \sqrt{\tilde{\mu}} \langle N_{\tilde{\mu}} e^{i\sqrt{2\pi}\phi_+}\rangle_{m=0}=\sqrt{\mu}\langle N_{\mu} e^{i\sqrt{2\pi}\phi_+}\rangle_{m=0}=\sqrt{\mu}\,e^{-\frac{i\theta}{2}}\,.
\end{equation}
This leads to the action
\begin{equation}\label{eq_EFT_non_Abelian}
    S_\text{EFT}\simeq S_\text{WZW}+ \int d^2x\,\tilde{c}\,
    m\sqrt{\mu}\Re\left(e^{-i\theta/2}\text{Tr}\left[\normord{U}\right]\right)\,.
\end{equation}
We see that, for $\theta \neq \pi$, turning on identical masses for the fermions results in a perturbation linear in $U$, which gaps out the theory; this phenomenon is analogous to what happens in the QCD pion Lagrangian when we turn on identical quark masses. The difference is that in $d=2$ there is no spontaneous symmetry breaking, and the $m=0$ theory is a strongly coupled WZW model. As a result, the mass acquired by the pions at $m\neq 0$ cannot be calculated semi-classically. 

Nevertheless, we can acquire some intuition from \eqref{eq_EFT_non_Abelian}. Since the operator $\normord{U}$ has conformal dimension $\frac{1}{2}$ and coefficient $ m\sqrt{\mu}\cos\frac{\theta}{2}$ (since $\text{Tr}(U)=\text{Tr}(U^\dagger)$ for $N_f=2$), from dimensional analysis we infer that the pion mass scales as
 \begin{equation}
  M_\pi \sim \left( m \sqrt{\mu} \cos \frac{\theta}{2} \right)^\frac{2}{3} \sim \left( m \cos \frac{\theta}{2} \right)^\frac{2}{3} g^\frac{1}{3} \,,
  \end{equation} 
where we made use of \eqref{eq_eta_mass}. This is not the end of the story however, as we will see.

The $2\to 2$ scattering of pions near threshold is well-described by a short-distance interaction, due to the absence of long range forces. In the real world, this leads to formation of a broad $I=0$ resonance called $\sigma$, which is centered around $450$ MeV\@.  In the two-flavor Schwinger model, by analogy with QCD, we expect the short-distance potential to be attractive in the $I=0$ isospin channel and repulsive in the $I=2$ one~\cite{Weinberg:1966kf}.\footnote{Treating the action~\eqref{eq_EFT_non_Abelian} semiclassically as in the pion Lagrangian, one would find the $2$-to-$2$ amplitude $\mathcal{M}^{AB\to CD}
=
\delta^{AB}\delta^{CD}\,A(s,t,u)
+\delta^{AC}\delta^{BD}\,A(t,s,u)
+\delta^{AD}\delta^{BC}\,A(u,t,s)$, with
$A(s,t,u)=s-m_\pi^2$. Projecting on the $I=0$ and $I=2$ channels at threshold then we have $
\mathcal{M}^{I=0}=7m_\pi^2$ and $\mathcal{M}^{I=2}=-2m_\pi^2$.}  In one spatial dimension, a two-body attractive contact interaction is enough to create a non-trivial bound-state also in the absence of long-range forces, as is well-known from the study of the delta-function potential in non-relativistic quantum mechanics.\footnote{In modern language, the non-relativistic EFT describing the scattering of two pions near threshold always admits two different two-body contact interactions, corresponding to delta-function potentials in the different isospin channels. These are relevant operators in non-relativistic counting for $d<3$ and may lead to bound states if they are attractive.} We therefore expect that the spectrum also includes a pion-pion bound state $\sigma$, which is an isospin singlet and is $C$-even and $P$-even at $\theta=0$. Microscopically, the $\sigma$ particle may be interpreted at strong coupling as a tetraquark state.\footnote{Since the $\sigma$-particle has the same quantum numbers as the next-to-lightest isosinglet meson at weak-coupling, its wave function
at generic $m/g$
is a nontrivial mixture of two- and four-quark components~\cite{Harada:1993va, Delphenich:1997ex}.}

Remarkably, the above qualitative expectations are corroborated by exact integrability results. To carry out the following calculations, it is convenient to revert to Abelian bosonization~\cite{Coleman:1974bu,Mandelstam:1975hb} and trade the pion matrix $U$ for a compact scalar $\phi_-$:\footnote{
\label{footnote_periodicity}
With this normalization the periodicity of the scalars translates into the following identifications
\begin{equation*}
\begin{split}
(\phi_+,\phi_-)& \sim (\phi_+\pm\sqrt{2\pi},\,\phi_-)\sim (\phi_+,\,\phi_-\pm\sqrt{2\pi}) \\
&\sim \left(\phi_+\pm\sqrt{\pi/2},\,\phi_-\pm\sqrt{\pi/2}\right)\sim \left(\phi_+\pm\sqrt{\pi/2},\,\phi_-\mp\sqrt{\pi/2}\right)\,.
\end{split}
\end{equation*}} 
\begin{equation}\label{eq_bosonized}
\begin{split}
\mL&=\frac12 \pd_\mu \phi_+ \pd^\mu \phi_+-\frac{\mu^2}{2 }\min_{n\in\mathds{Z}}\left(\phi_++\frac{\theta +2\pi n}{2\sqrt{2\pi }}\right)^2\\
&+\frac12 \pd_\mu \phi_- \pd^\mu \phi_-
+\frac{e^{\gamma_E}}{\pi}m\,\sqrt{\mu\mu_-} \,N_{\mu}\cos\left(\sqrt{2\pi}\phi_+\right)
N_{\mu_-}\cos\left(\sqrt{2\pi}\phi_-\right)
\,,
\end{split}
\end{equation}
where $\mu_-$ is an arbitrary mass scale, whose value is unphysical due to renormal-ordering. Integrating out the massive field $\phi_+$ systematically as in~\cite{Dempsey:2023gib}, we obtain the following Lorentzian Lagrangian 
\begin{equation}\label{eq_EFT_2flavor}
\begin{split}
    \mL_\text{EFT}&\simeq \frac{1-\delta}{2}\pd_\mu \phi_- \pd^\mu \phi_- +\cos\left(\frac{\theta}{2}\right)\frac{e^{\gamma_E}}{\pi}m\sqrt{\mu\,\mu_-} N_{\mu_-}\cos(\sqrt{2\pi}\phi_-)\\
    &-\frac{e^{2\gamma_E}}{4\pi}\delta\, \mu_-^2N_{\mu_-}\cos(\sqrt{8\pi}\phi_-)\,.
\end{split}
\end{equation}
As we will explain in detail in the next section, $\delta$ multiplies a marginal bilinear in the $\text{SU}(2)\times \text{SU}(2)$ Noether currents, and it is given by
\begin{equation}\label{eq_EFT_delta}
\delta=-\frac{e^{\gamma_E} m^2}{2\pi \mu^2}I_s(\theta)\,,
\end{equation}
where
\begin{equation}\label{eq_Is}
    I_s(\theta)=2\pi \int_0^{\infty}d\xi\,\xi^2\left(\frac{e^{K_0(\xi)}+e^{-K_0(\xi)}\cos\theta }{2}-\cos^2\frac{\theta}{2}\right) \simeq 5.55418 - 4.52885\cos\theta\,.
\end{equation}
At $\theta=0$ we find $I_s(0)\simeq 1.0253$. See Appendix~\ref{app_Large_m_theta0} for the derivation of~\eqref{eq_EFT_2flavor}.  As we will explain in detail in the next section, the Lagrangian~\eqref{eq_EFT_2flavor} neglects contributions of order $\mO(m^3)$ as well as irrelevant operators which are generated at order $\mO(m^2)$ and at subleading order in $g$.

For $\theta\neq \pi$ and at leading order in $m$, we can set $\delta=0$ and we are left with an integrable $\text{SO}(3)$-symmetric sine-Gordon theory. The spectrum of the sine-Gordon theory consists precisely of an isotriplet $\vec{\pi}$ and a heavier isosinglet $\sigma$.  In Appendix \ref{app_Large_m_theta0} we use integrability to compute the isotriplet mass
\begin{equation}\label{eq_Mpi}
\begin{split}
    M_\pi &= \frac{2^\frac{7}{6} e^\frac{\gamma_E}{3} \Gamma \left(\frac{1}{6}\right)}{\Gamma \left(\frac{2}{3}\right) \Gamma \left(\frac{1}{4}\right)^\frac{4}{3}}g^{\frac{1}{3}}\left(m \cos\left(\frac{\theta}{2}\right)\right)^{\frac{2}{3}}
    \left(1 +\mathcal{O}\left(\frac{m^2}{g^2}\right)\right)\\
    &\simeq 2.008 \,g^{\frac{1}{3}}\left(m \cos\left(\frac{\theta}{2}\right)\right)^{\frac{2}{3}} \left(1 +\mathcal{O}\left(\frac{m^2}{g^2}\right)\right)\,,
\end{split}
\end{equation}
which agrees with \cite{Smilga:1996pi}.  We also compute the $\sigma/\pi$ mass ratio: 
\begin{equation}\label{eq_ratio}
    \frac{M_\sigma^2}{M_{\pi}^2}= 3-\frac{\pi  e^{\gamma_E}m^2}{\sqrt{3}g^2} I_s(\theta)+\mathcal{O}\left(\frac{m^\frac{10}{3}}{g^\frac{10}{3}}\right)\,,
\end{equation}
where we incorporated the $\mO(m^2)$ corrections from the terms of order $\delta$ in~\eqref{eq_EFT_2flavor}.  The leading order result for the ratio~\eqref{eq_ratio} was already suggested in~\cite{Coleman:1976uz} based on the semiclassical analysis of the sine-Gordon theory~\cite{Dashen:1974cj}. The correction~\eqref{eq_ratio} to the sigma/pion mass ratio is a new result to the best of our knowledge; we derive it in Appendix~\ref{app_Large_m_theta0}, making use of the perturbative approach to deformed integrable models developed in~\cite{Delfino:1996xp}. Schematically, denoting by $\mO$ the operator multiplied by $\delta$, the corrections to the masses arise analogously to standard quantum-mechanical perturbation theory, taking the form $\Delta M^2_{a} \sim \delta \langle a | \mO | a \rangle$, where $a=\pi,\sigma$ labels the particles. By dimensional analysis the one-particle matrix element scales as $\langle a | \mO | a \rangle \sim M_\pi^2$, since $\mO$ has dimension $2$ and $M_\pi \sim M_\sigma$ is the only mass scale. In practice, since the $\mO(m^2)$ perturbation is marginal, this calculation requires a nontrivial regularization procedure, but the final result is scheme-independent as we explain in Appendix~\ref{app_Large_m_theta0}. The $\mO(m^{10/3}/g^{10/3})$ corrections are due to higher dimensional operators, whose form will be discussed in the next section.

The next particle after $\pi$ and $\sigma$ is the $\eta$ meson, which has quantum numbers $I^{PG}=0^{--}$. It is the lightest $C$-odd state; therefore, it remains stable also for nonzero $m$ at $\theta=0$. Indeed, in the bosonized Lagrangian \eqref{eq_bosonized_NonAbelian}, charge conjugation acts as $\phi_+\to-\phi_+$, thus protecting the lightest excitation associated with the $\phi_+$ field. Furthermore, the $\pi$--$\eta$ interaction is attractive at the semiclassical level, as can be seen by expanding the last term in \eqref{eq_bosonized_NonAbelian}. This attraction is expected to produce an $\eta\pi$ bound state with quantum numbers $1^{+-}$. 
In fact, a state with these quantum numbers can be seen in Figure \ref{fig:mass_triplet_singlet_theta0}.

Interactions may similarly generate a stable $\eta\sigma$ bound state, furnishing an additional stable $0^{--}$ particle. These heavier states, however, lie beyond the regime of validity of the EFT \eqref{eq_EFT_2flavor} and thus cannot be analyzed reliably using the integrability techniques employed for the lowest-lying particles.

Near $\theta = \pi$ the strong coupling analysis described above breaks down since the coefficient of the leading relevant operator in~\eqref{eq_EFT_2flavor} vanishes. As we shall explain in Section~\ref{sec_theta_Pi}, this breakdown occurs only in an exponentially small region near $\theta=\pi$.

\subsection{Resolution of Coleman's second puzzle}

Let us conclude our analytical arguments by explaining the resolution to Coleman's second puzzle regarding the comparison of the $\theta = 0$ spectrum between weak and strong coupling.  As discussed above, at weak coupling we have, in order, the following particle excitations (for which we list the values of $I^{PG}$): 
\begin{itemize}
     \item $1^{-+}$ (the triplet with $n=0$ in \eqref{eq_E_bound}, identified as $\pi$);  
     \item $0^{--}$ (the singlet with $n=0$, identified as $\eta$);  
     \item $1^{+-}$ and $0^{++}$ (the triplet and singlet (identified as $\sigma$) with $n=1$); 
     \item $1^{-+}$ (the triplet with $n=2$);  
     \item $0^{--}$ (the singlet with $n=2$); 
     \item $1^{+-}$ and $0^{++}$ (the triplet and singlet with $n=3$), 
     \item etc.
\end{itemize}
At strong coupling, there are
fewer states: 
\begin{itemize}
    \item $1^{-+}$ (the $\pi$) 
    \item a $\pi\pi$ $0^{++}$ bound state (the $\sigma$) 
    \item $0^{--}$ (the $\eta$)
    \item perhaps an $\eta \pi$ $1^{+-}$ bound state
    \item perhaps an $\eta \sigma$ $0^{--}$ bound state (which could also be called an $\eta \pi\pi$ bound state)
\end{itemize}
While the quantum numbers $1^{-+}$ of the lowest states (the pions) match between weak and strong coupling, Coleman was puzzled about the fact that the quantum numbers of the second lowest state are different ($0^{--}$ at weak coupling and $0^{++}$ at strong coupling).
The puzzle arose from the expectation that the lightest mesonic bound states of two fermions are always parity-odd.  But, as we have seen above, in contrast to the weakly-coupled theory, at strong coupling the next-to-lightest particle is the $\sigma$ particle and it is \emph{not} a conventional meson.  Rather, it is best understood as a bound state of two mesons analogous to a tetraquark state.
By comparing the weak-coupling and strong-coupling pictures, we see that there should be a level crossing at some finite $m/g$ between the pseudoscalar meson $\eta$ and the scalar, tetraquark-like $\sigma$. In the next section, we provide strong numerical evidence for such a level crossing.

\subsection{Numerical results for arbitrary coupling}

To obtain our numerical results we consider a Hamiltonian lattice version of the model \eqref{eq_Lag} using staggered fermions as in~\cite{Kogut:1974ag,Banks:1975gq, Steinhardt:1977tx, Banuls:2016gid, Itou:2023img, Itou:2024psm, Funcke:2023lli}:
\begin{equation}\label{lattice_H}
\begin{split}
H=&\frac{g^2 a}{2}\sum_{n} \left(L_n+\frac{\theta}{2\pi}\right)^2+\sum_{\alpha=1}^2 \sum_{n} m_{\text{lat}, \alpha } (-1)^n\cdag_{n,\alpha}c_{n,\alpha}\\
&-\frac{i}{2a}\sum_{\alpha=1}^2\sum_{n}\left(
\cdag_{n,\alpha} U_n c_{n+1,\alpha}-\cdag_{n+1,\alpha}U_n^\dagger c_{n,\alpha}\right)\,,
\end{split}
\end{equation}
where $a$ is the lattice spacing and $n$ labels the lattice site. In the above equation $\cdag_{n,\alpha}$ and $c_{n,\alpha}$ are the fermionic creation and annihilation operators, respectively, at site $n$, while the gauge field variables $U_n$, $L_n$ live on the link $(n,n+1)$. The canonical commutation/anti-commutation relations are
\begin{equation}
[L_n,U_m]=\delta_{nm}U_m\,,\qquad
\{c_{n,\alpha},\cdag_{m,\beta}\}=\delta_{nm}\delta_{\alpha\beta}\,.
\end{equation}
We work in conventions where Gauss's law enforces
\begin{equation}
(L_n-L_{n-1})\ket{\text{phys}}=Q_n\ket{\text{phys}}\,,\qquad
Q_n=\sum_{\alpha=1}^{2}\left(\cdag_{n,\alpha}c_{n,\alpha}-\frac{1-(-1)^n}{2}\right)
\end{equation}
for any physical state $\ket{\text{phys}}$.  As shown in~\cite{Dempsey:2022nys,Dempsey:2023gib}, with this choice of regularization of the electric charge $Q_n$ at site $n$, an improved choice of the lattice mass that preserves some continuum properties of discrete axial transformations is
\begin{equation}\label{eq:mass_shift}
m_{\text{lat}, \alpha}=m_\alpha-\frac{g^2 a}{4}\,.
\end{equation}
With this choice, the physical sector of the Hamiltonian specified by $Q=\sum_n Q_n=0$ satisfies
\begin{equation}
{\cal V} H_{m_1, m_2, \theta} {\cal V}^{-1}
= H_{-m_1, -m_2, \theta}\ ,
\end{equation}
where ${\cal V}$ is the operator that implements the unit lattice translation, which was constructed in 
\cite{Dempsey:2022nys,Dempsey:2023gib}.  In the continuum, this transformation maps to $\psi_\alpha\rightarrow\gamma^5\psi_\alpha$, and hence allows preserving a discrete subgroup of the axial symmetry on the lattice.\footnote{More precisely, the continuum limit of $\mathcal V$ is the product of an infinitesimal translation and a $\mathbb Z_2$ chiral rotation in accordance with the group structure of unit lattice translation $\mathbb Z$.} For comparison, we note that for $N_f=1$, the mass shift is 
$m_\text{lat}=m-\frac{g^2 a}{8}$ \cite{Dempsey:2022nys}. In this case, 
$  {\cal V} H_{m,\theta} {\cal V}^{-1}
= H_{-m, \theta + \pi}$, and the shift of $\theta$ implements an anomalous discrete axial transformation \cite{Schwinger:1962tp, Johnson:1963vz}.

To diagonalize the Hamiltonian, we use the infinite matrix product states construction for gauge theories recently developed in \cite{Dempsey:2025wia}. This approach allows us to approximate the ground state of the Hamiltonian directly on an infinite lattice, using the Variational Uniform Matrix Product State (VUMPS) algorithm~\cite{Zauner-Stauber:2017eqw}. Furthermore, it is possible to estimate the energy of excited states using the quasiparticle ansatz \cite{Haegeman:2011lcd} for both particle and solitonic excitations. We refer the reader to~\cite{Dempsey:2025wia} for further details on the numerical implementation of the ground state and particle excitations. More details on the numerical implementation for solitonic excitations will appear in forthcoming work. All numerical results reported are continuum values: quantities are calculated using the Hamiltonian \eqref{lattice_H} over an array of finite lattice spacings $a$ and extrapolated to the continuum limit $a\rightarrow 0$. Below we discuss our results.

\begin{figure}[ht]
    \centering
    \includegraphics[width=.8\linewidth]{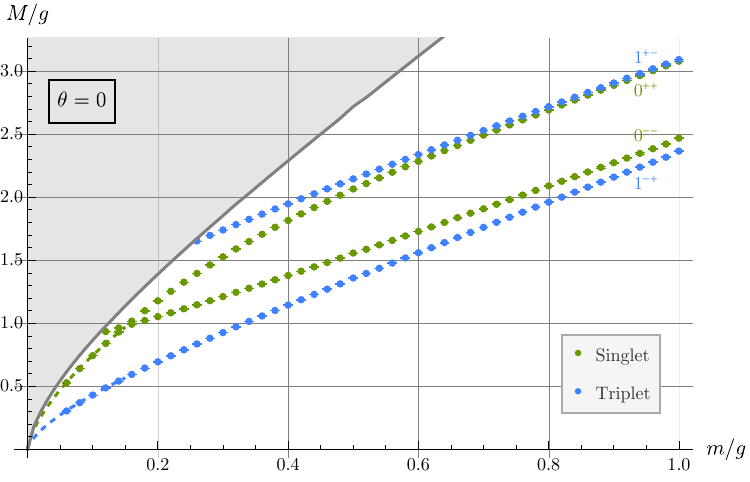}
    \caption{The masses of the two lightest isotriplets and the two lightest isosinglets at $\theta = 0$. We infer the $P$ and $G$ quantum numbers from our weak coupling analysis, and label each state with its $I^{PG}$. The two-pion continuum is shaded in gray. The dashed blue line at small $m/g$ is the strong-coupling prediction \eqref{eq_Mpi}, and the dashed green line is the same value multiplied by the ratio \eqref{eq_ratio}. Both agree very well with the lattice data. We see that there is a level crossing in the isosinglet sector at $m/g\approx 0.15$: above this value, the lightest singlet has negative parity, while below this value it has positive parity (at very strong coupling, this is the $\sigma$ meson, which is a bound state of two pions).\label{fig:mass_triplet_singlet_theta0}}
\end{figure}

In Figure~\ref{fig:mass_triplet_singlet_theta0} we show our numerical results for the four lightest states, two isotriplets and two isosinglets, as a function of $m/g$  at $\theta=0$. In the weak coupling region, the first isosinglet is heavier than the first isotriplet, in agreement with \eqref{eq_stsplitting}. The second singlet-triplet splitting is much smaller, and there the singlet is the lighter state.
Around $m/g\simeq 0.15$ we observe a level crossing between the positive and negative parity isosinglets.\footnote{Incidentally, the level crossing point is not too far from the naive estimate $m/g\sim 0.11$ obtained by equating the integrability result for $M_{\sigma}$ at strong coupling, which follows from~\eqref{eq_Mpi} and~\eqref{eq_ratio}, with the $m=0$ result for the $\eta$-mass~\eqref{eq_eta_mass}.} This level crossing, which was first observed with light-cone methods \cite{Harada:1993va}, solves Coleman's second puzzle.  In the strong coupling regime~$m/g\lesssim 0.15$, 
the lightest isosinglet is the positive parity state $\sigma$. 

Note that our numerical algorithm does not distinguish states with different $P$ and $G$ quantum numbers. In Fig.~\ref{fig:mass_triplet_singlet_theta0}, these quantum numbers are assigned by comparing the numerical spectrum with the weak-coupling analysis. For the same reason, in the strong-coupling regime we cannot reliably identify the $\eta$ or other heavier $G$-odd states 
within the two-pion continuum. Similarly, the lightest $1^{+-}$ state is below the $\eta\pi$ threshold where it can be resolved numerically. Its continuation to strong coupling is likely to 
be the stable bound state discussed in Section \ref{sec:analytical}.

Note also that the first singlet-triplet mass splitting in Figure~\ref{fig:mass_triplet_singlet_theta0} decreases for $m/g\gtrsim 0.2$ as the mass increases, in qualitative agreement with the non-relativistic prediction \eqref{eq_stsplitting}, which states that it should vanish as $g (g/m)^{5/3}$. Another notable feature is that the singlet-triplet splitting for the excited states is smaller than that between the two lowest states at weak coupling; this is consistent with the non-relativistic wave-function of the excited state being odd, so that the energy of the singlet does not get corrected by the contact interaction~\eqref{eq_DeltaH_NR} (see Appendix~\ref{app_weak_coupling_NR} for further details). 

\begin{figure}[t]
    \centering
    \includegraphics[width=0.65\linewidth]{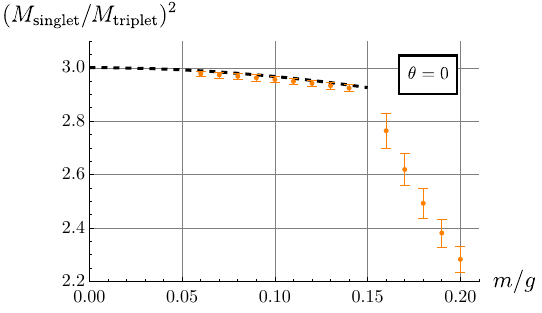}
    \caption{The mass ratio between the lightest singlet and triplet states. The dashed curve is the theoretical prediction~\eqref{eq_ratio}.}
    \label{fig:mass_ratio_theta0}
\end{figure}

The numerical results for $M_{\pi}$ and $M_{\sigma}$ in Figure~\ref{fig:mass_triplet_singlet_theta0} are in excellent agreement with the theoretical predictions~\eqref{eq_Mpi} and~\eqref{eq_ratio} at strong coupling. In Figure~\ref{fig:mass_ratio_theta0} we also plot the ratio of the masses of the lightest singlet and triplet states at $\theta = 0$, and compare it with the theoretical prediction~\eqref{eq_ratio}, again finding good agreement up to the level-crossing discussed above.

\begin{figure}[t]
    \centering
    \begin{subfigure}{0.49\textwidth}
    \includegraphics[width=\linewidth]{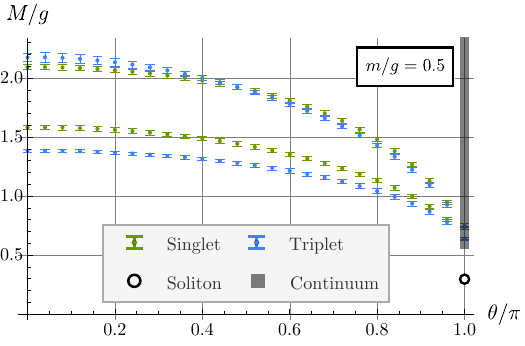}
    \caption{}
    \label{fig:mass_degA}
    \end{subfigure}
    \hfill
    \begin{subfigure}{0.49\textwidth}
        \centering
    \includegraphics[width=\linewidth]{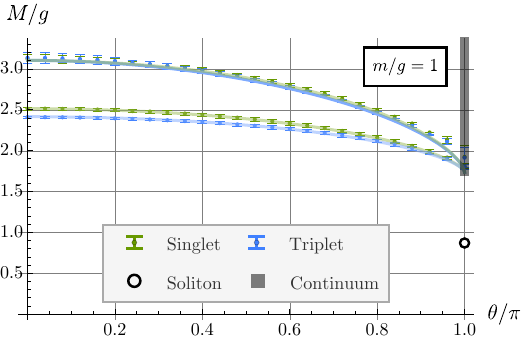}
    \caption{}
    \label{fig:mass_degB}    
    \end{subfigure}
    \caption{The mass of the two lightest isotriplets and isosinglets at $m/g=0.5$ (Figure \ref{fig:mass_degA}), and $m/g=1$ (Figure \ref{fig:mass_degB}). In both cases, the first singlet and triplet have negative parity, while the second singlet and triplet have positive parity. The continuous lines in Figure~\ref{fig:mass_degB} are the numerical results obtained solving the two-particle 't Hooft equation discussed in Appendix~\ref{app_tHooft}. At $\theta \to \pi$, all of these particle masses are expected to approach the two-soliton continuum threshold; we have marked the soliton mass and shaded the two-soliton continuum.}
    \label{fig:mass_deg}
\end{figure}

In Figure~\ref{fig:mass_deg}, we show the masses of the two lowest isotriplets and isosinglets, which are all below the continuum threshold, as functions of $\theta$ at $m/g = 0.5$, Figure~\ref{fig:mass_degA}, and $m/g=1$, Figure~\ref{fig:mass_degB}. Until a few years ago, numerical studies of masses as functions of $\theta$ struggled to reach $\theta = \pi$ (see, for example, \cite{Fukaya:2003ph}). More recently, results for the lowest singlet and triplet states covering the entire range of $\theta$ have been obtained in \cite{Itou:2023img, Itou:2024psm}. Here, we achieve even higher precision and obtain results for two additional stable mesons using the new method of \cite{Dempsey:2025wia}.

As discussed above, the singlet-triplet mass ratio approaches unity as $\theta \to \pi$ at weak coupling. Notably, Figure~\ref{fig:mass_deg} shows that a similar behavior persists at relatively strong coupling, with all states becoming degenerate at $\theta = \pi$. We will discuss the strong-coupling interpretation of this phenomenon in the next section. 

In interpreting Figure~\ref{fig:mass_deg} near $\theta=\pi$, a comment about the numerical implementation is in order: the quasiparticle ansatz is a matrix product state construction that models a bound state as a localized  perturbation of the approximate vacuum state. The excitation spectrum is then extracted using the Rayleigh-Ritz method with the quasiparticle ansatz as the finite basis. This implies that the energies obtained using the quasiparticle ansatz can only be trusted when the true energy eigenstates are localized in this sense. In particular, that means that the energy of scattering and very loosely bound states cannot be accurately determined using the quasiparticle ansatz. As mentioned in the Introduction and demonstrated in the weak coupling limit, the theory deconfines at $\theta=\pi$ and the bound states decay into soliton-antisoliton pairs. These scattering states cannot be modeled by the quasiparticle ansatz which explains why the masses in Figure~\ref{fig:mass_deg} do not all converge to the two soliton threshold as one would expect from deconfinement. By continuity the data points near $\theta=\pi$ corresponds to very loosely bound states and similarly poorly captured by the quasiparticle. Nevertheless, the energies near $\theta=\pi$ do serve as a proxy for the true masses as evidenced by the fact that they get very close to the correct value of the two soliton threshold.

To obtain a quantitative description of the results in Figure~\ref{fig:mass_deg}, in Appendix~\ref{app_tHooft} we discuss a relativistic 't~Hooft equation obtained by restricting the lightcone Hamiltonian to the two-fermion sector of the theory. While, strictly speaking, this approach is only justified at weak coupling, similarly to the non-relativistic Hamiltonian~\eqref{eq_H_NR}, it allows one to solve for the spectrum in a manifestly relativistic framework, accounting for some loop corrections. We therefore expect it to extrapolate more reliably than a purely non-relativistic treatment to moderate values of $m/g$. The numerical results at $m/g = 1$ in Figure~\ref{fig:mass_degB} are in excellent agreement with the predictions of the 't~Hooft equation. 

A larger discrepancy is observed at $m/g = 0.5$, and for this reason we do not display the 't~Hooft equation results in Figure~\ref{fig:mass_degA}. This discrepancy is explained by the fact that the next-to-lightest singlet state evolves from a purely mesonic particle at weak coupling into a nontrivial pion-pion bound state at strong coupling. We therefore expect its wave function to be a nontrivial mixture of mesonic and tetraquark components for $m/g \lesssim 1$, and accordingly a treatment based on a two-fermion approximation cannot be accurate in this regime.

\section{Unequal masses for \texorpdfstring{$\theta\neq \pi$}{θ≠π}}
\label{isobreak}

In this section, we address Coleman's first puzzle (see Section~\ref{threepuzzles} for the description of the puzzle). This puzzle was discussed in \cite{Delphenich:1997ex, Georgi:2020jik, Albandea:2025dhu}, but our treatment will be different. We consider unequal fermion masses, $m_1\neq m_2$, and show that the splittings within the $\text{SO}(3)$ multiplets of bound states are suppressed in the strong coupling limit.  As discussed in the previous section, when $m_1 = m_2$, the spectrum at strong coupling consists of a pion triplet of mass $M_\pi$ and a $\sigma$ particle with mass $M_\sigma \approx \sqrt{3} M_\pi$ (see \eqref{eq_ratio}).  Once we turn on a small fermion mass difference $m_1 - m_2 \neq 0$, we find that, at $\theta=0$, all these masses acquire corrections, and in addition the triplet of pions gets split, with\footnote{This differs from the corresponding $\theta=0$ 
formula in \cite{Albandea:2025dhu}, which gives $\frac{(m_1-m_2)^2}{g^\frac{4}{3}(m_1+ m_2)^\frac{2}{3}}$.
We also disagree with the
claim in \cite{Georgi:2020jik} that the mass difference is exponentially small.} 
\begin{equation}\label{eq_ratio_dm_intro}
\frac{M_{\pi^\pm}^2}{M_{\pi^0}^2}-1\propto  \frac{(m_1-m_2)^2}{g^2}\,. 
\end{equation}
At $\theta\neq 0$, the difference $M^2_{\pi^\pm}/M^2_{\pi^0}-1$ also scales as $1/g^2$, but with a more involved dependence on the masses and $\theta$.

\subsection{Strong coupling EFT and its symmetries}

The scaling quoted in \eqref{eq_ratio_dm_intro} follows from a general symmetry analysis.   In general, the two-flavor Schwinger model has $8$ real mass parameters that can be combined into a $2 \times 2$ complex-valued matrix $\widetilde{M}$, with the most general Lagrangian density being\footnote{As we will see shortly, due to the anomalous axial transformation, $\theta$ and $\arg \det \widetilde{M}$ are not individually physical, but only the combination $\arg \det \widetilde{M}-\theta$ is.}
\begin{equation}\label{eq_Lag_general_masses}
 \mL=
    i\bar{\psi}^\alpha\slashed{D}\psi_\alpha-\left( \widetilde{M}_{\alpha}{}^{\beta}\bar{\psi}^\alpha P_L\psi_\beta + (\widetilde{M}^\dagger)_{\alpha}{}^{\beta}\bar{\psi}^\alpha P_R\psi_\beta \right)-\frac{1}{4g^2}F_{\mu\nu}^2-\frac{\theta}{4\pi}\varepsilon^{\mu\nu}F_{\mu\nu}\,,
\end{equation}
where $P_{L}=(1-\gamma^5)/2$ and $P_{R}=(1+\gamma^5)/2$ are the chiral projectors. 
Without loss of generality, by appropriate field rotations we could focus on a Hermitian mass matrix of the form
\begin{equation} \label{tildeMDef}
    \widetilde{M}=m\,\mathds{1}+\delta m\,\sigma^3\,, \qquad m = \frac{m_1+m_2}{2} \,, \qquad \delta m = \frac{m_1 - m_2}{2} \,,
\end{equation}
in which case we obtain the theory with two mass parameters~\eqref{eq_Lag} we have been discussing.  However, it is beneficial to keep $\widetilde{M}$ general for now.

As in the previous section, it is useful to use bosonization and write the analog of \eqref{eq_bosonized_NonAbelian} with  arbitrary mass parameters.  This generalization can be obtained by replacing $m \text{Tr}(:U:)$ in \eqref{eq_bosonized_NonAbelian} by $\text{Tr}(\widetilde{M} :U:)$.  If we further shift the field $\phi_+$ by $\phi_+ \to \phi_+ - \theta/ (2 \sqrt{2 \pi})$, we can write the bosonized action as
\begin{equation}\label{eq_bosonized_NonAbelian2}
 \begin{split}
   S =S_{\text{WZW}} 
   +\int d^2x \, &\biggl[
    \frac12 \pd_\mu \phi_+ \pd^\mu \phi_+ 
    -\frac{g^2}{\pi} \min_{n\in\mathds{Z}}
    \left(\phi_++\frac{2\pi n}{2\sqrt{2\pi }}\right)^2 \\
  {}&+\tilde{c}\,\sqrt{\tilde{\mu}}
     \Re  \left( \,N_{\tilde{\mu}}
     e^{i\sqrt{2\pi}\phi_+}
     \text{Tr}\left(e^{-\frac{i\theta}{2}}\widetilde{M} :U: \right) \right) \biggr]
\,,
 \end{split}
\end{equation}
where $S_{\text{WZW}}$ is the action of the $\text{SU}(2)_1$ WZW model in \eqref{SWZW}.  We thus see that there is some redundancy in the parameters $\widetilde{M}$ and $\theta$ because the bosonized action depends only on the physical mass matrix
\begin{equation}\label{eq_M_matrix_def}
    M \equiv e^{-\frac{i\theta}{2}}\widetilde{M} \,.
\end{equation}
The origin of this redundancy is the axial anomaly.  

When $M = 0$, the theory has $\text{SO}(4) \cong \left[ \text{SU}(2)\times\text{SU}(2)\right]/\mathbb{Z}_2$ symmetry\footnote{As mentioned before, we mod out by  $\mathbb{Z}_2$ because the diagonal center element $g_L = g_R = -\mathds{1}$ preserves both $U$ and (further below) $M$.} that leaves $\phi_+$ invariant and acts on $U$ by $U \to g_L U g_R^\dagger$, where $g_L$ and $g_R$ are unitary matrices in the fundamental representations of the second and first $\text{SU}(2)$ factors, respectively, in the product $\text{SO}(4) \cong \left[ \text{SU}(2)\times\text{SU}(2)\right]/\mathbb{Z}_2$. When $M \neq 0$, the term on the second line of \eqref{eq_bosonized_NonAbelian2} breaks this $\text{SO}(4)$ symmetry.  However, we can view the matrix $M$ as a \emph{spurion}:  the full $\text{SO}(4)$ symmetry of the massless theory is then restored, provided that we assign a transformation rule $M\rightarrow g_R Mg_L^\dagger$ to the matrix $M$.

The basic physical phenomenon underlying Coleman's first puzzle follows from the structure of the low-energy EFT that is obtained by integrating out $\phi_+$ in \eqref{eq_bosonized_NonAbelian2}.  This EFT takes the form
 \begin{equation} \label{EffAction}
     S_\text{EFT} = S_\text{WZW} + \int d^2x\, \left( {\cal L}_1 + {\cal L}_2 + {\cal L}_3 + \cdots \right) \,,
 \end{equation}
where ${\cal L}_n$ is an expression of degree $n$ in $M$.  The question is what terms we can write down at each order in $M$.

The symmetry which restricts the possible terms is the $\text{SO}(4)$ symmetry mentioned above, as well as charge conjugation $C$ and parity $P$.  These are all symmetries of the full theory \eqref{eq_bosonized_NonAbelian2}, assuming that $M$ is treated as a spurion field.  Under $\text{SO}(4)$, we have $(U, M) \to (g_L U g_R^\dagger, g_R M g_L^\dagger)$, where $g_L$ and $g_R$ were mentioned above, and under $P$ and $C$ we have $(U,M)\stackrel{P}{\longrightarrow}( U^\dagger,M^\dagger)$ and $(U,M)\stackrel{C}{\longrightarrow}( U^*,M^*)$, respectively.  It will be easier to classify the possible terms in the EFT if we decompose $U$ and $M$ into three real four-vectors $\vec{u}$, $\vec{m}_R$, and $\vec{m}_I$ as
\begin{align}
U &=u^4+i\sum_{A=1}^3\sigma_A u^A\,,\quad
\vec{u}\cdot\vec{u}=1\,,\\
M&=(m^4_R+i\,m^4_I)+i\sum_{A=1}^3 (m^A_R+i\,m^A_I)\sigma_A\,.
\end{align}
These four-vectors transform in the fundamental (vector) representation of $\text{SO}(4)$.  The discrete $P$ and $C$ symmetries mentioned above act by
\begin{equation}\label{eq_discrete_UM}
\begin{aligned}
(\vec{u},\vec{m}_R,\vec{m}_I)&\stackrel{P}{\longrightarrow}(P\vec{u},P\vec{m}_R,-P\vec{m}_I)\,,
\qquad P=\text{diag}(1,-1,-1,-1)\,,\\
(\vec{u},\vec{m}_R,\vec{m}_I)&\stackrel{C}{\longrightarrow}(C\vec{u},C\vec{m}_R,-C\vec{m}_I)\,,
\qquad C=\text{diag}(-1,1,-1,1)\,.
\end{aligned}
\end{equation}

The terms ${\cal L}_n$ in the effective action \eqref{EffAction} are linear combinations of the operators of the $\text{SU}(2)_1$ WZW model, which is a CFT\@.  The low-dimension scalar operators of this CFT are, in order of increasing scaling dimension:  
 \begin{itemize}
     \item ${\cal O}_{1/2}^a$, $a = 1, \ldots, 4$, with scaling dimension $\Delta = 1/2$ transforming in the fundamental ($4$-dimensional) representation of $\text{SO}(4)$.  This operator is what the WZW variable $U$ flows to.  It is a Kac-Moody primary of both the left and right copies of the $\text{SU}(2)_1$ Kac-Moody algebra;
     \item ${\cal O}_{2}^{ab}$ with scaling dimension $\Delta = 2$ transforming in the rank-$2$ traceless symmetric tensor representation of $\text{SO}(4)$.  It is a Virasoro primary and a Kac-Moody descendant of the identity operator.  In particular, it can be thought of as a normal ordered product of the two $\text{SU}(2)$ currents, with their Lorentz indices contracted so as to make a Lorentz scalar;\footnote{Note that the currents are chiral, hence there is a unique Lorentz invariant contraction $\normord{J_L^A \bar J_R^B}$. This operator can be recast as an $\text{SO}(4)$ rank-two traceless symmetric tensor using the self-dual and anti-self-dual 't Hooft symbols $\eta^A_{ab}$ and $\bar{\eta}^A_{ab}$ as $\mO_{2\,ab}
\equiv
\normord{J_L^A \bar J_R^B}
\left(
\eta^A_{ac}\,\bar\eta^B_{cb}
+
\eta^A_{bc}\,\bar\eta^B_{ca}
\right)$.}
     \item ${\cal O}_{5/2}^{a}$ with scaling dimension $\Delta = 5/2$ transforming in the fundamental representation of $\text{SO}(4)$. The operator ${\cal O}_{5/2}^{a}$ is a Virasoro descendant of ${\cal O}_{1/2}^a$;
     \item etc.
 \end{itemize}
Under $P$,  these operators transform as
 \begin{equation}
  \begin{split} \label{PTransf}
     {\cal O}^a_{1/2} \stackrel{P}{\longrightarrow} P^{aa'}   {\cal O}^{a'}_{1/2} \,, \qquad
     {\cal O}^{ab}_{2} \stackrel{P}{\longrightarrow} P^{aa'} P^{bb'}   {\cal O}^{a'b'}_{2} \,, \qquad
     {\cal O}^{a}_{5/2} \stackrel{P}{\longrightarrow} P^{aa'}   {\cal O}^{a'}_{5/2} \,, 
  \end{split}
 \end{equation}
etc., where the $P$ matrix was defined in \eqref{eq_discrete_UM}.  The transformation under $C$ is similar to \eqref{PTransf}, with $P$ replaced with $C$.

It is then straightforward to write down all operators allowed by $\text{SO}(4)$, $P$, and $C$ symmetry.  At first order in $M$, we have: 
 \begin{equation} \label{L1}
     {\cal L}_1 = a_1\mu^{1/2} \vec{m}_R \cdot \vec{{\cal O}}_{1/2} + 
       a_2 \mu^{-3/2}\vec{m}_R \cdot \vec{{\cal O}}_{5/2} + \cdots \,,
 \end{equation}
where $a_1$, $a_2$, \ldots are constant coefficients, the powers of $\mu = \sqrt{\frac{2}{\pi}} g$ (see \eqref{eq_eta_mass}) are fixed by dimensional analysis, and the ellipses denote higher dimension operators. We will derive $\mL_1$ later, and we will find that only the first coefficient is non-zero:
\begin{equation}\label{L1_coeff}
    a_1=\tilde c \,,\quad
    a_2=a_3=\ldots=0\,.
\end{equation}

At second order in $M$, we have 
 \begin{equation} \label{L2}
     {\cal L}_2 = b_1 \mu^{-2} m_R^a m_R^b {\cal O}_2^{ab}
      + b_2 \mu^{-2} m_I^a m_I^b {\cal O}_2^{ab} + \cdots \,,
 \end{equation}
with coefficients $b_1$, $b_2$, \ldots  At third order in $M$, we have
 \begin{equation}\label{L3}
   {\cal L}_3 = c_1 \mu^{-3/2}\vec{m}_R^2 \vec{m}_R \cdot \vec{{\cal O}}_{1/2} + c_2 \mu^{-3/2} \vec{m}_I^2 \vec{m}_R \cdot \vec{{\cal O}}_{1/2} + c_3 \mu^{-3/2} \vec{m}_R \cdot \vec{m}_I \vec{m}_I \cdot \vec{{\cal O}}_{1/2} + \cdots \,,
 \end{equation}
with coefficients $c_1$, $c_2$, \ldots, and so on.

\subsection{Resolution of Coleman's first puzzle}

The resolution to Coleman's first puzzle follows from the fact that ${\cal L}_1$ preserves an $\text{SO}(3)$ subgroup of $\text{SO}(4)$, namely the subgroup that fixes the vector $\vec{m}_R$.  Because this $\text{SO}(3)$ is preserved, the pions must transform under it, and thus all three pions must have the same mass at leading order at strong coupling.

Let us now quantify the breaking of the degeneracy. This breaking arises from the terms involving $m_I$ at higher orders in the expansion of the EFT in powers of $M$.  These terms break $\text{SO}(3)$ to $\text{O}(2)$.  The $\text{O}(2)$ ensures that $M_{\pi^+}$ and $M_{\pi^-}$ remain degenerate.

In general, an isospin breaking interaction mediated by a dimension $\Delta$ operator with Wilson coefficient $\lambda$ induces a mass splitting of the order $M^2_{\pi^{\pm}}-M^2_{\pi^0}\sim \lambda (M_{\pi})^{\Delta} $, where $M_{\pi}\sim m^{2/3}g^{1/3}$ is the leading order pion mass (which is the only mass scale in absence of the perturbation).  The leading isospin breaking contribution is then induced by the term proportional to $b_2$ in~\eqref{L2}, which leads to
\begin{equation}\label{eq_ratio_pi_correction}
      \frac{M_{\pi^\pm}^2}{M_{\pi^0}^2}  - 1 = {\cal O}\left(\frac{M^2}{g^2}\right) \,.
 \end{equation}
Note that the terms in~\eqref{L3} simply rescale and rotate by an amount $\mO(M^2/g^2)$ the coefficient of the leading operator in~\eqref{L1}, hence they do not contribute to the mass splitting.

Note that when $\theta = 0$, the choice of masses in \eqref{tildeMDef} gives $\vec{m}_R = (0, 0,0, m)$ and $\vec{m}_I = (0, 0, -\delta m, 0)$, which implies
 \begin{equation}
   {\cal L}_2 = b_1 m^2 {\cal O}_2^{44} + b_2 (\delta m)^2 {\cal O}_2^{33} + \ldots
 \end{equation}
As mentioned above, the term proportional to $b_2$ is the leading $\text{SO}(3)$-breaking term, and therefore $\text{SO}(3)$-breaking effects will be proportional to $(\delta m)^2$, proving \eqref{eq_ratio_dm_intro}.  When $\theta \neq 0$, the analysis is more complicated.

\subsection{Additional comments}

Two additional comments are in order. The first comment we would like to make concerns a comparison of the effective theory \eqref{EffAction} with the chiral Lagrangian in QCD\@.  We can obtain the leading correction that comes from integrating out $\phi_+$ in \eqref{eq_bosonized_NonAbelian2} by simply setting $\tilde{\mu}=\mu$ and $\phi_+ = 0$:
 \begin{equation}\label{eq_EFT_derived_general}
     S_\text{EFT} = S_\text{WZW} + \int d^2x\, ({\cal L}_{\text{mass}} + \cdots) \,, \qquad 
     \mL_{\text{mass}} =\tilde{c}\sqrt{\mu}\Re\text{Tr}\left(M :U:\right) \,.
 \end{equation}
The term ${\cal L}_{\text{mass}}$ is a more explicit expression for the first term in ${\cal L}_1$ in \eqref{L1}, and~\eqref{L1_coeff} follows from it. This form of the effective action is analogous to that in QCD, 
 \begin{equation}
     S_\text{EFT}^{\text{(QCD)}} = \int d^4x\, \left[ \frac{f_\pi^2}{4} \tr (\partial_\mu U \partial^\mu U^\dagger) 
      + \frac{\lambda f_\pi^3}{2} \Re \tr (M U)  + \cdots \right] \, ,
 \end{equation}
 where at large number of colors $f_\pi\sim \sqrt{N_c}$ and
 $\lambda \sim 1/\sqrt{N_c}$.
Here, $U$ is related to the pion fields via
$U=\exp( i \vec \sigma \cdot \vec \pi/f_\pi)$.
Thus, the WZW term is analogous to the standard kinetic term in the chiral Lagrangian, and the mass term $\mL_\text{mass}$ is analogous to the pion mass term in the QCD chiral Lagrangian, $\sim f^3_\pi \Re\text{Tr}\left(M U\right)$.  (In QCD with two flavors $m_u$ and $m_d$, the mass matrix is $M = e^{-i\theta/2}\left(\frac{m_u+m_d}{2} \mathds{1} + \frac{m_u-m_d}{2} \sigma^3 \right)$, in analogy with \eqref{tildeMDef}.)

As in the Schwinger model, in QCD with two flavors the pion mass term preserves an $\text{SO}(3)$ subgroup of $\text{SO}(4)$.  In QCD, the first isospin breaking term compatible with the symmetries arises at order $\mO(M^2)$ and reads~\cite{Gasser:1983yg}
\begin{equation}\label{eq_iso_breaking_QCD}
    \delta \mL_\text{mass}^{\text{(QCD)}} \supset b f^2_{\pi}\left(\Tr(M U-U^\dagger M^\dagger)\right)^2=b f^2_\pi(\vec{u}\cdot\vec{m}_I)^2 \,,
\end{equation}
where $b$ is an $O(1)$ dimensionless Wilson coefficient.
Since $M_{\pi^\pm}^2\sim\lambda  f_{\pi} m$ and the pion kinetic term is proportional to $f^2_{\pi}$, the operator~\eqref{eq_iso_breaking_QCD} implies that the $\pi^{\pm}/\pi^0$ mass ratio scales as $M_{\pi^\pm}^2/M_{\pi^0}^2-1\propto 1/f_{\pi}$, and in particular at $\theta=0$ one finds
\begin{equation}
    \frac{M_{\pi^0}^2}{M_{\pi^\pm}^2}-1
    \propto b\frac{\delta m^2}{\lambda\,f_{\pi}\,m} \,, \qquad \text{at $\theta = 0$}\,.
\end{equation}
As mentioned above, in the two-flavor Schwinger model the chiral Lagrangian is replaced by the $\text{SU}(2)_1$ WZW model, which is a strongly coupled CFT\@. This implies a slightly different structure of isospin violating effects. Indeed, due to the truncation of the fusion algebra, the OPE $U\times U$ contains only the vacuum module, and the operator analogous to the QCD one in \eqref{eq_iso_breaking_QCD} is the $b_2$ term in \eqref{L2}, which is marginal and leads to~\eqref{eq_ratio_pi_correction}.

The second comment concerns the leading dependence of the $\pi$ and $\sigma$ masses on $m_1$ and $m_2$.  An immediate consequence of the EFT analysis is that if we keep only the leading mass correction (namely the first term in ${\cal L}_1$), all physical quantities depend only on $|\vec{m}_R|$.  For the choice of masses in \eqref{tildeMDef} that we are interested in, 
 \begin{equation}
     \vec{m}_R = \left(0, 0, -(\delta m) \sin \frac{\theta}{2}, m \cos \frac{\theta}{2} \right) \,. 
 \end{equation}
The magnitude of this vector is 
 \begin{equation}\label{eq_def_mt}
     \widetilde m \equiv \sqrt{m^2 \cos^2 \frac{\theta}{2} + (\delta m)^2 \sin^2 \frac{\theta}{2}} \,.
 \end{equation}
Then, we infer from~\eqref{eq_Mpi} (by replacing $m\cos\frac{\theta}{2}$ with $\tilde{m}$) that the pion masses (which, as discussed above, are degenerate at leading order) are
\begin{equation}\label{eq_Mpi_dm}
\begin{split}
    M_{\pi} &\approx \frac{2^\frac{7}{6} e^\frac{\gamma_E}{3} \Gamma \left(\frac{1}{6}\right)}{\Gamma \left(\frac{2}{3}\right) \Gamma\left(\frac{1}{4}\right)^\frac{4}{3}}g^{\frac{1}{3}}\widetilde{m}^\frac{2}{3}
     \approx 2.008 \,g^{\frac{1}{3}}\widetilde{m}^\frac{2}{3} \,, \qquad \text{as $\widetilde m \to 0$} \,.
    \end{split}
\end{equation}
The non-trivial leading-order dependence of $M_\pi$ on $m_1$, $m_2$, and $\theta$ in \eqref{eq_Mpi_dm} is entirely predicted by our symmetry analysis.  The same argument implies that the ratio between $M_\sigma$ and $M_\pi$ approaches $\sqrt{3}$,
 \begin{equation} \label{MspRatio}
     \frac{M_\sigma}{M_\pi} \approx \sqrt{3} \,, \qquad \text{as $\widetilde m \to 0$} \,.
 \end{equation}
We emphasize that both \eqref{eq_Mpi_dm} and \eqref{MspRatio} hold for any $m_1$, $m_2$, and $\theta\neq\pi$!

\subsection{Explicit computations using Abelian bosonization}

We may go a step further and compute explicitly the Wilson coefficients of the marginal perturbations at order $1/g^2$, and thus the corrections to the $\pi^0$-$\pi^{\pm}$ and $\sigma$-$\pi^{\pm}$ mass ratios as explained below~\eqref{eq_ratio}. We do this via Abelian bosonization in Appendix~\ref{app_Large_m_theta0}\@. It is convenient to define the coefficients
\begin{equation}
    \delta_2=\frac{e^{\gamma_E}(\delta m)^2}{2\pi\mu^2}I_s(\pi-\theta)\,,\qquad
    \delta_3=\frac{e^{\gamma_E} m\,\delta m\,\sin\theta}{2\pi\mu^2}I_3\,,
\end{equation}
where $I_s(\theta)$ is defined in~\eqref{eq_Is} and
\begin{equation}
     I_3=2\pi\int_0^\infty d\xi\,\xi^2\left(1-e^{-K_0(\xi)}\right)\simeq 9.05771\,.
\end{equation}
Note that $\delta_3=0$ for $\theta=0$. In terms of these, we obtained the following result for the $\pi^0/\pi^{\pm}$ mass ratio
\begin{equation}\label{eq_pi_pi_ratio_dm}
\begin{split}
    \frac{M_{\pi^0}^2}{M_{\pi^{\pm}}^2} &=1-\frac{4\pi}{3\sqrt{3}}(2\delta_2\cos^2\nu -2 \delta\sin^2\nu+\delta_3 \sin (2 \nu ))+\mathcal{O}\left(\frac{M^\frac{10}{3}}{g^\frac{10}{3}}\right)\\
    &\stackrel{\theta=0}{=}1-\frac{2\pi e^{\gamma_E}(\delta m)^2 }{3 \sqrt{3} \,g^2}I_s(\pi)+\mathcal{O}\left(\frac{M^\frac{10}{3}}{g^\frac{10}{3}}\right)\,,
    \end{split}
\end{equation}
where $\delta$ is given in~\eqref{eq_EFT_delta}, $\tilde{m}$ is defined in~\eqref{eq_def_mt}, $I_s(\pi)\simeq 10.08304$, and $\nu$ is obtained from
\begin{equation}\label{eq_def_mt_nu}
    \tan\nu\equiv\frac{\delta m}{m }\tan\frac{\theta}{2}\,.
\end{equation}

\begin{figure}
    \centering
    \includegraphics[width=0.7\linewidth]{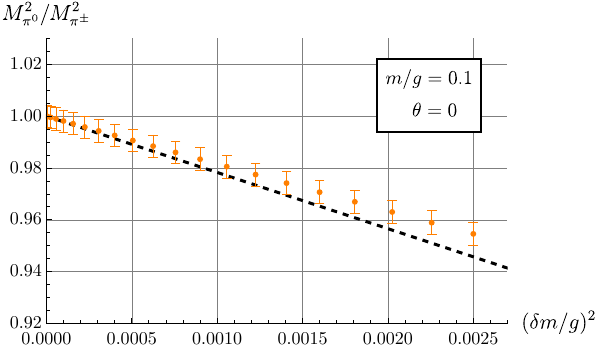}
    \caption{The $\pi^0/\pi^{\pm}$ squared mass ratio at $\theta=0$ as a function of $(\delta m/g)^2$. When we consider unequal masses on the lattice, as in \eqref{tildeMDef}, we see that the lightest state---an isotriplet for equal masses---breaks into $\pi^0$ and $\pi^\pm$, with the $\pi^0$ being lighter. For small $\delta m$, their mass ratio is consistent with \eqref{eq_pi_pi_ratio_dm} (black dashed line). Note that at the rightmost point of the plot, $m_1/m_2 = 3$, and yet the mass splitting is only $\sim 2\%$.}
    \label{fig:isospin_breaking}
\end{figure}

Our numerical results at $\theta=0$, which are plotted in Figure~\ref{fig:isospin_breaking}, are in good agreement with the formula~\eqref{eq_pi_pi_ratio_dm}. In particular, the plot clearly illustrates Coleman's first puzzle in that the splitting is at the percent-level despite one flavor being three times more massive than the other.

We also computed the correction to the $\sigma/\pi^{\pm}$ mass ratio
\begin{equation}\label{eq:sigma_pi_ratio}
\begin{aligned}
    \frac{M_{\sigma}^2}{M_{\pi^{\pm}}^2}&=3+\frac{4\pi}{3\sqrt{3}}((2+\cos2\nu)\delta-(2-\cos2\nu)\delta_2+ \delta_3 \sin(2\nu)) +\mathcal{O}\left(\frac{M^\frac{10}{3}}{g^\frac{10}{3}}\right)\\
    &\stackrel{\theta=0}{=}3-\frac{\pi e^{\gamma_E}}{\sqrt{3}}\left(\frac{m^2}{g^2}I_s(0)+\frac{(\delta m)^2}{3g^2}I_s(\pi)\right)+\mathcal{O}\left(\frac{M^\frac{10}{3}}{g^\frac{10}{3}}\right)\,,
    \end{aligned}
\end{equation}
where $I_s(0)\simeq 1.02533$.

In~\eqref{eq_pi_pi_ratio_dm} and~\eqref{eq:sigma_pi_ratio}, the $\mathcal{O}\left(M^{10/3}/g^{10/3}\right)$ corrections receive contributions both from dimension-$4$ operators at order $\mathcal{O}(M^2)$ and from the dimension-$5/2$ operator at order $\mathcal{O}(M^3)$. Their contributions are estimated analogously to the discussion below~\eqref{eq_ratio}, using the fact that an operator of dimension $\Delta$ with coefficient $\lambda$ contributes to the mass ratios as $\sim \lambda\, M_{\pi}^{\Delta-2}$.

The results of this section hold as long as the diagonal mass term is not too small, since for $m= 0$ the $\delta m$ perturbation is equivalent to an $\text{SO}(3)$-symmetric mass term at $\theta=\pi$, which leads to a different mass spectrum as we discuss in the next section.

\section{The equal mass theory at \texorpdfstring{$\theta =\pi$}{θ=π}}\label{sec_theta_Pi}

In this section, we discuss the equal mass two-flavor Schwinger model at $\theta=\pi$. In~\cite{Dempsey:2023gib}, it was proposed that charge conjugation symmetry $C$ is spontaneously broken for all values of the couplings, and that the mass gap becomes exponentially small in the strong-coupling regime. Here, we show that these claims are supported by exact integrability results at strong coupling. For any value of $g/m$, the lowest-mass states of the theory are deconfined solitons interpolating between the two vacua. These are Coleman's half-asymptotic particles \cite{Coleman:1976uz}, which correspond to the fundamental fermions and naturally form isospin doublets. 
The mass gap in the trivial topological sector is twice the mass of the soliton.
Therefore, our answer to Coleman’s third puzzle is that, for $\theta=\pi$ and $m_1=m_2$, there is no transition to a different phase of the theory as $g$ is increased.

\subsection{Analytical results at weak and strong coupling}

The $\theta=\pi$ theory with $m_1=m_2$ is related by an axial transformation to the $\theta=0$ theory with $m_1=-m_2$.
As the phase diagram in Figure~\ref{fig:Phase_Diagram} shows, in this case the system admits two $C$-breaking vacua. This is easy to see in the weak coupling limit, where we can integrate out the fermions and we are left with pure Maxwell theory 
\begin{equation}
    \mL=\frac{1}{2g_{\text{eff}}^2}F_{01}^2-\frac{1}{2}F_{01}+\mathcal{O}\left(\frac{F_{01}^4}{m^6},\frac{(\pd F_{01})^2}{m^4}\right)\,,
\end{equation}
where we defined the effective coupling
\begin{equation}\label{eq_g_eff}
    \frac{1}{g_{\text{eff}}^{2}}=\frac{1}{g^2}\left(1+\frac{g^2}{3\pi m^2}+\mathcal{O}\left(\frac{g^4}{m^4}\right)\right)\,.
\end{equation}
Setting $\theta = \pi$ is equivalent to inserting a half-integer-charge Wilson line stretching from $x=- \infty$ to $x=+ \infty$. This allows for two degenerate vacua with opposite values of the electric field:
\begin{equation}\label{eq_C_breaking_bkd}
    \langle F_{01}\rangle=\pm\frac{g_\text{eff}^2}{2}\left(1+\mathcal{O}\left(\frac{g^4}{m^4}\right)\right)\,.
\end{equation}
There are solitons interpolating between the two vacua which can be interpreted as elementary particles with charge of magnitude $2|\!\braket{F_{01}}\!|/g^2$.  These configurations were called \emph{half-asymptotic} particles in~\cite{Coleman:1976uz}, and they correspond to the lowest-energy excitations of the theory at $\theta=\pi$ (and, in fact, the only stable excitations). At weak coupling, these solitons are the charged fermions $\psi_\alpha$ suitably dressed by the electric field. The lowest charge-neutral states of the theory, i.e.~those which preserve identical boundary conditions at $x\rightarrow\pm\infty$, are two-particle states, forming a continuum starting at $2M_{\text{sol}}$, where $M_{\text{sol}}$ is the mass of a single soliton. Note that the solitons form isospin doublets, and hence transform in a projective representation of the $\text{SO}(3)$ internal symmetry. This is a consequence of the anomalies discussed in Section~\ref{sec:setup}~\cite{Komargodski:2017dmc,Cordova:2019jnf}.

At weak coupling, we can compute $M_\text{sol}$ as the pole of the loop-corrected fermion propagator. In $d=2$ there are some subtleties involving loops with photon propagators. Namely, using the standard unphysical Feynman gauge leads to infrared divergences \cite{Das:2012qz}, which are removed only after resumming an infinite series of nested rainbow graphs. This problem can be circumvented by using a physical gauge of the form $n^\mu A_\mu = 0$ for some fixed vector $n^\mu$. Usually $n^\mu$ is assumed to be space-like, in particular $n^\mu = (0,1)$ corresponds to the Coulomb gauge. However, in $d=2$ it is also useful to consider a light-like $n^\mu$ as we discuss below. The IR divergences in momentum space are then regularized using the principal value prescription for the photon propagator. In Coulomb gauge, $A_1 = 0$ the principal value prescription is equivalent to considering a linear interaction potential $\sim |x^1|$ in position space~\cite{Coleman:1976uz}.
\begin{figure}[t]
\centering
    \begin{subfigure}[t]{0.4\textwidth}
        \centering
        \includegraphics[width=4cm]{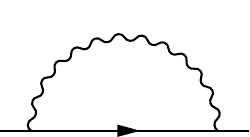}
        \caption{}\label{subfig::diag_a}
    \end{subfigure}~\begin{subfigure}[t]{0.4\textwidth}
        \centering
        \includegraphics[width=4cm]{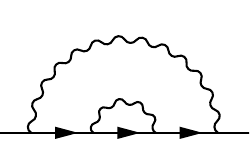}
        \caption{}\label{subfig::diag_b}
    \end{subfigure}\\~\\~\\
    
    \begin{subfigure}[t]{0.4\textwidth}
        \centering
        \includegraphics[width=4cm]{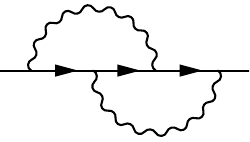}
        \caption{}\label{subfig::diag_c}
    \end{subfigure}~\begin{subfigure}[t]{0.4\textwidth}
        \centering
        \includegraphics[width=4cm]{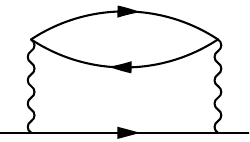}
        \caption{}\label{subfig::diag_d}
    \end{subfigure}
    \caption{Feynman diagrams for the mass correction.}
    \label{fig::feyn_diag_mass}
\end{figure}

In $d=2$ a convenient choice is to work in the lightcone gauge $A_- = 0$, where $A_- = (A_0 - A_1)/\sqrt{2}$. This gauge choice preserves Lorentz invariance and often leads to significant simplifications \cite{tHooft:1974pnl,Bergknoff:1976xr,Harada:1993va}. The relevant self-energy diagrams up to two-loop order are shown in Figure~\ref{fig::feyn_diag_mass}. We compute these in Appendix~\ref{app_Pert_Corr}, and extract the fermion pole mass as
    \begin{equation}\label{eq_m_shift_main}
M_\text{sol}=m \left (1 -\dfrac{g^2}{2\pi m^2} + \frac{(3 \pi ^2-4)g^4}{96 \pi^2m^4}+ \mathcal{O}\left( \dfrac{g^6}{m^6} \right) \right ).
\end{equation}
The leading correction of order $\frac{g^2}{m^2}$ is well known and was derived already in~\cite{Coleman:1976uz}. The two-loop correction is instead a new result to the best of our knowledge.

Let us note that the theory remains in a $C$-broken phase for arbitrary values of the coupling \cite{Dempsey:2023gib}. In the strong coupling regime, this can be understood semiclassically by analyzing the potential of the bosonized action \eqref{eq_bosonized} \cite{Dempsey:2023gib}, as we review below. This vacuum structure is closely analogous to the Dashen $CP$-breaking phenomenon in $N_f=2$ QCD~\cite{Dashen:1970et,Smilga:1998dh}. 
To appreciate this analogy, consider the non-Abelian formulation of the low energy EFT~\eqref{eq_EFT_non_Abelian}.
As in the $\text{SU}(2) $ $d=4$ pion Lagrangian, the leading mass perturbation in~\eqref{eq_EFT_non_Abelian} vanishes at $\theta=\pi$:
\begin{equation}
    e^{-\frac{i\pi}{2}}\,\Tr(U) + \text{c.c.} = 0\,,
\end{equation}
since $\Tr(U) = \Tr(U^\dagger)$. This is not an accident. Indeed, while the transformations~\eqref{eq_discrete_UM} do not leave invariant the mass matrix $M=e^{-\frac{i\pi}{2}}m\,\mathds{1}$ at $\theta=\pi$, the action is instead invariant under the modified discrete symmetries $P'=e^{2\pi iQ_3^L}P$ and $C'=e^{2\pi iQ_3^L}C$, where $e^{2\pi iQ_3^L}=-\mathds{1}$ is a chiral rotation. These thus act as
\begin{equation}
    U\stackrel{P'}{\longrightarrow}-U^\dagger\,,\quad
    U\stackrel{C'}{\longrightarrow}-U^*\,.
\end{equation}
Therefore, both $P'$ and $C'$ flip the sign of $\text{Tr}(U)=\text{Tr}(U^\dagger)$ at $\theta=\pi$, preventing a linear term also at higher orders in the mass.

In QCD, the pion potential is therefore governed by $\mO(m^2)$ operators, which generically lead to two degenerate vacua related by $CP$~\cite{Smilga:1998dh}.\footnote{Let us review this phenomenon. Spurion analysis shows that the leading nontrivial contribution to the potential at $\theta=\pi$ is
$V(U)\propto \ell_7 m^2(e^{-i\theta}\Tr(U)-e^{i\theta}\Tr(U^\dagger))^2
\stackrel{\theta=\pi}{=}-\ell_7 m^2\Tr(U)^2$~\cite{Smilga:1998dh}, where $\ell_7$ is a Wilson coefficient.
It is expected that $\ell_7>0$. Parameterizing $U=\cos\alpha\,\mathds{1}+i \sin\alpha\, \vec{\sigma}\cdot \hat{n}$ with $\hat{n}^2=1$, the $\theta=\pi$ potential is thus minimized at  $\alpha=0\,,\pi$, which are exchanged by $CP$.}  Similarly, in the $k=1$ WZW theory we need to analyze higher order terms in the mass. As discussed in detail in the previous section, due to the truncation of the fusion algebra the next-to-leading operator invariant under the preserved $\text{SO}(3)$ is a bilinear $J^A_L\bar{J}^A_R$ in the conserved currents, which is generated at order $\mO(m^2)$. This is marginal and cannot be studied semiclassically.  Nevertheless, this operator generates a mass gap at the quantum level, as we discuss in detail below, and leads to spontaneous $C$-breaking.

\subsection{Resolution of Coleman's third puzzle}

Before discussing the mass spectrum quantitatively, we point out that the $C$-breaking scenario described above provides a natural resolution of Coleman's third puzzle. In~\cite{Coleman:1976uz}, by applying the semiclassical analysis of~\cite{Dashen:1974cj} to the $\theta=\pi$ sine-Gordon EFT~\eqref{eq_EFT_2flavor}, Coleman argued that, at infinite coupling, the spectrum consists of massive isospin doublets with vanishing gauge charge. This appears to be in contradiction with the requirement that all neutral bound states must carry integer isospin. Coleman therefore hypothesized the existence of a phase transition between the deconfined weak-coupling regime and an exotic confined strong-coupling regime, with neutral doublet bound states. Coleman's puzzle is instead resolved by interpreting the lightest massive states as \emph{half-asymptotic} particles, i.e.~solitons interpolating between the two vacua.  At weak coupling, these solitons are nothing but the fundamental fermions of the theory. The vanishing of the gauge charge simply reflects the fact that the difference in the electric field expectation value between the two vacua decreases as the theory approaches the conformal point at $m = 0$.

Let us discuss in greater detail how the half-asymptotic particles \emph{lose} their charge as $m/g\to 0$.
Note first that, since $g_\text{eff}^2<g^2$, the charge difference between the two vacua decreases as we increase the coupling $g^2/m^2$ at weak coupling, due to~\eqref{eq_C_breaking_bkd}. That this phenomenon persists as we increase the coupling can be seen from the following semiclassical argument using the bosonized description~\eqref{eq_bosonized}. The classical potential is
\begin{equation}\label{eq_bosonized_pot_2f}
    V(\phi_+,\phi_{-})=\frac{\mu^2}{2 }\min_{n\in\mathbb{Z}}\left(\phi_++\frac{\theta +2\pi n}{2\sqrt{2\pi }}\right)^2-\frac{e^{\gamma_E}}{\pi} m\mu\cos\left(\sqrt{2\pi}\phi_+\right)\cos\left(\sqrt{2\pi}\phi_-\right)\,.
\end{equation}
We have removed normal ordering indicators, since our analysis will be purely qualitative. Let us study the minima of this potential for $\theta=\pi$. Note first that, within the first period $0\leq \mp\phi_{\pm}<\sqrt{2\pi}$, we may set $n=0$. Both at weak coupling (small $\mu$) and at strong coupling, we find two global minima. At weak coupling these are given by
\begin{equation}\label{eq_vacua_weak}
\begin{cases} 
 \displaystyle
\phi_-=0\ & \displaystyle\phi_+=-\frac{2\pi\,\mu }{8 \sqrt{2 \pi } \,e^{\gamma_E} m}+\mathcal{O}\left(\frac{\mu^2}{m^2}\right) \,, \\[0.6em]
 \displaystyle
\phi_-=\sqrt{\frac{\pi}{2}} &
\displaystyle\phi_+=-\sqrt{\frac{\pi }{2}}+\frac{2\pi\,\mu }{8 \sqrt{2 \pi } \,e^{\gamma_E} m}+\mathcal{O}\left(\frac{\mu ^2}{m^2}\right)\,.
\end{cases}
\end{equation}
For large $\mu$ the two vacua are instead classically described by
\begin{equation}\label{eq_vacua_strong}
\begin{cases} \displaystyle
    \phi_-=0\ &\displaystyle\phi_+= -\sqrt{\frac{\pi}{8}}+\frac{\sqrt{2 \pi } e^{\gamma_E} m}{\pi\,\mu }+\mathcal{O}\left(\frac{m^3}{\mu^3}\right)\,,\\[0.6em]\displaystyle
    \phi_-=\sqrt{\frac{\pi}{2}} &
    \displaystyle\phi_+=-\sqrt{\frac{\pi}{8}}-\frac{ \sqrt{2 \pi } e^{\gamma_E} m}{\pi\,\mu }+\mathcal{O}\left(\frac{m^3}{\mu^3}\right)\,.
\end{cases}
\end{equation}
The two vacua are exchanged by $C$, which can be taken to act as\footnote{More precisely, here we use both the canonical action $(\phi_+,\phi_-)\stackrel{C}{\rightarrow}-(\phi_+,\phi_-)$, with the periodicity $(\phi_+,\phi_-) \sim \left(\phi_+\pm\sqrt{\pi/2},\,\phi_-\pm\sqrt{\pi/2}\right)$ to remain within the $n=0$ sector of~\eqref{eq_bosonized_pot_2f}, see footnote~\ref{footnote_periodicity}.} 
\begin{equation}
    \phi_{\pm}\rightarrow\mp\sqrt{\frac{\pi}{2}}-\phi_{\pm}\,.
\end{equation}

We thus see that, as the coupling is increased from weak to strong, the vacua move from $\phi_+=0,-\sqrt{\pi/2}$ toward the midpoint $\phi_+=-\sqrt{\pi/8}$, while $\phi_-$ remains unchanged. The solitons between the two vacua are classically described by solutions to the equations of motion interpolating between the minima. Independently of the detailed form of these solutions, the $\text{SO}(3)$ Cartan charge $Q_3$ and the gauge charge $Q_{\text{gauge}}$ correspond to winding charges in the bosonized description:
\begin{align}
Q_3 &=\frac{1}{\sqrt{2\pi}}\int dx \,\pd_x\phi_-=\pm\frac12\,,
\\[0.8em]
Q_{\text{gauge}} &=\sqrt{\frac{2}{\pi}}\int dx \,\pd_x \phi_+ =\pm\begin{cases}\displaystyle
1-\frac{\mu }{2 e^{\gamma_E} m}+\mathcal{O}\left(\frac{\mu ^2}{m^2}\right) &\text{for }\mu\ll m\,, \\[1em]
\displaystyle
\frac{4e^{\gamma_E} m}{\pi \,\mu }+\mathcal{O}\left(\frac{m^3}{\mu^3}\right)
& \text{for }\mu\gg m\,.
\end{cases}
\end{align}
We therefore find that the solitons transform as spin-$\frac{1}{2}$ under isospin, which is a projective representation of the $\text{SO}(3)$ flavor symmetry, as predicted by the anomaly~\cite{Komargodski:2017dmc,Cordova:2019jnf}. Their gauge charge instead decreases progressively as $\mu/m$ is increased. As in other instances of charge fractionalization, the fact that the gauge charge is not quantized is not problematic: the only charge-neutral states consist of soliton-antisoliton pairs over a single $C$-broken vacuum.

\begin{figure}
    \centering
\includegraphics[width=0.7\linewidth]{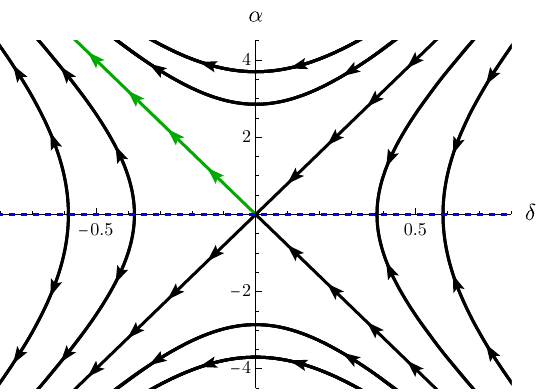}
    \caption{The one-loop RG flow diagram in terms of the sine-Gordon parameters $\alpha$ and $\delta$ in \eqref{eq_EFT_2flavor_theta_Pi}. The blue line $\alpha=0$ is the fixed line of $c=1$ CFTs. The diagonal $\alpha=-8\delta$ RG flow is $\text{SO}(3)$-preserving. The RG flow towards the gapped phase takes place along the green line.}
    \label{fig:BKT}
\end{figure}

Let us now provide a quantitative description of the mass spectrum at strong coupling. As noted above, for $\theta=\pi$ the coefficient of the first relevant perturbation in the low-energy EFT vanishes. It is convenient to rely on the Abelian formulation~\eqref{eq_EFT_2flavor}, which at $\theta=\pi$ reads
\begin{equation}\label{eq_EFT_2flavor_theta_Pi}
    \mL_\text{EFT}\simeq \frac{1-\delta}{2}\pd_\mu \phi_- \pd^\mu \phi_- +\frac{e^{2\gamma_E}}{32\pi}\alpha\, \mu_-^2N_{\mu_-}\cos(\sqrt{8\pi}\phi_-)\,,
\end{equation}
where we momentarily distinguished the wave-function renormalization and the coefficient of the potential. As shown in~\eqref{eq_EFT_2flavor}, $\text{SO}(3)$ invariance relates the couplings as 
\begin{equation}\label{eq_alpha_delta_SU2}
    \alpha=-8\delta+\mathcal{O}(\delta^2)\,,
\end{equation}
and $\delta$ is given in~\eqref{eq_EFT_delta}.  The couplings $\alpha$ and $\delta$ multiply classically marginal operators and acquire nontrivial beta functions at loop level, describing the well known BKT RG phase diagram \cite{Kosterlitz:1974nba,Jose:1977gm}. The most general form of the $2$-loop beta functions of $\alpha$ and $\delta$ is~\cite{Amit:1979ab}: 
\begin{equation}\label{eq_beta_2loop}
    \beta_\alpha=2\alpha \delta+ A \alpha^3\,,\qquad\beta_{\delta}=\frac{1}{32}\alpha^2+B\alpha^2\delta\,,
\end{equation}
where $A$ and $B$ can be determined at two-loop order and are scheme dependent, but the combination
\begin{equation}\label{eq_A_B_cond}
    2A+B=\frac{3}{32}
\end{equation}
is scheme-independent. The schematic form of the RG flow lines that follows from the one-loop beta functions (setting $A= B = 0$) is shown in Figure~\ref{fig:BKT}. Along the special $\text{SO}(3)$-preserving locus~\eqref{eq_alpha_delta_SU2} the two-loop ambiguities drop out, in agreement with the general result for a single running coupling~\cite{Weinberg:1996kr}, and we obtain\footnote{To show this explicitly, let us assume that the $\text{SO}(3)$-preserving locus lies at
\begin{equation}\label{eq_alpha_delta_rel}
\alpha=-8\delta+c\,\delta^2+\mathcal{O}\left(\delta^3\right)\,,
\end{equation}
for some $c$. The relation~\eqref{eq_A_B_cond} and the condition that RG preserves~\eqref{eq_alpha_delta_rel} imply
\begin{equation}
A=\frac{8-c}{256}\,,\qquad
B=\frac{c+4}{128}\,.
\end{equation}
Plugging back into $\beta_{\delta}$ we obtain~\eqref{eq_beta_delta}, which is independent of $c$, as expected.}
\begin{equation}\label{eq_beta_delta}
    \beta_{\delta}=2\delta^2+2\delta^3
     +\mathcal{O}(\delta^4)\, .
\end{equation}
 Since the RG flow takes place for $\delta<0$, the theory is asymptotically free \cite{Dempsey:2023gib}, just like QCD \cite{Gross:1973id, Politzer:1973fx}. Therefore, the 
RG running~\eqref{eq_beta_delta} implies the existence of an exponentially small mass gap at strong coupling. Using the relation~\eqref{eq_EFT_delta} between $\delta$ and $m/g$, we obtain the following result for the strong coupling scale of the theory
\begin{equation}\label{eq_2loop_m}
\Lambda_\text{IR}\simeq  \frac{m}{\sqrt{\pi \,A_s}} e^{-A_s\frac{ g^2}{m^2}}\,,\qquad
A_s=\frac{2 \,e^{-\gamma_E}}{I_s(\pi)}\simeq 0.111\,,
\end{equation} 
where we identified the UV matching scale at which the EFT running begins with the mass of the heavy scalar $\mu=\sqrt{2/\pi}\,g$; different choices of the UV matching scale of the form $\mu = c \,g$, with $c$ a numerical constant, only affect the $\mO(1)$ numerical prefactor in~\eqref{eq_2loop_m}. 

Let us note that the two-loop correction in the beta function~\eqref{eq_beta_delta} plays an important role in our arguments. In the one-loop approximation, the IR scale is $\sim g e^{-A_s\frac{ g^2}{m^2}}$ \cite{Dempsey:2023gib}, but the two-loop correction multiplies this by the factor $(-\delta)^{b_1/b_0^2}$, where $b_0=2$ and $b_1=2$ are the one-loop and two-loop beta function coefficients. Using \eqref{eq_EFT_delta}, we see that this factor is $\sim \left (\frac{m^2}{g^2}\right )^{1/2}$.  We therefore conclude that, for strong coupling, the solitons have a mass of order~$\Lambda_\text{IR}$, although the exact numerical prefactor cannot be determined from these considerations alone.

It is instructive to contrast our findings so far with other systems in which an exponentially small mass gap is generated via dimensional transmutation, such as in the Bardeen-Cooper-Schrieffer (BCS) theory and QCD\@. In those and other similar cases, the long RG flow implies that the massive states of the theory are best understood as nontrivial bound states, and are therefore qualitatively different from the \emph{fundamental fields} that describe the UV theory. By contrast, in the two-flavor Schwinger model at $\theta = \pi$, the dynamically generated mass scale is not associated with any nontrivial binding mechanism. Instead, the spectrum remains qualitatively similar to that at weak coupling, consisting of half-asymptotic doublet particles, which are the fermions. The exponential smallness of the fermion mass at strong coupling should follow from resumming the self-energy corrections \eqref{eq_m_shift_main}.  It would be interesting to analyze this mechanism using resurgence methods. 

Remarkably, the semi-quantitative discussion above is once again corroborated by integrability considerations at strong coupling. However, the application of the integrability results is rather subtle due to the marginal nature of the coupling.  We review the salient features of this analysis below.

Let us first notice that the $\text{SO}(3)$-symmetric sine-Gordon model with marginal perturbation, describing the low energy EFT~\eqref{eq_EFT_2flavor} at $\theta=\pi$, is related to another well studied theory: the SU$(2)$ Thirring model.
The latter consists of two Dirac fermions $\psi_\alpha$, $\alpha =1,2$, with Lagrangian 
\begin{equation}\label{eq_Thirring}
\mL=i\bar{\psi}^\alpha \slashed{\pd}\psi_\alpha -\lambda\left(J^\mu J_\mu+j^{\mu}_A j_{A\mu}\right)\,,
\end{equation}
where
\begin{equation}
J^\mu=\frac12\bar{\psi}^\alpha \gamma^\mu\psi_\alpha \,,\qquad
j^{\mu}_A=\frac12\bar{\psi}^\alpha \gamma^\mu (\sigma_A)_{\alpha}{}^{\beta}\psi_\beta\,,
\end{equation}
with $\sigma_A$ being the Pauli matrices. 
As we review in appendix \ref{app_Large_m_thetaPi}, the SU$(2)$ Thirring model describes a free decoupled compact massless scalar and an $\text{SO}(3)$-symmetric sine-Gordon model with marginal coupling given by~\cite{Amit:1979ab}\footnote{More precisely, the Thirring model is a $\mathbb Z_2$ quotient of the product of the free compact boson and the sine-Gordon sector and has a unique vacuum. This difference in global structure does not affect the perturbative RG equations or the mass scale extracted below. See Appendix~\ref{app_Large_m_thetaPi} for details.
}

\begin{equation}\label{eq_Thirring_coupling}
    \delta=-\frac{\lambda}{\pi}+\mathcal{O}(\lambda^2)\,.
\end{equation}
The two-loop beta function of the SU$(2)$ Thirring model is \cite{Bondi:1989nq}
\begin{equation}\label{eq_beta_lambda}
    \beta_{\lambda}=- \frac{2}{\pi} \lambda^2+ \frac {2}{\pi^2} \lambda^3+\mathcal{O}(\lambda^4)\, ,
\end{equation}
which agrees with \eqref{eq_beta_delta}. Importantly for us, the precise relation between the sine-Gordon coupling and the SU$(2)$ Thirring model coupling is scheme-dependent beyond leading order, and requires matching the two theories at loop level.

The SU$(2)$ Thirring model has been solved by Bethe ansatz in~\cite{Belavin:1979pq,Andrei:1979sq}.\footnote{More precisely,~\cite{Andrei:1979sq} studied the chiral Gross-Neveu model, which can be shown to coincide with the SU$(2)$ Thirring model for $N_f=2$ after using Fierz rearrangement identities~\cite{Loebbert:2016cdm}.} Setting aside the decoupled free boson, the lowest excitation spectrum consists of pairs of spin-$\frac{1}{2}$ doublet particles, which are interpreted as solitons between the two vacua in the original sine-Gordon model. Their mass is exponentially small in the coupling, in agreement with the RG prediction~\eqref{eq_2loop_m}.

We may now use integrability to compute the soliton mass at strong coupling. To this aim we need to express the SU$(2)$ Thirring coupling in terms of $\delta=-\frac{e^{\gamma_E} m^2}{2\pi \mu^2}I_s(\pi)$ in a given scheme, and use the relations between the particle masses and perturbative parameters obtained, for instance,  in~\cite{Forgacs:1991nk}. In Appendix~\ref{app_Large_m_thetaPi} we determine the SU$(2)$ Thirring coupling in the Pauli-Villars regularization to subleading order $\mO(\delta^2)$ by matching  its finite isospin density partition function with a perturbative calculation in the complete Schwinger model. Comparing with the results of~\cite{Forgacs:1991nk}, we determine that the mass of the solitons at strong coupling to subexponential accuracy reads:
\begin{equation}\label{eq_kink_mass_integrability}
    M_{\text{sol}}\simeq K\, m\, e^{-A_s\frac{ g^2}{m^2}}\left(1+\mathcal{O}\left(\frac{m^2}{g^2}\right)\right),\qquad
    K=0.9945(2)\,.
\end{equation}
The gap of the two-particle continuum is therefore twice~\eqref{eq_kink_mass_integrability}.

Let us finally note that since the solitons transform as isospin doublets, the gauge-invariant two-particle states are fourfold degenerate. This implies that the lowest isosinglet and isotriplet mesons at $\theta<\pi$ must become degenerate as $\theta\to\pi$, as we observed in Figure~\ref{fig:mass_deg}, and more generally that all bound-states deconfine and approach the two-soliton threshold at $\theta=\pi$. These degeneracies can be shown explicitly at weak coupling; see Appendix~\ref{app_weak_coupling_NR}.  At strong coupling, the $\sigma$–$\pi$ mass ratio~\eqref{eq_ratio} indeed decreases as $\theta$ is increased from $0$ to $\pi$, but the analysis of the previous section cannot be trusted all the way to $\theta=\pi$. More precisely, the $\theta\neq\pi$ analysis breaks down when the leading-order $\pi$ mass derived from~\eqref{eq_EFT_2flavor} becomes comparable to the strong-coupling scale induced by the marginal perturbation, namely when 
\begin{equation}
    M_{\pi}\sim \Lambda_{IR}\quad\implies\quad
    |\pi-\theta|\sim \sqrt{\frac{m}{g}}\,e^{-A_s\frac{ 3g^2}{2m^2}}\,.
\end{equation}

\subsection{Numerical results for arbitrary coupling}

\begin{figure}[t]
    \centering
    \includegraphics[width=0.7\linewidth]{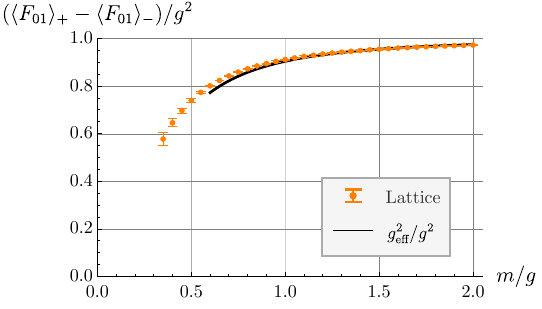}
    \caption{The difference in the electric field expectation value between the two vacua in the $C$-broken phase at $\theta = \pi$. The lattice values agree well with \eqref{eq_C_breaking_bkd} at large mass.}
    \label{fig:soliton_charge}
\end{figure}

Let us now present our numerical results. In Figure~\ref{fig:soliton_charge}, we show the difference in the expectation value of the electric field in the two vacua for different values of $m/g$.  The physical interpretation of this quantity is the electric charge $Q_\text{sol}$ of the soliton interpolating between the two vacua. As expected, $Q_\text{sol}$ decreases as we increase the coupling, and it is in good agreement with the weak coupling prediction~\eqref{eq_C_breaking_bkd}. At strong coupling, $Q_\text{sol}$ decreases exponentially to zero.

\begin{figure}[t]
    \centering
    \includegraphics[width=0.7\linewidth]{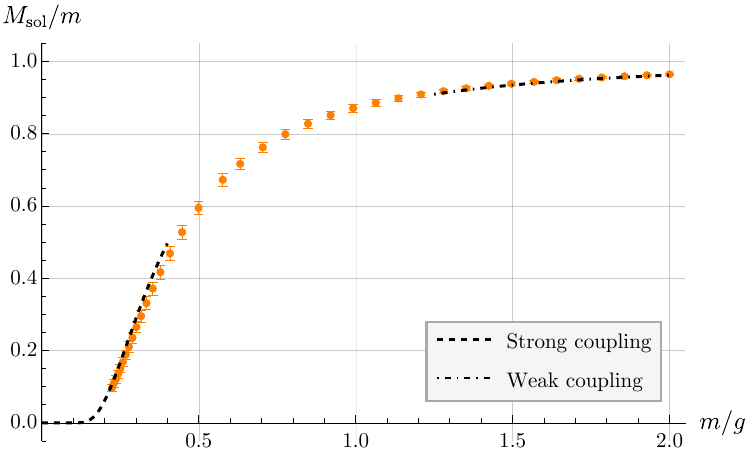}
    \caption{The numerical mass of the lowest soliton state (the half-asymptotic particle) at $\theta=\pi$.  The dashed line is the two-loop strong coupling result \eqref{eq_kink_mass_integrability}, and the dash-dotted line is the weak coupling result \eqref{eq_m_shift_main}. Both are consistent with lattice data within error bars.}
    \label{fig:kink_mass}
\end{figure}

In Figure~\ref{fig:kink_mass}, we plot the mass of the soliton as a function of $m/g$. This plot indicates the onset of the strong coupling behavior, showing an exponential decay of the mass gap. The lattice data appears consistent with the theoretical estimate \eqref{eq_kink_mass_integrability} at strong coupling, as well as the prediction \eqref{eq_m_shift_main} at weak coupling.

\begin{figure}[t]
        \centering
    \includegraphics[width=0.7\linewidth]{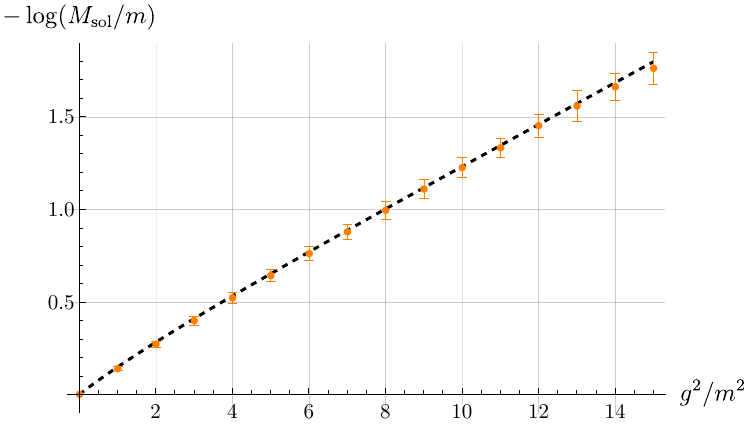}
    \caption{Logarithmic plot of the mass of the lowest soliton state. The dashed line is the [2,1] Pad\'e approximant that interpolates between the weak~\eqref{logkink_weak} and the strong coupling \eqref{logkink_strong} results.}
    \label{fig:kink_massB}
\end{figure}

In Figure~\ref{fig:kink_massB} we plot
$ -\log (M_{\text{sol}}/m) $ versus $g^2/m^2$. We compare the numerical results with the two-sided [2,1] Pad\'e approximant interpolating between the weak coupling result
\begin{equation}\label{logkink_weak}
  -\log \left ( \frac{M_{\text{sol}}}{m}\right ) =\dfrac{g^2}{2\pi m^2} -
   \frac{(3 \pi ^2-16)g^4}{96 \pi^2m^4}
   + \mathcal{O}\left( \dfrac{g^6}{m^6} \right)
\end{equation}
and the strong coupling formula
\begin{equation}\label{logkink_strong}
-\log \left ( \frac{M_{\text{sol}}}{m}\right ) = A_s \dfrac{g^2}{m^2} - \log K  + \mathcal{O}\left( \dfrac{m^2}{g^2} \right )
\,.
\end{equation}
Requiring the approximant to match the slope in the strong coupling limit gives
\begin{equation} \label{Pade}
    -\log \left ( \frac{M_{\text{sol}}}{m}\right )\Bigg\vert_{\text{Pad\'e}}= \dfrac{g^2}{m^2}\frac{(16-3\pi^2)A_s\frac{g^2}{m^2} + 48\pi A_s - 24}{(16-3\pi^2)\frac{g^2}{m^2}+96 \pi^2 A_s-48\pi}
    \,.
\end{equation}
Expanding this expression at strong coupling and comparing with \eqref{logkink_strong} gives $K = 0.853$ which is smaller than the exact result $\approx 0.99$ (we hope that including higher orders of perturbation theory can improve the agreement). The Pad\'e continuation and the numerical data are in good agreement, as shown in Figure \ref{fig:kink_massB}.

\section{Unequal masses for \texorpdfstring{$\theta=\pi$}{θ=π}: the return of confinement}
\label{confret}

The one-flavor massive Schwinger model with $\theta=\pi$ is not confining for $g\ll m$ due to the presence of half-asymptotic particles \cite{Coleman:1976uz}. However, it undergoes a confining quantum phase transition at $g_*\approx 3m$, which is related to the restoration of the charge conjugation symmetry $C$ \cite{Coleman:1976uz, Byrnes:2002nv,ArguelloCruz:2024xzi, Fujii:2024reh}.
A surprising property of the two-flavor model with $\theta=\pi$, which is relevant to Coleman's third puzzle, is that there is no similar confinement transition for $m_1=m_2$. As explained in \cite{Dempsey:2023gib}, and in this paper, for any $g$ this SO$(3)$ symmetric theory is in the deconfined phase characterized by spontaneous breaking of $C$. 

\begin{figure}[t]
    \centering
    \includegraphics[width=0.6\linewidth]{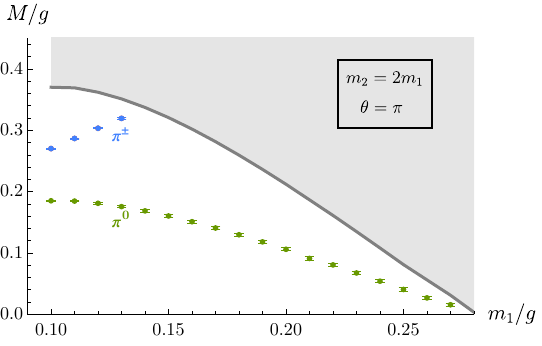}
    \caption{At $\theta = \pi$ with $m_2 = 2m_1$, we see a $\pi^0$ meson (green dots) that becomes massless at a $\mathbb{Z}_2$ Ising critical point with $g_*\approx 3.57 m_1$. For $m_1/g\lesssim 0.14$, we also observe the $\pi^+$ and $\pi^-$ mesons (blue dots) below the two $\pi^0$ continuum (gray shaded region).}
    \label{fig:heavylight}
\end{figure}

For completeness, let us discuss what happens for 
$m_2 \neq  m_1$ at $\theta = \pi$. 
For any $m_1 \neq m_2$, we expect a confining transition to occur at some critical value $g_*$ where the theory exits the $C$ broken region of the phase diagram. The critical value $g_*$ is obtained from the intersection of an Ising line with the straight line where $m_2/m_1$ is constant.\footnote{In the $\theta=0$ phase diagram in Figure~\ref{fig:Phase_Diagram}, one has to consider the line with $m_2/m_1$ fixed and negative.} For $g>g_*$, bound states can form. This is easy to understand for the case $m_2\gg m_1$, where the second flavor is much heavier and can be integrated out \cite{Dempsey:2023gib}. Then $g_*\approx 3m_1$ and the confining transition reduces to that in the one-flavor model.
For $g> g_*$ the spectrum contains the ``light-light" bound state $\pi^0$, as well as heavier mesons, by continuity with the rest of the phase diagram. As $m_2/m_1$ is reduced, additional bound states can become visible in the strong coupling phase.  In Figure~\ref{fig:heavylight}, we show numerical evidence for this phenomenon: with the mass ratio fixed to $m_2/m_1 = 2$, the confining transition takes place at $g_*\approx 3.57 m_1$. For $g> g_*$, the $\pi^0$ bound state appears, and for $g> 7.14 m_1$, the $\pi^+$ and $\pi^-$ mesons become visible below the two $\pi^0$ continuum.

When $m_2/m_1$ gets close to $1$, $g_*$ becomes much bigger than $3 m_1$. Indeed, a nontrivial value of $m_1-m_2=2\delta m$ results in a relevant perturbation $\sim \delta m \sqrt{\mu}\Re\text{Tr}(:U:)$ in the strong coupling EFT\@. Therefore, as remarked in~\cite{Dempsey:2023gib}, the $C$-breaking phase persists only in an exponentially small region specified by $|\delta m|^{2/3}g^{1/3}\lesssim M_{\text{sol}}$. For $\delta m \ll m$, 
where $m=(m_1+m_2)/2$,
we infer the following scaling for the critical coupling
\begin{equation}
\label{critcoupl}
    g_{*}^2
    \propto m^2\log\left(\frac{m^2}{\delta m^2}\right)\, .
\end{equation}
Within the $C$-broken region at $g^2<g_*^2$, a nonzero value of $\delta m$ induces a splitting between the doublet masses of order $\delta M_{\text{sol}}\sim |\delta m|^{2/3}g^{1/3}$ at strong coupling. In the phase with $g>g_*$ there are bound states, so confinement returns. Note that the $\theta=\pi$ theory with $m_2>m_1>0$ is equivalent to the $\theta=0$ model with $m=\frac{m_2-m_1}{2}$ and $\delta m=\frac{m_2+m_1}{2}$. We can therefore apply the results of Section~\ref{isobreak} for the spectrum in the strong coupling regime provided $g\gg g_*$.

\section*{Acknowledgments}

We thank Curt Callan, Grisha Tarnopolsky, Yifan Wang, Edward Witten, and Peter Zoller for useful discussions. 
This work was supported in part by the Simons Foundation Grant No.~917464 (Simons Collaboration on Confinement and QCD Strings), the US Department of Energy under Award No. DE-SC0007968, and 
by the US National Science Foundation Grant PHY-2209997 and PHY-2609862. RD is supported by a Pappalardo Fellowship in Physics at MIT.

\appendix

\section{The equal mass \texorpdfstring{$\theta\neq \pi$}{θ≠π} model at weak coupling}\label{app_weak_coupling_NR}

\subsection{Quantum mechanics in a linear potential}

The meson spectrum at weak coupling is determined by the non-relativistic Hamiltonian~\eqref{eq_H_NR}. In this Appendix, we review the eigenfunctions of this Hamiltonian.

To simplify the time-independent Schr\"odinger equation
\begin{equation}
    -\frac{1}{m} \frac{d^2 \psi}{dx^2} +\frac{g^2}{2}\left(|x|-\frac{\theta}{\pi}x\right)\psi(x)=E\,\psi(x)\,,
\end{equation}
it is convenient to define dimensionless variables
\begin{equation}\label{eq_app_dimless_vars}
    y \equiv \frac{g^\frac{2}{3} m^\frac{1}{3}}{2^\frac{1}{3}}x\,,\qquad z \equiv \frac{(4m)^\frac{1}{3}}{g^\frac{4}{3}}E\,.
\end{equation}
The time-independent Schr\"odinger equation then becomes
\begin{equation}
    - \frac{d^2 \psi}{dy^2} +\left(|y|-\frac{\theta}{\pi}y\right)\psi(y) =z\,\psi(y)\,.
\end{equation}
This equation is closely related to the Airy equation, and it can be solved by gluing together Airy functions in a way that makes the eigenfunction $\psi$ vanish as $y\rightarrow\pm\infty$:
\begin{equation}
    \psi(y)=\begin{cases}
    C_R\, \text{Ai}\left(y \left(1-\frac{\theta }{\pi }\right)^\frac{1}{3}-z\left(1-\frac{\theta }{\pi}\right)^{-\frac{2}{3}}\right) & y>0 \\
    C_L\, \text{Ai}\left(-y \left(1+\frac{\theta }{\pi }\right)^\frac{1}{3}-z\left(1+\frac{\theta }{\pi}\right)^{-\frac{2}{3}}\right) & y<0\,.
    \end{cases}
\end{equation}
The ratio $C_L/C_R$ and the  dimensionless eigenenergy $z$ are determined from the continuity conditions at $y=0$:
\begin{equation}
    \psi(0^+)=\psi(0^-)\quad\text{and}\quad
    \frac{d\psi}{dy}\Bigg|_{y = 0^+} = \frac{d\psi}{dy}\Bigg|_{y = 0^-}\,.
\end{equation}
Below we discuss the cases $\theta=0$ and $0<\theta<\pi$ separately.  We denote the value of $z$ corresponding to the $n$th excited state in the theory with topological angle $\theta$ by $z_n(\theta)$.
\begin{itemize}
    \item $\bm{\theta=0}$. At $\theta=0$, the system has parity symmetry and thus the eigenfunctions are either even or odd. In this case the solution simplifies. For even level we have $C_L=C_R$ and thus
\begin{equation}
    \psi_{2n}(y)=C\,\text{Ai}\left(|y| -z_{2n}(0)\right)\,,
\end{equation}
while for odd level we have instead $C_L=-C_R$, and so
\begin{equation}
    \psi_{2n+1}(y)=C\;\text{sgn}(y)\;\text{Ai}\left(|y| -z_{2n+1}(0)\right)\,,
\end{equation}
where $z_n(0)$ with $n=0,1,2,\ldots$ denotes the eigenfrequencies at $\theta=0$. These values are fixed by the conditions
\begin{equation}
 \begin{aligned}
    \text{even $n$}:& \qquad \text{Ai}'\left( -z_{2n}(0)\right)=0 \\
    \text{odd $n$}:&  \qquad  \text{Ai}\left( -z_{2n+1}(0)\right)=0 \,.
 \end{aligned}
\end{equation}
For the first few states we find
\[
\begin{array}{c|cccccc}
n & 0 & 1 & 2 & 3 & 4 & 5 \\ \hline
\text{parity} 
& \text{even} & \text{odd} & \text{even} & \text{odd} & \text{even} & \text{odd} \\[4pt]
z_n(0) 
& 1.01879 & 2.33811 & 3.24820 & 4.08795 & 4.82010 & 5.52056 
\end{array}
\]
\item $\bm{0<\theta<\pi}$. Let us set
\begin{equation}
    \alpha_{\pm}=1\pm\frac{\theta}{\pi}\,.
\end{equation}
The continuity conditions now read
\begin{equation}
    \begin{aligned}
    C_R \text{Ai}\left(-z_n(\theta)\alpha_-^{-\frac{2}{3}}\right) &= C_L \text{Ai}\left(-z_n(\theta)\alpha_+^{-\frac{2}{3}}\right)\,,\\
    C_R\alpha_-^\frac{1}{3}\text{Ai}'\left(-z_n(\theta)\alpha_-^{-\frac{2}{3}}\right) &= -C_L\alpha_+^\frac{1}{3}\text{Ai}'\left(-z_n(\theta)\alpha_+^{-\frac{2}{3}}\right)\,.
\end{aligned}
\end{equation}
Eliminating $C_L/C_R$, we find
\begin{equation}\label{eq_app_cond_z}
    \alpha_-^\frac{1}{3}\frac{\text{Ai}'\left(-z_n(\theta)\alpha_-^{-\frac{2}{3}}\right)}{\text{Ai}\left(-z_n(\theta)\alpha_-^{-\frac{2}{3}}\right)}+\alpha_+^\frac{1}{3}\frac{\text{Ai}'\left(-z_n(\theta)  \alpha_+^{-\frac{2}{3}}\right)}{\text{Ai}\left(-z_n(\theta)\alpha_+^{-\frac{2}{3}}\right)}=0\,.
\end{equation}
We plot the numerical solutions for $z_0(\theta)$ and $z_1(\theta)$ in Figure~\ref{fig:app_z_NR_QM}. All eigenvalues approach zero for $\theta\rightarrow \pi$, since in this limit the potential flattens on one side and the theory deconfines.
\end{itemize}

\subsection{Isosinglet-isotriplet splitting from perturbation theory} \label{App:SingletTriplet}

The energy of the isosinglet states receives a contribution from the contact term~\eqref{eq_DeltaH_NR}:
\begin{equation}
    \Delta E_n=\langle\psi_n|\Delta H_{\text{singlet}}|\psi_n\rangle\,,\qquad \Delta H_\text{singlet} = \frac{g^2}{2m^2}\delta(x)\,.
\end{equation}
Evaluating this, we find 
\begin{equation}\label{eq_app_DeltaE_QM}
   \Delta E_n=
   \frac{g^\frac{8}{3}}{2^\frac{7}{3}m^\frac{5}{3}}\Delta_n(\theta)\,,
\end{equation}
where 
\begin{equation}\label{eq_app_Delta}
    \Delta_n(\theta)=\frac{2\alpha_-\alpha_+A_-^2A_+^2}{2z_n(\theta)A_-^2A_+^2+\alpha_+\alpha_-^{\frac{2}{3}}(A_+A'_-)^2+\alpha_-\alpha_+^{\frac{2}{3}}(A_-A_+')^2}\,,
\end{equation}
where we used the shorthand notation
\begin{equation}
    A_{\pm}=\text{Ai}\left(-z_n(\theta)\alpha _{\pm}^{-\frac{2}{3}}\right)\,,\quad
    A_{\pm}'=\text{Ai}'\left(-z_n(\theta)\alpha _{\pm}^{-\frac{2}{3}}\right)\,.
\end{equation}
For $\theta=0$, this expression simplifies to
\begin{equation} \label{GotDeltan}
    \Delta_n(0)=\begin{cases}
        \displaystyle\frac{1}{z_n(0)} & \text{for even $n$} \,,\\[0.6em]
        0 & \text{for odd $n$} \,.
    \end{cases}
\end{equation}
We plot the first four $\Delta_n(\theta)$ in Figure~\ref{fig:app_Delta_NR_QM}. Note that the singlet-triplet splitting in the first excited level is smaller than for the ground-state at all values of $\theta$. Singlets and triplets become degenerate for $\theta\rightarrow \pi$ since the theory deconfines at $\theta=\pi$.

\subsection{'t Hooft equation}\label{app_tHooft}

There is a different way to obtain a systematic weak-coupling expansion for the bound state masses, using a manifestly relativistic framework, as follows. We start by quantizing the theory in lightcone coordinates $x^{\pm} = \frac{x^0 \pm x^1}{\sqrt{2}}$, with $x^+$ considered as a ``time'' coordinate and $x^-$ considered as a ``space'' coordinate, in the gauge $A_- = 0$.   \cite{Harada:1993va,Burkardt:1997bw,Eller:1986nt}.  For explicit calculations we use $\gamma^0 = \sigma_2$ and $\gamma^1 = i\sigma_1$, as given in Section~\ref{sec:setup}. We represent Dirac fermions as
\begin{equation}
    \psi_\alpha = \dfrac{1}{2^{\frac{1}{4}}}\begin{pmatrix}
    \xi_\alpha\\\zeta_\alpha
\end{pmatrix} \,.
\end{equation}
With the choices above, the $\zeta_\alpha$ component of the fermions and the $A_+$ component of the gauge field are non-dynamical (their $x^+$ derivatives do not appear in the action) and can be integrated out.  When integrating out $A_+$ using its equations of motion, note that $\theta$-term gives an additional contribution $A_+^{(b)} =\frac{\theta g^2}{2\pi}x^-$ corresponding to a uniform background field of strength $\frac{\theta g^2}{2\pi}$ as in \cite{Coleman:1976uz}. The resulting light-cone momentum $P^+$ and light-cone Hamiltonian $P^-$ are as follows:
\begin{equation}\label{eq_Lightcone_Ham}
 \begin{aligned}
    P^+ &= i\int dx^- \xi^{\dagger\alpha} \pd_- \xi_\alpha \,, \\
    P^- &= \dfrac{1}{2}\int dx^- \left[  m^2 \xi^{\dagger\alpha} \dfrac{1}{i\pd_-} \xi_\alpha + g^2 \xi^{\dagger\alpha} \xi_\alpha \dfrac{1}{(i\pd_-)^2} \xi^{\dagger\beta} \xi_\beta - \theta\dfrac{ g^2}{\pi} x^- \xi^{\dagger\alpha} \xi_\alpha  \right] \,.
 \end{aligned}
\end{equation}

The field $\xi_\alpha$ is then expanded in terms of the standard creation and annihilation operators:
\begin{equation}
    \xi_\alpha = \int_0^{\infty} \dfrac{dp}{\sqrt{2\pi}} \left[ b_{\alpha}(p) e^{-i p x^-} + d_{\alpha}^\dagger (p) e^{i p x^-} \right] \,;
\end{equation}
the only non-trivial commutation relations are
\begin{equation}
    \{b^{\dagger\alpha}(p),b_\beta(k)\} = \{d^\alpha(p),d_\beta^\dagger(k)\} =\delta_\beta^\alpha \delta(p-k) \,.
\end{equation} Note that the lightcone momentum is always positive. For this reason, the Fock vacuum $\ket{0}$ is the interacting ground state \cite{Brodsky:1997de}. Note that the lightcone momentum operator $P^+$ admits a very simple expression in terms of these operators:
\begin{equation}\label{eq_Lightcone_Mom}
    P^+ = \int_0^\infty dp\,p\left[b^{\dagger\alpha}(p) b_\alpha(p) + d^\dagger_\alpha(p) d^\alpha(p)\right].
\end{equation}

In this approach, the gauge-invariant states have to be charge-neutral, as the gauge field is integrated out. Since at weak coupling the mesons are fermion-antifermion bound states, we can restrict to the two-particle sector.  In this sector, we can either form flavor singlets or adjoints.  A singlet state with light-cone momentum $P^+$ is of the form 
\begin{equation}
    \ket{\psi_s}= \int_0^1dt\, \chi_s(t) b^{\dagger\alpha}\left(tP^+\right) d^\dagger_{\alpha}\left((1-t)P^+\right)\ket{0}\,,
\end{equation}
where $\chi_s(t)$ is its wavefunction.  An adjoint state with light-cone momentum $P^+$ is of the form
 \begin{equation}
   \ket{\psi_a(T)}= T_\alpha^{\;\beta}\int_0^1dt\, \chi_a(t) b^{\dagger\alpha}\left(tP^+\right) d^\dagger_{\beta}\left((1-t)P^+\right)\ket{0}\,,
 \end{equation}
where $\chi_a(t)$ is its wavefunction, and $T$ is a traceless hermitian matrix.  In the two-particle sector, the lightcone Hamiltonian \eqref{eq_Lightcone_Ham} reduces to
\begin{equation}\label{eq_2to2_Ham}
\begin{aligned}
    &P^-_{2\to2} = \dfrac{1}{2}\left( m^2 - \dfrac{g^2}{\pi} \right)\int_0^\infty dp \,\dfrac{b^{\dagger\alpha}(p) b_\alpha(p)+d_\alpha^{\dagger}(p) d^\alpha(p)}{p}\\
    &+\dfrac{g^2}{2\pi}\int_0^\infty dp\, dp'\, dq\, dq'\,\delta(p+p'-q-q' )\left[ \dfrac{b^{\dagger\alpha}(p)d_\beta^\dagger(p')b_\alpha(q)d^\beta(q')}{(p-q)^2} -\dfrac{b^{\dagger\alpha}(p) d^\dagger_\alpha(p') b_\beta(q) d^\beta(q')}{(p+p')^2}\right]\\
    &+i\theta \dfrac{g^2}{2\pi} \int_0^\infty dp\,dq\,\delta'(p-q)\left[b^{\dagger\alpha}(p) b_\alpha(q) - d^\dagger_\alpha(p) d^\alpha(q)\right]
    \end{aligned}
\end{equation}
after transforming to the normal-ordered form. Here, we ignored the terms that take us away from the two-particle sector, as well as those that annihilate all two-particle states. The divergences that naively arise from inverting the gauge kinetic term $(i\pd_-)^2$ are treated using the principal value prescription \cite{tHooft:1974pnl}. We give a more detailed explanation in Appendix \ref{app_Pert_Corr}.

Using the expression for $P^+$ in \eqref{eq_Lightcone_Mom} and the approximate expression for $P^-$ in \eqref{eq_2to2_Ham}, we can now diagonalize the mass operator $M^2 = 2 P^+ P^-$ \cite{Burkardt:1997bw}.  For the  singlet states, one obtains the eigenvalue equation
\begin{equation}\label{eq_mes_mass_s}
    M^2 \chi_s(t) = \left( m^2 - \dfrac{g^2}{\pi} \right)\dfrac{ \chi_s(t)}{t(1-t)} - \dfrac{g^2}{\pi} \mathcal{P}\int_0^1 ds\, \dfrac{\chi_s(s)}{(t-s)^2} + \dfrac{g^2N_f}{\pi} \int_0^1 ds\,\chi_s(s) - i\theta\dfrac{g^2}{\pi} \pd_t \chi_s(t)\,,
\end{equation}
while for the adjoint states one obtains
\begin{equation}\label{eq_mes_mass_a}
    M^2 \chi_a(t) = \left( m^2 - \dfrac{g^2}{\pi} \right)\dfrac{ \chi_a(t)}{t(1-t)} - \dfrac{g^2}{\pi} \mathcal{P}\int_0^1 ds\, \dfrac{\chi_a(s)}{(t-s)^2}  - i\theta\dfrac{g^2}{\pi} \pd_t \chi_a(t)\,.
\end{equation}
In these expressions, we state the principal value integration explicitly. Note that the penultimate term in \eqref{eq_mes_mass_s} vanishes for parity-even singlets at $\theta = 0$, since these singlets have wave functions that are antisymmetric with respect to $s=1/2$. The formerly discussed non-relativistic Hamiltonian \eqref{eq_H_NR} with the singlet contact term \eqref{eq_DeltaH_NR} can be obtained by the change of variables $t = \frac{1}{2} + \frac{k}{2m}$, followed by a Fourier transform in $k$ and an expansion in powers of $1/m$.

The equations~\eqref{eq_mes_mass_s} and~\eqref{eq_mes_mass_a} resum all planar diagrams without fermion loops \cite{tHooft:1974pnl}. For instance, these equations incorporate the one-loop correction to the fermion mass.  We can further improve this approach by incorporating the loop corrections to the coupling \eqref{eq_g_eff} at energies much below the fermion mass and the two-loop mass shift \eqref{eq_m_shift_main} (see Appendix~\ref{app_Pert_Corr}):
\begin{equation}\label{eq_renorm_thooft}
 \begin{aligned}
    M^2 \chi_s(t) &= m_{\text{pole}}^2\dfrac{ \chi_s(t)}{t(1-t)} - \dfrac{g_{\text{eff}}^2}{\pi} \mathcal{P}\int_0^1 ds\, \dfrac{\chi_s(s)}{(t-s)^2} + \dfrac{g^2N_f}{\pi} \int_0^1 ds\,\chi_s(s) - i \theta \dfrac{g_{\text{eff}}^2}{\pi}  \pd_t \chi_s(t)\,, \\
      M^2 \chi_a(t) &= m_{\text{pole}}^2\dfrac{ \chi_a(t)}{t(1-t)} - \dfrac{g_{\text{eff}}^2}{\pi} \mathcal{P}\int_0^1 ds\, \dfrac{\chi_a(s)}{(t-s)^2}  - i \theta \dfrac{g_{\text{eff}}^2}{\pi}  \pd_t \chi_a(t)\,, 
 \end{aligned}
\end{equation}
where $g_{\text{eff}}^2$ and $m_{\text{pole}}$ are given, respectively, in~\eqref{eq_g_eff} and~\eqref{eq_m_shift_main}. 

We solve this equation numerically using a basis of Chebyshev polynomials, as discussed in \cite{Hanson:1976ey,Kochergin:2024quv}. Due to the Lorentz covariance of this approach, we expect the results to extrapolate well also at moderate coupling values. The results we find for the masses of the four lightest particles at $g/m=1$ are plotted in Figure~\ref{fig:mass_degB} and are in excellent agreement with the numerical lattice Hamiltonian results. Note that all the states become degenerate at $\theta = \pi$.

To estimate the error of this approach, we consider subleading loop contributions. The first such effect is the renormalization of the contact term. Note that the gauge coupling entering the contact term is evaluated at a scale $s\simeq 2m$, so its loop corrected value cannot be computed by the simple replacement $g\rightarrow g_{\text{eff}}$, as for the potential term. Indeed, to determine the coefficient of the contact term at subleading order one should account for the vacuum polarization diagrams both in the s- and t-channel, as well as for the one-loop vertex correction and the four-fermion box diagrams in the $2$-to-$2$ amplitude.
Since the leading order contribution from the contact term according to \eqref{eq_stsplitting} scales as $\Delta M/g \sim (g/m)^{5/3}$, the contribution from the one-loop corrections that we are neglecting is at most of order $\Delta M/g \sim (g/m)^{11/3}$. Hence, we expect the predictions for $M/g$ obtained using~\eqref{eq_renorm_thooft} to be accurate up to order $\mathcal{O}\left( (g/m)^{11/3} \right)$ at least.

The two-loop mass shift in $m_\text{pole}$ is a correction of order $\mathcal{O}\left( (g/m)^{3} \right)$, so its inclusion is warranted. At higher orders, one needs to include additional terms in the potential, corresponding to higher-derivative contact interactions $\sim\delta^{(n)}(x)$. It would be interesting to develop a systematic procedure to all orders, analogous to the well studied EFT matching in non-relativistic QED~\cite{Caswell:1985ui}.

Finally, there are powerful methods to study equations like \eqref{eq_renorm_thooft} analytically \cite{Fateev:2009jf,Ziyatdinov:2010vg,Litvinov:2024riz,Artemev:2025cev}. For instance, a systematic expansion was developed in \cite{Ziyatdinov:2010vg} for $\theta = 0$ and without the contact term. Incorporating the loop corrections discussed above into that approach, we find:
\begin{equation}
\begin{aligned}
    \dfrac{M_{a,1}}{g} &= 2\dfrac{m}{g} + \frac{z_0}{4^\frac{1}{3}} \dfrac{g^\frac{1}{3}}{m^\frac{1}{3}}-\frac{1}{\pi } \frac{g}{m} + \frac{4-z_0^3}{40\; 2^{\frac{1}{3}}z_0}\frac{g^\frac{5}{3}}{m^\frac{5}{3}}- \dfrac{2^{\frac{1}{3}}z_0}{36 \pi}\frac{g^{\frac{7}{3}}}{m^{\frac{7}{3}}} \\
    &+\left(\frac{11 z_0^3}{5600}-\frac{1}{200 z_0^3}-\frac{1}{12\pi^2}+\frac{27}{400}\right)\frac{g^3}{m^3} + \mathcal{O}\left ( \frac{g^\frac{11}{3}}{m^\frac{11}{3}}\right)\,,\\
    \dfrac{M_{a,2}}{g} &= 2\dfrac{m}{g} + \frac{z_1}{4^\frac{1}{3}} \dfrac{g^\frac{1}{3}}{m^\frac{1}{3}}-\frac{1}{\pi } \frac{g}{m}-\frac{z_1^2}{40 \sqrt[3]{2}} \frac{g^\frac{5}{3}}{m^\frac{5}{3}}- \dfrac{2^{\frac{1}{3}}z_1}{36 \pi} \frac{g^\frac{7}{3}}{m^\frac{7}{3}}\\
    & + \left(\frac{11 z_1^3}{5600}-\frac{1}{12 \pi ^2}+\frac{59}{560}\right)\frac{g^3}{m^3}+ \mathcal{O}\left ( \frac{g^\frac{11}{3}}{m^\frac{11}{3}}\right)\,.
\end{aligned}
\end{equation}
Here, $M_{a,1}$ and $M_{a,2}$ are the masses of the two lightest adjoint states, and $z_n = z_n(0)$ are the same as in the non-relativistic calculation~\eqref{eq_app_cond_z}. Since we are neglecting the contact term, these expressions are not applicable for isosinglets. It would be interesting to extend these results to $\theta \neq 0$ and nonzero contact term cases. Note that a naive substitution of the loop-corrected quantities into the non-relativistic result would differ from this expansion already at $\left(g/m \right)^{5/3}$, so the relativistic corrections discussed above provide an important improvement.

\section{Mass spectrum at strong coupling from Abelian bosonization and integrability}\label{app_Large_m_theta0}

In this appendix we use Abelian bosonization and integrability to study the mass spectrum of the theory at strong coupling. We work for arbitrary real values of the masses
\begin{equation}
    m_1=m+\delta m\,,\quad m_2=m-\delta m\,.
\end{equation}
The results in Section~\ref{sec_equal_m_theta_Not_Pi} are obtained specializing to $m_1=m_2$.

\subsection{The low energy EFT at strong coupling}

Via Abelian bosonization~\cite{Coleman:1974bu,Mandelstam:1975hb}, we can rewrite the Lagrangian density~\eqref{eq_Lag} in terms of two $\sqrt{\pi}$-periodic scalars as
\begin{equation}
    \mL=\frac{1}{2}\sum_{\alpha=1}^2\pd_\mu \phi_\alpha \pd^\mu \phi_\alpha-\frac{1}{\sqrt{\pi}}F_{01}\left(\sum_{\alpha=1}^2\phi_\alpha+\frac{\theta}{2\sqrt{\pi}}\right)+\frac{1}{2g^2}F_{01}^2+\frac{e^{\gamma_E}}{2\pi}\sum_{\alpha=1}^2 m_{\alpha}\tilde{\mu}N_{\tilde{\mu}}\cos (2\sqrt{\pi}\phi_\alpha)\,,
 \end{equation}
where $N_{\tilde{\mu}}$ denotes normal ordering with respect to an arbitrary mass $\tilde{\mu}$, and $\gamma_E$ is the Euler-Mascheroni constant. It is convenient to integrate out the electric field and introduce the linear combinations
\begin{equation}
    \phi_{\pm}=\frac{\phi_1\pm\phi_2}{\sqrt{2}}\,.
\end{equation}
After Wick-rotating to Euclidean signature, the Lagrangian density becomes
\begin{equation}\label{eq_bosonized_app}
\begin{split}
\mL &=\frac12 \pd_\mu \phi_+ \pd^\mu \phi_++\frac{g^2}{\pi }\min_{n\in\mathds{Z}}\left(\phi_++\frac{\theta +2\pi n}{2\sqrt{2\pi }}\right)^2
\\
&+\frac12 \pd_\mu \phi_- \pd^\mu \phi_--\frac{e^{\gamma_E}}{\pi}\sqrt{\mu\,\mu_-}\left(m\,\mO^{(1)}_+\mO^{(1)}_--\delta m\,\mO^{(2)}_+\mO^{(2)}_-\right)
\,,
\end{split}
\end{equation}
where we have been careful about the global quantization of the electric field~\cite{Komargodski:2020mxz}, and we defined
\begin{gather}
\mO^{(1)}_{+}=N_{\mu}\cos(\sqrt{2\pi}\,\phi_+)\,,
\qquad
\mO^{(1)}_{-}=N_{\mu_-}\cos(\sqrt{2\pi}\,\phi_-)
\,,\\
\mO^{(2)}_{+}=N_{\mu}\sin(\sqrt{2\pi}\,\phi_+)\,,
\qquad
\mO^{(2)}_{-}=N_{\mu_-}\sin(\sqrt{2\pi}\,\phi_-)
\,.
\end{gather}
For $m=\delta m =0$, the fields $\phi_+$ and $\phi_-$ decouple and become free, with $\phi_+$ having mass $\mu=\sqrt{\frac{2}{\pi}}g$, while $\phi_-$ being massless.

To analyze the limit of small but nonzero mass, it is convenient to integrate out the massive scalar $\phi_+$ and work within an EFT for the light field $\phi_-$. Up to order $\mO(1/\mu^2)$ this can be done treating perturbatively the mass term and performing the path-integral over $\phi_+$, as in~\cite{Smilga:1998dh,Dempsey:2023gib}. Explicitly, the effective Lagrangian density reads
\begin{equation}\label{eq_app_EFT_pre}
\begin{aligned}
    \mL_\text{EFT} &= \frac{1}{2}(\pd\phi_-)^2
    -\frac{e^{\gamma_E}}{\pi}\sqrt{\mu\,\mu_-}\left(m\,
    \left \langle \mO^{(1)}_+ \right \rangle \mO^{(1)}_--\delta m\,\left \langle\mO^{(2)}_+ \right \rangle\mO^{(2)}_-\right)\\
    &{}-\frac{e^{2\gamma_E}\mu\,\mu_-}{2\pi^2}m^2\int d^2y \left \langle\mO_+^{(1)}(x)\mO_+^{(1)}(y) \right \rangle_c\,\mO_-^{(1)}(x)\mO_-^{(1)}(y)\\
    &{}-\frac{e^{2\gamma_E}\mu\,\mu_-}{2\pi^2}\delta m^2\int d^2y \left \langle\mO_+^{(2)}(x)\mO_+^{(2)}(y)\right \rangle_c\,\mO_-^{(2)}(x)\mO_-^{(2)}(y)\\
    &{}+\frac{e^{2\gamma_E}\mu\,\mu_-}{2\pi^2}m\,\delta m\int d^2y\left( \left \langle\mO_+^{(1)}(x)\mO_+^{(2)}(y) \right \rangle_c\,\mO_-^{(1)}(x)\mO_-^{(2)}(y)
    +(1\leftrightarrow2)\right)\\
    &{}+\mathcal{O}(m^{3})\,,
    \end{aligned}
\end{equation}
where the expectation values are evaluated in the $m_1 = m_2 =0$ theory, and the subscript $c$ denotes connected correlators.  In \eqref{eq_app_EFT_pre} and below, by ${\cal O}(m^n)$ we denote generically all terms of total order at least $n$ in $m_1$ and $m_2$.

To evaluate the first line of~\eqref{eq_app_EFT_pre}, we focus on the $n=0$ sector and set $\phi_+=-\theta/(2\sqrt{2\pi})$, so that we find
\begin{equation}
    \mL_\text{EFT}=\frac{1}{2}\pd_\mu \phi_- \pd^\mu \phi_--\frac{e^{\gamma_E}}{\pi}\tilde{m}\sqrt{\mu\,\mu_-}\,N_{\mu_-}\cos\left(\sqrt{2\pi}\,\phi_--\nu\right)+\mathcal{O}(m^2)\,,
\end{equation}
where $\tilde{m}$ and $\nu$ are related to $m$, $\delta m$, and $\theta$ by
\begin{equation}\label{eq_app_mt_nut}   \tilde{m}\cos\nu=m\cos\frac{\theta}{2}\,,\qquad
    \tilde{m}\sin\nu=\delta m\sin\frac{\theta}{2}\,.
\end{equation}
At the next order, it is convenient to discuss each line of~\eqref{eq_app_EFT_pre} separately. The evaluation of the second line proceeds as in~\cite{Smilga:1998dh,Dempsey:2023gib}, as follows. The connected correlator gives
\begin{equation}
    \langle\mO^{(1)}_+(x)\mO_+^{(1)}(y)\rangle_c=\frac12\left(e^{2\pi G_{\mu}(x-y)}+\cos\theta\, e^{-2\pi G_{\mu}(x-y)}\right)-\cos^2\left(\frac{\theta}{2}\right)\,,
\end{equation}
where $G_{\mu}(x)=\frac{1}{2\pi}K_0(\mu|x|)$ is the massive propagator for a scalar field of mass $\mu$. We can also use renormal-ordering to rewrite
\begin{equation}
    \begin{split}
    \mO_-^{(1)}(x)\mO_-^{(1)}(y) 
    &=\frac {e^{-2\pi G_{\mu_-}(x-y)}}{2}N_{\mu_-}\cos\left(\sqrt{2\pi}(\phi_-(x)+\phi_-(y))\right)\\
    &+\frac{ e^{2\pi G_{\mu_-}(x-y)}}{2}N_{\mu_-}\cos\left(\sqrt{2\pi}(\phi_-(x)-\phi_-(y))\right)\,.
    \end{split}
\end{equation}
To compute the integral in the second line of~\eqref{eq_app_EFT_pre} all we have to do then is to set $y=x+\xi/\mu$ and expand at large $\mu$ using
\begin{equation}
    \mO_-^{(1)}(x)\mO_-^{(1)}(y)=\frac{e^{\gamma_E}|\xi|\mu_-}{4\mu}\left(N_{\mu_-}\cos(\sqrt{8\pi}\phi_-(x))-\frac{4\pi\,e^{-2\gamma_E}}{|\xi|^2\mu_-^2}(\xi^\mu \pd_\mu\phi_-(x))^2\right)+\mathcal{O}(\mu^{-2})\,.
\end{equation}
Eventually we find
\begin{equation}\label{eq_app_Left_21}
\begin{aligned}
    \mL_\text{EFT}^{\text{2nd line}} 
    &=-\frac{\delta}{2}\pd_\mu \phi_- \pd^\mu \phi_-+\delta\frac{e^{2\gamma_E}}{4\pi}\mu_-^2 N_{\mu_-}\cos(\sqrt{8\pi}\,\phi_-)+\mathcal{O}(\mu^{-3})\,,
    \end{aligned}
\end{equation}
where
\begin{equation}\label{eq_app_Is}
\begin{split}
    \delta &=-\frac{e^{\gamma_E} m^2}{2\pi \mu^2}I_s(\theta)\,, \\[0.4em]
    I_s(\theta) &= \pi \int_0^{\infty}d\xi\,\xi^2\left(e^{K_0(\xi)} - 1 +(e^{-K_0(\xi)} - 1) \cos\theta \right)\approx 5.55418 - 4.52885\cos\theta\,.
\end{split}
\end{equation}

The third line of~\eqref{eq_app_EFT_pre} is identical to~\eqref{eq_app_Left_21} upon the replacement $(m,\theta,\phi_-)\rightarrow (\delta m ,\theta-\pi,\phi_--\sqrt{\pi/8})$.  Therefore we find
\begin{equation}\label{eq_app_Left_22}
\begin{split}
    \mL_\text{EFT}^{\text{3rd line}} 
    &=\frac{\delta_2}{2}\pd_\mu \phi_- \pd^\mu \phi_-+\delta_2\frac{e^{2\gamma_E}}{4\pi}\mu_-^2 N_{\mu_-}\cos(\sqrt{8\pi}\,\phi_-)+\mathcal{O}(\mu^{-3})\,,
    \end{split}
\end{equation}
where
\begin{equation}\label{eq_app_delta2}
    \delta_2=\frac{e^{\gamma_E} \delta m^2}{2\pi \mu^2}I_s(\pi-\theta)\,.
\end{equation}

To compute the fourth line in \eqref{eq_app_EFT_pre}, we use
\begin{equation}
    \langle \mO_+^{(1)}(x)\mO_+^{(2)}(y)\rangle_c=
    \langle \mO_+^{(2)}(x)\mO_+^{(1)}(y)\rangle_c=\frac{\sin\theta}{2}\left(1-e^{-2\pi G_{\mu}(x-y)}\right)
\end{equation}
and the renormal ordering identity
\begin{equation}
\begin{aligned}
    \mO_-^{(1)}(x)\mO_-^{(2)}(y)+\mO_-^{(2)}(x)\mO_-^{(1)}(y)&=e^{-2\pi G_{\mu_-}(x-y)}N_{\mu_-}\sin\left(\sqrt{2\pi}(\phi_-(x)+\phi_-(y))\right)\,.
\end{aligned}
\end{equation}
Setting again $y=x+\xi/\mu$ and expanding at large $\mu$, we find
\begin{equation}\label{eq_app_Left_23}
    \begin{aligned}
    \mL_\text{EFT}^{\text{4th line}} 
    &=\delta_3\frac{e^{2\gamma_E}}{4\pi}\mu_-^2 N_{\mu_-}\sin(\sqrt{8\pi}\,\phi_-)+\mathcal{O}(\mu^{-3})\,,
    \end{aligned}
\end{equation}
where
\begin{equation}\label{eq_app_delta3}
 \begin{aligned}
   \delta_3 &= \sin\theta\frac{e^{\gamma_E}m\,\delta m}{2\pi \mu^2}I_3\,,\\
    I_3 &= 2\pi \int_0^\infty d\xi \,\xi^2\left(1-e^{- K_0(\xi)}\right)\simeq 9.05771\,.
 \end{aligned}
\end{equation}

Putting everything together, we conclude that the low energy Euclidean Lagrangian density takes the following form up to (and including) terms of order $1/\mu^2$:
\begin{equation}\label{eq_app_EFT_final}
\begin{aligned}
    \mL_\text{EFT}&\simeq\frac{1-\delta+\delta_2}{2} \pd_\mu \phi_- \pd^\mu \phi_- -\frac{e^{\gamma_E}}{\pi}\tilde{m}\sqrt{\mu\,\mu_-}\,N_{\mu_-}\cos(\sqrt{2\pi}\,\phi_--\nu)\\
    &+\frac{e^{2\gamma_E}\mu_-^2}{4\pi}\left((\delta+\delta_2) N_{\mu_-}\cos(\sqrt{8\pi}\,\phi_-)+\delta_3N_{\mu_-}\sin(\sqrt{8\pi}\,\phi_-)\right)\,.
    \end{aligned}
\end{equation}
In the following subsections we will use~\eqref{eq_app_EFT_final} to compute the mass spectrum at strong coupling via integrability.

\subsection{Leading order results}

To compute the mass spectrum at strong coupling, we compare the low energy EFT~\eqref{eq_app_EFT_final} with the results in the integrability literature. It is useful to set 
\begin{equation}
    \phi_-=\frac{\phi}{\sqrt{1-\delta+\delta_2}}+\frac{\nu}{\sqrt{2\pi}}\,,
\end{equation}
and write the potential in terms of the canonically normalized vertex operators\footnote{These vertex operators are defined such that $\langle 0| : e^{ia\phi(x)}:: e^{-i a\phi(0)}:|0\rangle \approx |x|^{-\frac{a^2}{2\pi}}$ as $|x|\rightarrow 0$.}
\begin{equation}\label{eq_app_normalization}
    \mO_a \equiv : e^{ia\phi}:=\left(\frac{e^{\gamma_E}\mu_-}{2}\right)^{\frac{a^2}{4\pi}}N_{\mu_-}e^{ia\phi} \,.
\end{equation}
In this notation, we have
\begin{equation}
    \mL_\text{EFT} =\mL_0+\delta\mL\,,
\end{equation}
where 
\begin{equation}\label{eq_int_action}
    \begin{aligned}
    \mL_0 &=\frac{1}{2}\pd_\mu \phi \pd^\mu \phi 
    -2\mu_Z :\cos(\beta\phi):\,,\\
    \delta\mL &=\frac{\delta+\delta_2}{\pi}:\cos(2\beta\phi+2\nu):
    + \frac{\delta_3}{\pi} :\sin(2\beta\phi+2\nu):\,.
\end{aligned}
\end{equation}
Here, we defined the following parameters
\begin{equation}\label{eq_int_pars}
    \mu_Z=\frac{\tilde{m}}{\pi }  \sqrt{\frac{\mu \, e^{\gamma_E }}{2}}+\mathcal{O}\left(\frac{1}{g^2}\right)\,,\qquad\beta=\sqrt{\frac{2\pi}{1-\delta+\delta_2}}\,.
\end{equation}

As is well known, $\mL_{0}$ describes the integrable sine-Gordon model.  The spectrum consists of a soliton-antisoliton pair with mass $M_S$ related to $\mu_Z$ by \cite{Zamolodchikov:1995xk} 
\begin{equation}\label{eq_soliton_mass}
    \mu_Z=\kappa(p)M_S^{\frac{2}{p+1}}\,,\qquad\kappa(p)=\frac{\Gamma \left(\frac{p}{p+1}\right)}{\pi\Gamma \left(\frac{1}{p+1}\right)}\left(\frac{\sqrt{\pi } \Gamma \left(\frac{p+1}{2}\right)}{2 \Gamma \left(\frac{p}{2}\right)}\right)^{\frac{2}{p+1}}\,,
\end{equation}
where
\begin{equation} \label{pDef}
    p=\frac{\beta^2}{8\pi-\beta^2}\,.
\end{equation}
Additionally there are a number of breathers $B_k$ (with $k = 1, 2, \ldots$ and $k < 1/p$), with masses
\begin{equation}\label{eq_app_ratio_SG}
    M_{B_k}=2\sin\left(\frac{k\pi p}{2}\right) M_S \,.
\end{equation}
Working at leading order in $\delta$, $\delta_2$, and $\delta_3$, we have $\beta=\sqrt{2\pi}$ and hence $p = 1/3$, and thus there are only two breathers of masses $M_{B_1} = M_S$ and $M_{B_2} = \sqrt{3} M_S \equiv M_\sigma$~\cite{Coleman:1976uz}. The first breather is degenerate with the soliton, thus completing the pion isotriplet, while the other is an isosinglet.  From~\eqref{eq_soliton_mass} and~\eqref{eq_int_pars}, we obtain the relations~\eqref{eq_Mpi} and~\eqref{eq_Mpi_dm} between the pion mass and the Lagrangian parameters.

\subsection{Subleading corrections to the mass ratios}

To study perturbations of an integrable model, we can use the method of~\cite{Delfino:1996xp} (see also \cite{Mussardo:2020rxh}). Let us review the formulas we need. Consider an integrable model perturbed by a (relevant) local operator ${\cal O}(x)$ so that the change in the action is $-\lambda\int d^2x\,  \mO(x)$ (in Lorentzian signature).  Let us denote the matrix elements of ${\cal O}(x)$ between one-particle states as
\begin{equation}
\,_{b}F^{\mO}_{a}(w',w)=\,^\text{out}\langle b(w')|\mO(0)|a(w)\rangle^\text{in}\,,
\end{equation}
where $\ket{a(w)}^\text{in}$ and $\ket{b(w')}^\text{out}$ represent in and out asymptotic states with masses $M_a$ and $M_b$, respectively, and momenta $p_a = M_a (\cosh w,\sinh w)$ and $p_b = M_b (\cosh w',\sinh w')$, respectively. Here, single-particle states are normalized using the usual relativistic normalization $\langle b(w')|a(w)\rangle=2\pi\delta_{ab} E\delta (\vec{p}{\,'}-\vec{p})=2\pi\delta_{ab}\delta(w'-w)$. Crossing symmetry relates one-particle matrix elements to two-particle form factors as \cite{Karateev:2019ymz}
\begin{equation}
\,_{b}F^{\mO}_{a}(w',w)=
\,^\text{out}\langle 0|\mO(0)|\bar{b}(w'+i\pi)a(w)\rangle^\text{in}\equiv
F^{\mO}_{\bar{b}, a}(w'+i\pi,w)\,,
\end{equation}
where $\bar{b}$ denotes the anti-particle.  Then it can be shown that, at leading order in the coupling $\lambda$, the squared mass of the asymptotic states in the perturbed theory differs from that in the unperturbed theory by an amount~\cite{Delfino:1996xp} 
\begin{equation}\label{eq_deltaM_pre}
\delta M^2_{a}=2 \lambda F_{\bar{a},a}^{\mO}(i\pi,0)\,.
\end{equation}
The expression~\eqref{eq_deltaM_pre} is similar to the standard result in Rayleigh-Schr\"odinger perturbation theory in non-relativistic quantum mechanics,\footnote{To see that~\eqref{eq_deltaM_pre} in fact coincides with the naive result $\delta E=\langle a(0)|\mO|a(0)\rangle_{\text{NR}}$, note that there is a factor of $2$ from $\delta M^2\simeq 2M\delta
M$ and that relativistic states differ from the non-relativistic ones in their normalization by $|a(w)\rangle= \sqrt{E_a}|a(w)\rangle_{\text{NR}}$, where $|a\rangle_{\text{NR}}$ is the single-particle state with non-relativistic normalization and $E_a\stackrel{w\rightarrow0}{\longrightarrow}M_a$ is its energy.} but its derivation requires a careful treatment of the vacuum energy and follows from the $S$-matrix in the presence of the perturbation.

To use~\eqref{eq_deltaM_pre}, it is convenient to rewrite the perturbation $\delta {\cal L}$ in~\eqref{eq_int_action} as
\begin{equation}
    \delta\mL=
    \frac{1}{\pi}\left(\bar{\delta}_1:\cos(2\beta\phi):+\,\bar{\delta}_2:\sin(2\beta\phi):\right)\,,
\end{equation}
where we work in Euclidean signature, and where
\begin{equation}
    \bar{\delta}_1=(\delta+\delta_2)\cos(2\nu)+\delta_3\sin(2\nu)\,,\quad
    \bar{\delta}_2=\delta_3\cos(2\nu)-(\delta+\delta_2)\sin(2\nu)\,.
\end{equation}
Because of the symmetry $\phi\rightarrow-\phi$ of the leading order action, we can neglect the term proportional to $\bar{\delta}_2$ at linear order in perturbation theory. Therefore, we can use~\eqref{eq_deltaM_pre} with
\begin{equation}\label{eq_app_pert}
    \mO= :\cos 2\beta\phi:
\,,\qquad \lambda=\frac{\bar{\delta}_1}{\pi} \,,
\end{equation}
in order to determine the corrections to the breather and soliton squared masses.

There is a small subtlety we need to address. For $\beta = \sqrt{2\pi}$ the perturbation is marginal and, correspondingly, the matrix elements of $\mO$ diverge.\footnote{Equivalently, if we instead set $\beta=\sqrt{2\pi/(1-\delta+\delta_2)}\simeq\sqrt{2\pi}(1+\delta/2-\delta_2/2)$ we would find that such matrix elements are $\propto 1/(\delta-\delta_2)$,  thus naively defying perturbation theory.} Thus, a regularization procedure is required. To this aim, we shall treat $\beta$ as a varying parameter analogously to the number of spacetime dimensions in dimensional regularization. As in dimensional regularization, taking $\beta<\sqrt{2\pi}$ makes the perturbation strictly relevant at intermediate steps, rendering perturbation theory finite.

We remark that this regularization breaks the $\text{SO}(3)$ symmetry at intermediate steps. On the other hand, the operator~\eqref{eq_app_pert} remains compatible with the periodicity of the leading order sine-Gordon potential at all steps.  In other words, the leading order potential and the perturbation are  mutually local for all values of $\beta$. This avoids the subtle changes in the vacuum structure and the nature of the soliton states discussed in~\cite{Delfino:1997ya} for perturbations which are non-mutually local with the leading order potential.

It turns out that, within this regularized approach, although the individual matrix elements diverge, the mass ratios remain finite to the order of interest. This is related to the existence of a consistent renormalization scheme, that we shall briefly explain at the end of this section. Therefore, using~\eqref{eq_soliton_mass},~\eqref{eq_app_ratio_SG}, and~\eqref{eq_deltaM_pre}, we find that the breather-soliton mass ratios are given by
\begin{equation}\label{eq_mass_ratio_pre2}
\begin{aligned}
    &\frac{M^2_{B_k}}{M^2_S}= 4 \sin^2 \left(\frac{\pi k}{6}\right)+\frac{8\pi k}{9}(\delta-\delta_2)\sin\left(\frac{\pi k}{3}\right)\\
    &+\frac{2\bar{\delta}_1}{\pi M^2_{S}}\lim_{\beta\rightarrow \sqrt{2\pi}}\left(F^{\mO}_{B_k,B_k}(i\pi,0)-4 \sin^2\left(\frac{\pi kp}{2}\right) F^{\mO}_{\bar{S},S}(i\pi,0)\right)+\mathcal{O}\left(\bar{\delta}_1^2,\,\bar{\delta}_1\delta,\,\delta^2\right)\,.
\end{aligned}
\end{equation}
Note that in the second line we can use the leading order value for the soliton mass. In practice the matrix elements are proportional to $M_S^2$ themselves as expected from dimensional analysis, as we shall see explicitly below. Consistency with the $\text{SO}(3)$ symmetry demands that for $\delta_2=\delta_3=0$ (which follows from $m_1 = m_2$ as per \eqref{eq_app_delta2} and \eqref{eq_app_delta3}), the first breather remains degenerate with the soliton, providing thus a nontrivial check of the procedure.

Let us now report the information we need about the form factors, following \cite{Lukyanov:1997bp} (see also \cite{Bajnok:2000ar}).  Due to the $\phi\leftrightarrow-\phi$ symmetry of the action, we may equivalently discuss form factors of vertex operators $\mO_a=:e^{i a\phi}:$ defined in \eqref{eq_app_normalization}.  Lorentz invariance implies that the two-particle form factor corresponding to the operator ${\cal O}_a(0)$ takes the following form~\cite{Lukyanov:1997bp} 
\begin{equation}\label{eq_form_fact_prod}
F^{\mO_a}_{a_1,a_2}(w_1,w_2)=\mathcal{G}_a \hat{F}^{\mO_a}_{a_1,a_2}(w_2-w_1)\,,
\end{equation}
where $\mathcal{G}_a = \left \langle 0 | {\cal O}_a(0) | 0 \right \rangle$ is the vacuum expectation value of the vertex operator, and $\hat{F}^{\mO_a}_{a_1,a_2}(w_2-w_1)$ is a momentum-dependent factor discussed below. Combining \eqref{eq_deltaM_pre} with \eqref{eq_app_pert} and \eqref{eq_form_fact_prod} and using the $\phi \to -\phi$ symmetry as described above, we find that the correction to the squared mass of a particle $P$ is 
 \begin{equation}
    \delta M^2_{P} 
    = \frac{2\bar \delta_1}{\pi} {\cal G}_{2 \beta} \hat F^{{\cal O}_{2 \beta}}_{\bar P, P}(-i \pi)
 \end{equation}
with $\beta = \sqrt{2 \pi}$. What remains to do is to determine the factor ${\cal G}_{2 \beta}$, as well as the momentum-dependent factor $\hat F^{{\cal O}_{2 \beta}}_{\bar P, P}(-i \pi)$ in the cases where $P$ is either a soliton or a breather.

Let us start with ${\cal G}_{2 \beta}$.  It takes the form \cite{Lukyanov:1996jj,Lukyanov:1997bp}\footnote{Note that our $\beta$ differs from \cite{Lukyanov:1996jj,Lukyanov:1997bp} by normalization $\beta_{\text{here}}=\sqrt{8\pi}\beta_{\text{\cite{Lukyanov:1996jj}}}$.} 
\begin{equation}\label{eq_Ga_def}
    \mathcal{G}_{2\beta}
    =\left(\frac{\sqrt{\pi} \Gamma \left(\frac{1+p}{2}\right)}{2\,\Gamma \left(\frac{p}{2}\right)}M_S\right)^{\frac{\beta^2}{\pi}} e^{g(\beta)}\,,
\end{equation}
where $p=\frac{\beta^2}{8\pi-\beta^2}$, as in \eqref{pDef}, and $g(\beta)$ is given by the following integral
\begin{equation}\label{eq_g_def}
    g(\beta)=\int_0^{\infty}\frac{dt}{t}\left(\frac{\sinh ^2(\frac{\beta^2  t}{2\pi})}{2 \sinh (t) \sinh \left(\frac{\beta ^2 t}{8\pi}\right) \cosh \left(\left(1-\frac{\beta ^2}{8\pi}\right) t\right)}-\frac{\beta^2}{\pi}\exp (-2 t)\right)\,.
\end{equation}
It is not too hard to see that the integrand behaves as $\sim \frac{e^{t \left(\frac{\beta^2}{\pi} -2\right)}}{t}$ for $t\rightarrow\infty$ when $\beta\approx \sqrt{2\pi}$. This implies that $\mathcal{G}_{2 \beta}$ diverges as $\sim \exp(\int_\#^{\infty} dt \,t^{-1})$ for $\beta=\sqrt{2\pi}$. On the other hand, it turns out that the function $\hat{F}^{\mO_{2 \beta}}$, defined in~\eqref{eq_form_fact_prod}, remains finite as $\beta\rightarrow \sqrt{2\pi}$. Below we discuss in some detail how to isolate the divergent contribution in $\mathcal{G}_{2 \beta}$.

To regulate~\eqref{eq_g_def}, we write the integrand isolating the divergent piece as
\begin{equation}
    g(\beta)=\int_0^{\infty} dt\,\tilde{g}_\text{fin}(\beta;t)+\int_{\frac12}^{\infty}\frac{dt}{t}e^{t (\frac{ \beta^2}{\pi} -2)}\,,
\end{equation}
where $\tilde{g}_\text{fin}$ is integrable at $\beta=\sqrt{2\pi}$,
\begin{equation}
\begin{split}
    \tilde{g}_\text{fin}(\beta;t) &=\frac{1}{t}\left(\frac{\sinh ^2(\frac{\beta^2  t}{2\pi})}{2\sinh t \sinh \left(\frac{\beta ^2 t}{8\pi}\right) \cosh \left(\left(1-\frac{\beta ^2}{8\pi}\right) t\right)}
    -\frac{\beta^2}{\pi}\exp (-2 t)\right)
    -\frac{e^{t (\frac{\beta^2}{\pi} -2)}}{t}\Theta\left(t-\frac12\right)\,,
\end{split}
\end{equation}
and the remaining term may be integrated explicitly
\begin{equation}
    \int_{\frac12}^{\infty}\frac{dt}{t}\exp\left(t \left(\frac{\beta^2}{\pi} -2\right)\right)=\Gamma\left(0,1-\frac{\beta^2}{2\pi}\right)\,.
\end{equation}
It is now straightforward to expand the answer as $\beta\rightarrow \sqrt{2\pi}$:
\begin{equation}
\begin{aligned}\label{eq_app_G2b}
    \mathcal{G}_{2\beta}=-\frac{1}{\varepsilon}\left(\frac{\pi M^2_S}{24 \sqrt{3}}\right)+\mathcal{O}(\varepsilon^0)\,,
\end{aligned}
\end{equation}
where we defineds $\varepsilon=\frac{\beta-\sqrt{2\pi}}{\sqrt{8\pi}}$ and where we used the following integral
\begin{align}\label{eq_app_g1_0}
    &\int_0^\infty dt\,\tilde{g}_\text{fin}(\sqrt{2\pi};t)= \log \left(\frac{2 \,\Gamma \left(\frac{1}{3}\right) \Gamma \left(\frac{1}{6}\right)}{3\, \Gamma \left(\frac{2}{3}\right) \Gamma \left(\frac{5}{6}\right)}\right)+\gamma_E\,.
\end{align}
Since all the matrix elements are proportional to $\mathcal{G}_{2\beta}$, the finite corrections in~\eqref{eq_app_G2b} do not contribute to the mass ratios and hence we do not report them.
To evaluate~\eqref{eq_app_g1_0}, we used that the integrand can be conveniently rewritten in terms of an infinite series as
\begin{equation}\label{eq_app_g1}
\begin{aligned}
    &\tilde{g}_\text{fin}(\sqrt{2\pi};t)=\frac{1}{t}\left(
    \frac{e^t+1}{e^t-e^{t/2}+1}-2e^{-2t}-\Theta\!\left(t-\frac12\right)\right) \\
    &=\frac{1-\Theta\!\left(t-\frac12\right)-e^{-2t}}{t}+\frac{1}{t}\left(\sum_{n=0}^\infty (-1)^n\left(e^{-(3n+1)t/2}+e^{-(3n+2)t/2}\right)-e^{-2t}\right)\,.
    \end{aligned}
\end{equation}
The integral of the first term is elementary. To integrate the second term in \eqref{eq_app_g1}, it is convenient to introduce a regulator $t^{\epsilon}$, with $\epsilon>0$, so that we can integrate each summand individually using
\begin{equation}
    \int_0^\infty dt\, t^{\epsilon-1}e^{-k\,t}=\Gamma(\epsilon)
    k^{-\epsilon}\,.
\end{equation}
Performing the sum and taking the limit $\epsilon\rightarrow 0$ we obtain~\eqref{eq_app_g1_0}.

We now discuss the momentum dependent contribution to the regular part $\hat{F}^{\mO_{a}}_{\bar P, P}(w)$ of the form factors, which has to be computed for each type of particle separately. For the soliton case, the required expression is given in full generality in~\cite{Lukyanov:1997bp}.  Specializing to the case of interest here, we find 
\begin{equation}
\begin{aligned}
    \hat{F}^{\mO_{2\beta}}_{\bar{S},S}(w)=-\frac{G(w)}{G(-i\pi)}\frac{4ie^{-\frac{w+i\pi}{2 p}}  \cot \frac{\pi p}{2} \cot (\pi p)}{p\sinh \frac{w+i\pi}{p}}
    \left( \cosh \frac{w}{2} \cot \frac{\pi p}{2} + \cosh \frac{3 w}{2} \cot \frac{3 \pi p}{2} \right)\,,
    \end{aligned}
\end{equation}
where again $p=\frac{\beta^2}{8\pi-\beta^2}$ as in \eqref{pDef}, and the function $G(w)$ is given in~\cite{Lukyanov:1997bp};  we will not need the explicit expression for $G(w)$ because we are interested in $w=-i\pi$.  Taking the limit $w\rightarrow -i\pi$, we find the following simple result
\begin{equation}\label{eq_F_soliton}
    \hat{F}^{\mO_{2\beta}}_{\bar{S},S}(-i\pi)=-2 \sqrt{3}+\frac{112}{9} \pi  \varepsilon+\mathcal{O}(\varepsilon^2)\,.
\end{equation}

The breather form factors are conveniently given in~\cite{Bajnok:2000ar}.\footnote{Note that we only have a unique vacuum, the $k=0$ one of~\cite{Bajnok:2000ar}, compared to the many vacua that emerge when the periodicity of the perturbation is not a multiple of the one of the unperturbed sine-Gordon potential.} We report them below
\begin{equation}
    \begin{aligned}
    \hat{F}_{\bar B_1,B_1}^{\mO_{2 \beta}}(-i\pi) &=-2\mathcal{N}\frac{\sin^2\left(2 \pi p\right)}{\sin\left(\frac{p\pi}{2}\right)}\exp\left(-\frac{1}{\pi}\int_0^{\pi p}\frac{dt\;t}{\sin t}\right)\,,\\ 
    \hat{F}_{\bar B_2,B_2}^{\mO_{2 \beta}}(-i\pi)&=-4\mathcal{N}^{2}\cot\left(\frac{\pi p}{2}\right)\sin^2\left(2 \pi p \right)\left(1+\frac{\cos ^2\left(2 \pi p \right)}{\cos \left(\pi p\right)}\right)\exp\left(-\frac{2}{\pi}\int_0^{\pi p}\frac{dt\;t}{\sin t}\right)\,,
\end{aligned}
\end{equation}
where
\begin{equation}
    \begin{aligned}
    \mathcal{N}&=\exp\left(\int_0^{\infty} dt\,\frac{4\sinh t\sinh(tp) \sinh \left(t(1+p)\right)}{t \sinh ^2(2 t)}\right)=\exp(\mathcal{N}_1+\varepsilon\mathcal{N}_2+\mathcal{O}(\varepsilon^2))\,,\\
    \mathcal{N}_1 &=\int_0^\infty dt\frac{4\sinh t\sinh(\frac{t}{3})\sinh (\frac{4t}{3})}{t \sinh^2(2t)}=\log \left(\frac{2}{3^{2/3}}\right)+\frac{\psi ^{(1)}\left(\frac{1}{6}\right)-\psi ^{(1)}\left(\frac{5}{6}\right)}{12 \sqrt{3} \pi }\,, \\
    \mathcal{N}_2 &=\int_0^\infty dt\frac{16(1+2\cosh \left(\frac{2 t}{3}\right)+2\cosh \left(\frac{4 t}{3}\right))}{9(1+2\cosh \left(\frac{2 t}{3}\right))\cosh^2t}=\frac{56\sqrt{3}\pi}{81}\,. \label{calNs}
\end{aligned}
\end{equation}
To derive result for $\mathcal{N}_1$ quoted in \eqref{calNs}, we expressed again the integrand as a series via
\begin{equation}
   \frac{4\sinh t\sinh(\frac{t}{3})\sinh (\frac{4t}{3})}{t \sinh^2(2t)}=\sum_{n=1}^\infty 4 ne^{-4n t} \frac{ \left(-\sinh \left(\frac{2 t}{3}\right)+\sinh \left(\frac{8 t}{3}\right)-\sinh (2 t)\right)}{t}\,,
\end{equation}
and then we integrated each term individually.  To derive the expression for $\mathcal{N}_2$ quoted in \eqref{calNs}, we used the substitution $x=\tanh (t/3)$, which results in an elementary integral.

Expanding the matrix elements as $\beta\rightarrow \sqrt{2\pi}$, we find
\begin{equation}
    \begin{aligned}
    \hat{F}_{B_1,B_1}^{\mO_{2\beta}}(-i\pi)&=-2\sqrt{3}+\frac{160}{9}\pi\varepsilon+\mathcal{O}(\varepsilon^2)\,,\\
    \hat{F}_{B_2,B_2}^{\mO_{2\beta}}(-i\pi)&=-6\sqrt{3}+\mathcal{O}(\varepsilon^2)\,,
\end{aligned}
\end{equation}
where we used
\begin{equation}
    \frac{1}{\pi}\int_0^{\pi p}\frac{dt\;t}{\sin t}=\frac{\sqrt{3}(\psi ^{(1)}\left(\frac{1}{6}\right)-\psi ^{(1)}\left(\frac{5}{6}\right))-6 \pi  \log (3)}{36\pi}+\frac{32\pi\varepsilon}{27 \sqrt{3}}+\mathcal{O}(\varepsilon^2)\,.
\end{equation}

We may now put together the above results and compute the mass ratios from~\eqref{eq_mass_ratio_pre2}. We find that the ratio of the first breather and the soliton mass is
\begin{equation}\label{eq_ratio2}
     \frac{M^2_{B_1}}{M^2_S}=1-\frac{4\pi}{3 \sqrt{3}}(\delta_2+\bar{\delta}_1-\delta)+\mathcal{O}(\delta_k^2)\,.
\end{equation}
We see that the correction vanishes for $\delta_2=\delta_3=0$ and $\bar{\delta}_1=\delta$, as required by $\text{SO}(3)$ invariance:  $M^2_{B_1}=M^2_S\equiv M_{\pi}^2$. From this we infer the result~\eqref{eq_pi_pi_ratio_dm} for the $\pi^0/\pi^{\pm}$ mass ratio for unequal fermion masses. We similarly find the ratio between the mass of the second breather and the soliton:
\begin{equation}\label{eq_ratio3}
    \frac{M^2_{B_2}}{M^2_S}=3-\frac{4\pi}{3 \sqrt{3}}(2\delta_2-\bar{\delta}_1-2\delta)+\mathcal{O}\left(\delta_k^2\right)\,.
\end{equation}
From this we obtain the results~\eqref{eq_ratio} and~\eqref{eq:sigma_pi_ratio} for the $\sigma/\pi^{\pm}$ mass ratio.

Let us finally note that the divergences that appear in the perturbative corrections to the individual particle masses can be consistently renormalized shifting the only coupling of the leading order theory, namely $\mu_Z$. Equivalently, by~\eqref{eq_soliton_mass}, this means that perturbation theory can be made finite reabsorbing the $1/\varepsilon$ poles in the soliton mass. To see this, note that the perturbed value of the soliton mass is
\begin{equation}
    M^2_{S,0}+\delta M^2_{S}=M^2_{S,0}+2\frac{\bar{\delta}_1}{\pi}\mathcal{G}_{2\beta} \hat{F}_{\bar{S},S}^{\mO_{2\beta}}(-i\pi)\simeq M^2_{S,0}\left(1
    +\frac{\bar{\delta}_1}{6\varepsilon}+\mathcal{O}(\bar{\delta}_1\varepsilon^0)\right)\,,
\end{equation} 
where we only wrote the divergent part and we distinguished between the leading order soliton mass $M^2_{S,0}$ and its real perturbed value. We can make the result finite for $\beta\rightarrow\sqrt{2\pi}$ by defining the renormalized mass
\begin{equation}
    M^2_{S,\text{ren}}=M^2_{S,0}\left(1+\frac{\bar{\delta}_1}{6\varepsilon}\right)\,.
\end{equation}
Note that the renormalized mass in general coincides with the physical mass only at leading order in $\bar{\delta}_1$.
Given this definition, since the mass ratios computed above are finite for $\varepsilon\rightarrow 0$, it follows that the breather masses are both finite when written in terms of the renormalized soliton mass
\begin{equation}
    \begin{aligned}
    &M^2_{B_1}=4\sin^2\left(\frac{\pi p}{2}\right)M^2_{S,0}+2\frac{\bar{\delta}_1}{\pi}\mathcal{G}_{2\beta} \hat{F}_{B_1,B_1}^{\mO_{2\beta}}(-i\pi)=M^2_{S,\text{ren}}\left(1+\mO\left(\bar{\delta}_1\varepsilon^0\right)\right)\,,\\[0.5em]
    &M^2_{B_2}=4\sin^2(\pi p)M^2_{S,0}+2\frac{\bar{\delta}_1}{\pi}\mathcal{G}_{2\beta} \hat{F}_{B_2,B_2}^{\mO_{2\beta}}(-i\pi)=3 M^2_{S,\text{ren}}\left(1+\mO\left(\bar{\delta}_1\varepsilon^0\right)\right)\,.
\end{aligned}
\end{equation}

Note that we cannot compute the perturbative correction to the physical soliton mass without specifying the finite part of the coupling at order $\mO(\bar{\delta}_1)$. This requires matching the two-flavor Schwinger model with the perturbed sine-Gordon theory (within this scheme) at order $\mO(m^3)$, which is beyond the scope of this work. This subleading correction however does not affect the mass ratios~\eqref{eq_mass_ratio_pre2}, which are hence scheme independent to this order, as shown above.

\section{Perturbative corrections to the fermion mass at weak coupling}\label{app_Pert_Corr}

In this Appendix we derive the weak coupling corrections to the fermion mass for the equal mass theory. These are especially relevant at $\theta = \pi$, as explained in Section~\ref{sec_theta_Pi}. The result arises using the diagrammatic expansion derived from the Lagrangian \eqref{eq_Lag}. The $\gamma$-matrices necessary for the calculation are defined as in sec.~\ref{sec:setup}.

As mentioned in the main text, we will use the lightcone gauge as in Appendix~\ref{app_tHooft}, which preserves Lorentz invariance and often leads to significant simplifications in $d=2$ \cite{tHooft:1974pnl,Bergknoff:1976xr,Harada:1993va}. The $\gamma$-matrices in lightcone coordinates take a very simple form:
\begin{equation}
    \gamma^+ = \begin{pmatrix}
        0&0\\
        \sqrt{2}i&0\\
    \end{pmatrix},\quad \gamma^{-} = -\begin{pmatrix}
        0&\sqrt{2}i\\
        0&0\\
    \end{pmatrix}\,.
\end{equation}
Because $A_- = 0$, each fermion propagator $S(\slashed{p})$ in an amputated diagram is surrounded by two vertices containing $\gamma^+$, so only the component $S^-$, proportional to $\gamma^{-}$, survives. Hence the only propagators we need in this approach are
\begin{equation}
    S^-(\slashed{p}) = \dfrac{i p^+ \gamma^-}{2p^+p^- - m^2 + i\epsilon}\,,\qquad G_{++}(p) = \dfrac{i g^2}{(p^+)^2}\,,
\end{equation}
where $G_{++}$ is the only nonzero component of the photon propagator $G_{\mu\nu}$ in this gauge. 
Infrared divergences are taken care of by treating the photon propagator as a distribution~\cite{Das:2012qz}:\footnote{More precisely~\eqref{eq_Pf_def} defines the finite part distribution, which is the derivative of the principal value (hence we denote it with the same symbol $\mathcal{P}$). More in detail, when we integrate $\mathcal{P} \frac{1}{x^2} $ against a test function $f(x)$ which is smooth around $x=0$, we have
\begin{align}\label{eq_Pf_def}
    \int_{-A}^{A} dx \, \mathcal{P}\frac{1}{x^2} f(x) \equiv  \lim_{\epsilon\rightarrow 0^+}\left[\int_{\epsilon}^{A} d x \frac{f(x)}{x^2} +
    \int^{-\epsilon}_{-A} dx  \frac{f(x)}{x^2}-2\frac{f(0)}{\epsilon}\right]\,,
\end{align}
for arbitrary $A>0$. 
}
\begin{equation}\label{eq_pr_val_presc}
   G_{++}(p) =  \mathcal{P} \dfrac{ig^2}{(p^+)^2} = \dfrac{ig^2}{2(p^+-i\epsilon)^2} + \dfrac{ig^2}{2(p^++i\epsilon)^2}\,.
\end{equation}

There are various ways to think of this prescription. First, after a Fourier transform it gives an interaction potential $\sim |x^--x'^-|$ between the points $x$ and $x'$. So the principal value naturally arises if we require the potential, or, in other words, the gauge field propagator, to be symmetric and vanish at $x^- = x'^-$. Note that the gauge field can be completely integrated out since it is non-dynamical with respect to the lightcone "time" $x^+$. This approach is used, for example, in \cite{Coleman:1976uz,Harada:1993va,Burkardt:1997bw}. 

An alternative way to justify the prescription~\eqref{eq_pr_val_presc} consists in integrating out the constant mode of the gauge field separately. This imposes the neutrality condition on the spectrum, and removes the zero mode $p^+ = 0$ when inverting the photon kinetic term $(p^+)^2$. This strategy is widely used in Discretized Light Cone Quantization (DLCQ) for various $1+1$-dimensional gauge theories \cite{Hornbostel:1988fb,Eller:1986nt, Dalley:1992yy}, where the momentum is discrete and the elimination of the zero mode is straightforward. In the continuum, this approach amounts to a symmetric cutoff of the integrals containing $\frac{1}{(p^+)^2}$ at $p^+ = \pm \epsilon$---this scheme was used in \cite{tHooft:1974pnl}. This regularization does not yet remove the divergences in the $\epsilon\rightarrow 0$ limit. However, as argued in \cite{Einhorn:1976uz}, infrared divergences eventually cancel for gauge-invariant observables, as confirmed by many available DLCQ calculations. The principal value prescription then provides a convenient way to remove the $\epsilon$-dependence and the unphysical infrared divergences.

We are now ready to compute the perturbative shift of the fermion propagator's pole:
\begin{equation}
    m_{\text{pole}} = m +\Delta m\,.
\end{equation}
The diagrams contributing to the fermion self-energy up to order $\mathcal{O}\left( \frac{g^4}{m^3} \right)$ are shown in Figure~\ref{fig::feyn_diag_mass} in the main text. In a theory with $N_f$ flavors, the diagram on Figure~\ref{subfig::diag_d} acquires a factor of $N_f$.

If we denote the sum of one-particle irreducible (1PI) diagrams by $i\Sigma(\slashed{p})$, the propagator obtained after a resummation of chains of such diagrams is given by
\begin{equation}
    S^-_R(\slashed{p}) = \dfrac{i p^+ \gamma^-}{2p^+ p^- - m^2 + \Tr(\Sigma(\slashed{p})\gamma^-)p^+ + i\epsilon}\,,
\end{equation}
from which the pole mass shift $\Delta m$ follows. Note that $\Sigma(\slashed{p}) \propto \gamma^+$ due to the structure of the diagrams in lightcone gauge. 

The amplitudes we need are evaluated in momentum space. Choosing a parametrization of the integrand which is symmetric under $q^- \to - q^-$, where $q$ is the fermion momentum, the diagram in Figure~\ref{subfig::diag_a} evaluates to
\begin{equation}
    i \Sigma^{(a)}(\slashed{p}) = 2g^2\gamma^{+} \mathcal{P}\int \dfrac{dq^+ dq^-}{(2\pi)^2} \dfrac{q^+ m^2}{(2 q^+ q^- - m^2 + i\epsilon)(2 q^+ q^- + m^2 - i\epsilon)(p^+-q^+)^2}\,.
\end{equation}
The $dq^-$ integral can be computed by closing the contour in the upper half-plane:
\begin{equation}
    i\Sigma^{(a)}(\slashed{p}) = -i\dfrac{g^2 \gamma^+}{4\pi} \mathcal{P}\int d q^+ \dfrac{1}{(\text{sign}\, q^+)(p^+-q^+)^2} = \dfrac{ig^2\gamma^+}{2\pi p^+}\,.
\end{equation}
This is the well-known result for the leading order electron mass correction, already derived in e.g. \cite{Coleman:1976uz}. In order to get the pole mass shift up to $O\left(\frac{g^4}{m^3} \right)$, it is enough to evaluate the rest of the diagrams on-shell. We will also take $p^+ > 0$ without loss of generality.

It is possible to see that for the diagram in Figure~\ref{subfig::diag_b} one can always close the integration contour avoiding all the poles, hence
\begin{equation}
    i\Sigma^{(b)}(\slashed{p}) = 0\,.
\end{equation}
For the diagram in Figure~\ref{subfig::diag_c}, we get:
\begin{equation}
\begin{aligned}
    \left. i \Sigma^{(c)}(\slashed{p}) \right|_{p^2=m^2} = \dfrac{8 i g^4\gamma^+}{(2\pi)^4}\mathcal{P}\int dq^+ dq^- dk^+ dk^- \dfrac{q^+ k^+ (q^++k^+-p^+)}{(2q^+ q^--m^2+i\epsilon) (2k^+ k^- - m^2+i\epsilon)}\\
    \times \dfrac{1}{(2(q^++k^+-p^+)(q^-+k^--p^-) - m^2 +i\epsilon)(p^+-q^+)^2(p^+-k^+)^2}\,.
\end{aligned}
\end{equation}
By closing the corresponding contours in the upper half-plane, we see that the integral over $dk^- dq^-$ is non-zero if $q^+,k^+ >0$ and $q^+ + k^+ < p^+$. After taking the residues, we obtain:
\begin{equation}
    \left.i\Sigma^{(c)}(\slashed{p}) \right|_{p^2=m^2} = -\dfrac{ig^4\gamma^+}{2\pi^2m^2 p^+}\int_0^1 d\xi \int_0^{1-\xi} d\eta \dfrac{\xi\eta(1-\xi-\eta)}{(\xi+\eta)(1-\eta)^3(1-\xi)^3} = -\dfrac{ig^4 \gamma^+}{64 m^2 p^+}\,.
\end{equation}

Lastly, the diagram in Figure~\ref{subfig::diag_d} is as follows:
\begin{equation}
\begin{aligned}
    \left. i \Sigma^{(d)}(\slashed{p}) \right|_{p^2=m^2} &= \dfrac{8iN_fg^4\gamma^+}{(2\pi)^4}\mathcal{P}\int dq^+ dq^- dk^+ dk^- \dfrac{(k^+-q^+) k^+ (p^+-q^+)}{(2(k^+-q^+) (k^--q^-)-m^2+i\epsilon)}\\
    &\times \dfrac{1}{(2(p^+-q^+)(p^--q^-) - m^2 +i\epsilon) (2k^+ k^- - m^2+i\epsilon)(q^+)^4}\,.
\end{aligned}
\end{equation}
Deforming the contour as before, we find that only the region $p^+ > q^+ > k^+ >0$ contributes. The integration over $dq^+ dk^+$ yields:
\begin{equation}
    \left.i \Sigma^{(d)}(\slashed{p})\right|_{p^2=m^2} = \dfrac{iN_f g^4 \gamma^+}{2\pi^2 m^2 p^+}\int_0^1 d\xi\,\mathcal{P}\int_0^1 d\eta \dfrac{\xi(1-\xi)(1-\eta)}{\left(1-\eta+\xi(1-\xi)\eta^2\right)\eta^2} =  -\dfrac{i N_f g^4 (3\pi^2 +16) \gamma^+}{384\pi^2 m^2 p^+}\,.
\end{equation}
Here we used the prescription \eqref{eq_pr_val_presc} for the $1/\eta^2$ term. After collecting all these terms, we find that in a theory with $N_f$ flavors
\begin{equation}\label{eq_m_shift}
    \Delta m = -\dfrac{g^2}{2\pi m} + \dfrac{g^4(N_f(3\pi^2+16)+6\pi^2-48)}{384\pi^2 m^3} + \mathcal{O}\left(\dfrac{g^6}{m^5}\right)\,,
\end{equation}
which for $N_f = 2$ leads to \eqref{eq_m_shift_main}.

Let us finally comment on a potential concern regarding the contribution of the background field $E_0 = g^2 \theta / (2\pi)$ induced by the $\theta$ term. As for the binding energy, this effect is consistently taken into account by matching the full theory onto a two-fermion bound-state equation (with the renormalized fermion mass that we computed above), which can be carried out systematically at weak coupling, as discussed in Appendix~\ref{app_weak_coupling_NR}. In particular, at $\theta = \pi$ the potential deconfines, and the pole mass coincides with the soliton mass, as expected. We have checked this agreement explicitly by computing the one-loop correction to the mass using the fermion propagator in a background field via Schwinger parametrization.

\section{The soliton mass at \texorpdfstring{$\theta=\pi$}{θ=π} at strong coupling from integrability}\label{app_Large_m_thetaPi}

As explained in the main text, the low energy theory~\eqref{eq_EFT_2flavor_theta_Pi} at the $\text{SO}(3)$-invariant point is integrable but rather subtle to study, since the marginal interaction requires specifying an appropriate regularization scheme. The integrability of this model is inherited from that of the SU$(2)$ Thirring model~\eqref{eq_Thirring}, which is equivalent to the sine-Gordon theory of interest and a free decoupled compact massless scalar via bosonization.

Let us review this identification in detail. We can bosonize the two Dirac fermions of the Thirring model \eqref{eq_Thirring} in terms of two compact scalars $\phi_\alpha\sim\phi_\alpha+\sqrt{\pi}$, $\alpha=1,2$. It is convenient to introduce the linear combinations
\begin{equation}
\phi_\pm=\frac{\phi_1\pm\phi_2}{\sqrt{2}}\, .
\end{equation}
The field $\phi_+$ bosonizes the $U(1)$ current and remains a decoupled massless scalar, whereas the interaction in the relative field $\phi_-$ gives the $SO(3)$-invariant sine-Gordon model. In the conventions of \eqref{eq_EFT_2flavor_theta_Pi}, the resulting Lagrangian takes the form
\begin{equation}
\mathcal L_{\rm Th}
=
\frac{1}{2}(\partial\phi_+)^2
+\frac{1-\delta}{2}(\partial\phi_-)^2
+\frac{e^{2\gamma_E}}{32\pi}
\alpha\,\mu_-^2 N_{\mu_-}
\cos\left(\sqrt{8\pi}\phi_-\right),
\qquad
\alpha=-8\delta+\mathcal O(\delta^2),
\end{equation}
with $\delta=-\lambda/\pi+ O(\lambda^2)$. 

There is, however, an important difference in the global structure between the Thirring model and the EFT \eqref{eq_EFT_2flavor_theta_Pi}. The compactification of the original bosons implies the following relations
\begin{equation}
(\phi_+,\phi_-)
\sim
\left(\phi_++\sqrt{\frac{\pi}{2}},
\phi_-+\sqrt{\frac{\pi}{2}}\right)
\sim
\left(\phi_++\sqrt{\frac{\pi}{2}},
\phi_--\sqrt{\frac{\pi}{2}}\right),
\end{equation}
in addition to the shifts $\phi_\pm\sim\phi_\pm+\sqrt{2\pi}$. Thus the Thirring model is more precisely a $\mathbb Z_2$ quotient of the product of the free compact boson and the sine-Gordon sector. In particular, although $\cos(\sqrt{8\pi}\phi_-)$ has period $\sqrt{\pi/2}$ and the sine-Gordon factor considered in isolation has two minima within the interval $\phi_-\sim\phi_-+\sqrt{2\pi}$, the two minima are exchanged by the above identification together with a half-period shift of the free field $\phi_+$. Consequently, they do not correspond to two distinct vacua of the full Thirring model. Correspondingly, the lowest $SU(2)$ doublet states, which in the Schwinger model are domain walls between the two vacua, or equivalently half-asymptotic particles, correspond to ordinary asymptotic particle states in the Thirring model.
This global distinction from the Schwinger-model EFT does not affect the local RG flow and in particular the mass of the $SU(2)$ doublet extracted from integrability, which is the quantity that we study below.

The original approach of~\cite{Andrei:1979sq} for solving the $\text{SU}(2)$ Thirring model using the Bethe ansatz uses a regulator that breaks Lorentz invariance at intermediate steps, making it difficult to compare its detailed predictions for the mass spectrum with our setup. To overcome this obstacle, we instead resort to the analysis of~\cite{Forgacs:1991nk}, where
the authors obtained the relation between the mass of the doublets and the RG scale $\Lambda_{\text{IR}}$ of the SU$(2)$ Thirring model using Pauli-Villars regularization (see also \cite{Evans:1995dn,VanAcoleyen:2002xt}). The result reads
\begin{equation}\label{eq_int_match}
    M_{\text{sol}}=\frac{e^{\frac{1}{4}}}{\sqrt{\pi }}\Lambda_{\text{IR}}\,.
\end{equation}
Here $\Lambda_{\text{IR}}$ is defined as the scale at which the perturbative coupling $\lambda$  in \eqref{eq_Thirring} diverges. Therefore we can express it in terms of $\lambda$ at some UV scale $\Lambda_{\text{UV}}$ using the beta-function~\cite{Forgacs:1991nk} 
\begin{equation}\label{eq_app_lambda_beta}
    \Lambda\frac{\pd \lambda (\Lambda)}{\pd \Lambda}=-\frac{2}{\pi}\lambda^2(\Lambda)+\frac{2}{\pi^2}\lambda^3(\Lambda)+\mathcal{O}\left(\lambda^4\right)\,,
\end{equation}
which yields
\begin{equation}
    \Lambda_\text{IR} \approx \frac{\Lambda_\text{UV} \,\sqrt{\lambda(\Lambda_\text{UV})}}{\sqrt{\pi}}\, e^{-\frac{\pi}{2\lambda(\Lambda_\text{UV})}}\quad\implies\quad
    M_{\text{sol}} \approx \frac{e^{\frac{1}{4}}\Lambda_\text{UV} \,\sqrt{\lambda(\Lambda_\text{UV})}}{\pi}e^{-\frac{\pi}{2\lambda(\Lambda_\text{UV})}}\,.
\end{equation}

Therefore we can determine the soliton mass by matching the SU$(2)$ Thirring coupling $\lambda$ of the low energy theory to the full theory one at a scale $\Lambda_{\text{UV}}\gg m$. It is important that such matching is done at two-loop accuracy, i.e.~including the $\mO(\lambda^2)$ term in~\eqref{eq_Thirring_coupling}, to obtain the spectrum to subexponential accuracy. The determination of the EFT coupling to this order is rather subtle since its precise definition depends on the renormalization scheme.

To this aim, we match the free energy at chemical potential $h$, such that $\mu\gg h\gg m$, for the isospin charge $Q_3$ in the full theory at order $\mathcal{O}(m^4)$, with the two-loop result in the SU$(2)$ Thirring model. The SU$(2)$ Thirring model free energy was computed in~\cite{Forgacs:1991nk} and reads
\begin{equation}
F(h)-F(0)=-\frac{h^2}{4\pi}
    \left(1-\frac{\lambda(h)}{\pi}
    +(2\log 2-1)\frac{\lambda^2(h)}{\pi^2}
    +\mO(\lambda^3)
    \right)\,.
\end{equation}
where $\lambda(h)$ is the running coupling at the scale $h$. Using the beta-function~\eqref{eq_app_lambda_beta}, we can re-express the result in terms of the coupling at the matching scale, which we take to be the heavy scalar mass $\mu$:
\begin{equation}\label{eq_F_thirring}
    F(h)-F(0)=-\frac{h^2}{4\pi}
    \left( 1-\frac{\lambda(\mu)}{\pi}+\left(
      2\log 2-1+\log\left(\frac{h^2}{\mu^2}\right) \right)\frac{\lambda^2(\mu)}{\pi^2}+\mO(\lambda^3)
    \right)\,.
\end{equation}
In terms of $\lambda(\mu)$, the scale that we should use in~\eqref{eq_int_match} reads
\begin{equation}
M_{\text{sol}}=\frac{e^{\frac{1}{4}}\mu \,\sqrt{\lambda(\mu)}}{\pi }
e^{-\frac{\pi}{2\lambda(\mu)}}
\,.
\end{equation}

Let us now discuss the calculation in the full theory. In the bosonized formulation, the Euclidean Lagrangian density at finite chemical potential $h$ reads
\begin{equation}
 \begin{aligned}
    \mL &=\frac{1}{2}\pd_\mu \phi_+ \pd^\mu \phi_+ +\frac{\mu^2}{2}\phi_+^2+\frac{1}{2}(\pd\phi_-)^2-\frac{h}{\sqrt{2\pi}}\pd_x\phi_- \\
    &{}-\frac{e^{\gamma_E}}{\pi}m\sqrt{\mu\,\mu_-}N_{\mu}\sin\left(\sqrt{2\pi}\phi_+\right)N_{\mu_-}\cos\left(\sqrt{2\pi}\phi_-\right) \,.
 \end{aligned}
\end{equation}
To eliminate the tadpole, we shift 
\begin{equation}
\phi_-\rightarrow \phi_-+\frac{h}{\sqrt{2\pi}}x\,.
\end{equation}
We also work with canonically normalized vertex operators for $\phi_-$ via
\begin{equation}
N_{\mu_-}\cos\left(\sqrt{2\pi}\phi_-\right)=\sqrt{\frac{2 e^{-\gamma_E}}{\mu_-}}:\cos\left(\sqrt{2\pi}\phi_-\right):\,.
\end{equation}
Overall, we recast the action in the form
\begin{equation}
    \mL=-\frac{h^2}{4\pi}+\frac{1}{2}(\pd\phi_+)^2+\frac{\mu^2}{2}\phi_+^2+\frac{1}{2}(\pd\phi_-)^2-2c 
\,m\,\mO(x)\,,
\end{equation}
where
\begin{equation}
\mO(x)=\sqrt{\mu}\sin(\sqrt{2\pi}\phi_+):\cos(\sqrt{2\pi}\phi_-+hx):\,,\qquad c=\frac{e^{\frac{\gamma_E}{2}}}{\sqrt{2}\pi}\,.
\end{equation}

We may now compute the free energy. For $h\gg m$ we can do this expanding perturbatively in $\mO(x)$:
\begin{equation}\label{eq_F_full}
\begin{aligned}
    &F(h)-F(0)=-\frac{h^2}{4\pi}-\frac{(2 cm)^2}{2}\int d^2x\left[\langle\mO(0)\mO(x)\rangle^{(h)}-\langle\mO(0)\mO(x)\rangle^{(0)}\right] \\
    &-\frac{(2c m)^4}{24}\int d^2x d^2y d^2z\left[\langle\mO(0)\mO(x)\mO(y)\mO(z)\rangle^{(h)}_c-\langle\mO(0)\mO(x)\mO(y)\mO(z)\rangle^{(0)}_c\right]+\mathcal{O}(m^6)\,,
\end{aligned}
\end{equation}
where the superscript denotes whether the correlator is evaluated $h=0$ or not, and the four-point connected correlator is
\begin{equation}
\begin{split}
&\langle\mO(0)\mO(x)\mO(y)\mO(z)\rangle^{(h)}_c =\langle\mO(0)\mO(x)\mO(y)\mO(z)\rangle^{(h)}-\langle\mO(0)\mO(x)\rangle^{(h)}\langle\mO(y)\mO(z)\rangle^{(h)}\\
&
-\langle\mO(0)\mO(y)\rangle^{(h)}\langle\mO(x)\mO(z)\rangle^{(h)}-\langle\mO(0)\mO(z)\rangle^{(h)}\langle\mO(x)\mO(y)\rangle^{(h)}\,.
\end{split}
\end{equation}
The subtraction of the disconnected contribution ensures that the integration in~\eqref{eq_F_full} is free of extensive divergences.

Since we do not consider the effect of irrelevant couplings in the EFT, we can neglect terms of order $h^4/\mu^2$ in~\eqref{eq_F_full}. In the following it will be convenient to work in units in which $\mu=1$.  For the first subleading order we then use
\begin{equation}
\begin{split}
    &\int d^2x(\langle\mO(0)\mO(x)\rangle^{(h)}-\langle\mO(0)\mO(x)\rangle^{(0)}) =\int d^2x\frac{\sinh\left(K_0(|x|)\right)}{2|x|}\left(\cos\vec{h}\cdot\vec{x}-1 \right)\\
    &=-\frac{1}{4}\int d^2x\frac{\sinh\left(K_0(|x|)\right)}{|x|}\left(\vec{h}\cdot\vec{x}\right)^2+\mathcal{O}(h^4)=-\frac{h^2}{8}I_s(\pi)+\mathcal{O}(h^4)\,,
\end{split}
\end{equation}
where $\vec{h}=(0,h)$.

The $\mO(m^4)$ contribution cannot be simply evaluated Taylor expanding the integrand in $h$. Indeed, the comparison with~\eqref{eq_F_thirring} implies that the result depends logarithmically on $h$. More precisely, we must have
\begin{equation}\label{eq_app_I4}
\begin{split}
    I_4(h) &\equiv\int d^2x\, d^2y\, d^2z\,\left(\langle\mO(0)\mO(x)\mO(y)\mO(z)\rangle^{(h)}_c-\langle\mO(0)\mO(x)\mO(y)\mO(z)\rangle^{(0)}_c\right)\\
    &=A h^2\log h^2+B h^2+\mathcal{O}(h^4)\,,
\end{split}
\end{equation}
where $A$ and $B$ will be determined below. 

Comparing~\eqref{eq_F_thirring} and~\eqref{eq_F_full} we match the SU$(2)$ Thirring coupling in Pauli-Villars regularization to the Schwinger model parameters as (restoring units)
\begin{equation}
    \sqrt{\lambda(\mu)}=\pi c\sqrt{I_s(\pi)}\,\frac{m}{\mu}+\frac{\pi m^3c^3(3\pi I_s^2(\pi)(2\log 2-1)-8 B)}{6\mu^3\sqrt{I_s(\pi)}}+\mathcal{O}\left(\frac{m^5}{\mu^5}\right)\,.
\end{equation}
We also obtain the following consistency condition
\begin{equation}\label{eq_app_A_consistency}
A=\frac{3\pi}{8}I_s(\pi)^2\simeq 119.77434\,.
\end{equation}
Putting everything together, we express the soliton mass as a function of $B$ as
\begin{equation}\label{eq_app_M_sol}
    M_{\text{sol}}\simeq  1.48569 \,e^{-0.00417455 B} \,\times m\, e^{-A_s\frac{ g^2}{m^2}}\left(1+\mathcal{O}\left(\frac{m^2}{g^2}\right)\right)\,.
\end{equation}
Below we confirm analytically the validity of~\eqref{eq_app_A_consistency} and express the coefficient $B$ in terms of a finite integral that can be computed numerically.

To compute the small $h$ limit of the integral~\eqref{eq_app_I4}, note first that the four-point point correlator factorizes into a product of the $\phi_+$- and $\phi_-$ correlators, that can be computed using Wick's theorem.
We find
\begin{equation}
\begin{split}
\langle\mO(x_1)\mO(x_2)\mO(x_3)\mO(x_4)\rangle^{(h)}_c
=&\,\frac{1}{32}H_4L_4\\
&-\frac14\sum_{(ij|kl)}
\frac{\sinh K_0(r_{ij})\sinh K_0(r_{kl})}{r_{ij}r_{kl}}
\cos[\vec{h}\cdot(\vec{x}_i-\vec{x}_j)]
\cos[\vec{h}\cdot(\vec{x}_k-\vec{x}_l)]\,.
\end{split}
\end{equation}
Here and below the sum runs over the three pairings
$(ij|kl)=(12|34),(13|24),(14|23)$, and
$r_{ij}\equiv|\vec{x}_i-\vec{x}_j|$. In the above we defined the heavy field's contribution to the correlator as
\begin{align}\label{eq:H-explicit}
 H_4&=
 e^{K_0(r_{12})+K_0(r_{34})}
 \cosh\!\bigl[K_0(r_{13})-K_0(r_{14})-K_0(r_{23})+K_0(r_{24})\bigr]\nonumber\\
 &-e^{K_0(r_{12})-K_0(r_{34})}
 \cosh\!\bigl[K_0(r_{13})+K_0(r_{14})-K_0(r_{23})-K_0(r_{24})\bigr]\nonumber\\
 &-e^{-K_0(r_{12})+K_0(r_{34})}
 \cosh\!\bigl[K_0(r_{13})-K_0(r_{14})+K_0(r_{23})-K_0(r_{24})\bigr]\nonumber\\
 &+e^{-K_0(r_{12})-K_0(r_{34})}
 \cosh\!\bigl[K_0(r_{13})+K_0(r_{14})+K_0(r_{23})+K_0(r_{24})\bigr]\,,
\end{align}
and similarly the light field's contribution reads
\begin{equation}
\begin{split}
L_4&=\frac{r_{12}r_{34}}{r_{13}r_{14}r_{23}r_{24}}\cos[\vec{h}\cdot(\vec{x}_1+\vec{x}_2-\vec{x}_3-\vec{x}_4)]\\
&+\frac{r_{13}r_{24}}{r_{12}r_{14}r_{23}r_{34}}\cos[\vec{h}\cdot(\vec{x}_1+\vec{x}_3-\vec{x}_2-\vec{x}_4)]\\
& +\frac{r_{14}r_{23}}{r_{12}r_{13}r_{24}r_{34}}\cos[\vec{h}\cdot(\vec{x}_1+\vec{x}_4-\vec{x}_2-\vec{x}_3)]\,.
 \end{split}
\end{equation}

To proceed, we note that by rotational invariance the result depends only on $\vec{h}^2=h^2$. We may then average the integrand over the direction of $\vec{h}$. Using
\begin{equation}\label{eq:J0-average}
 \frac1{2\pi}\int_0^{2\pi} d\alpha\,\cos(h|q|\cos\alpha)=J_0(h|q|)\,,
\end{equation} 
and the trigonometric identity $\cos(a)\cos (b)=\frac12\cos(a+b)+\frac12\cos(a-b)$ to simplify products of cosines, we arrive at
\begin{equation}
\begin{split}
&\overline{\langle\mO(x_1)\mO(x_2)\mO(x_3)\mO(x_4)\rangle^{(h)}_c}\\
&=\frac{H_4}{32}\overline{L}_4
-\frac18\sum_{(ij|kl)}
\frac{\sinh K_0(r_{ij})\sinh K_0(r_{kl})}{r_{ij}r_{kl}}
\left[
J_0(h|\vec{x}_i-\vec{x}_j+\vec{x}_k-\vec{x}_l|)
+J_0(h|\vec{x}_i-\vec{x}_j-\vec{x}_k+\vec{x}_l|)
\right]\,,
\end{split}
\end{equation}
where we defined
\begin{equation}
\begin{split}
\overline{L}_4
&=\frac{r_{12}r_{34}}{r_{13}r_{14}r_{23}r_{24}}
J_0(h|\vec{x}_1+\vec{x}_2-\vec{x}_3-\vec{x}_4|)\\
&+\frac{r_{13}r_{24}}{r_{12}r_{14}r_{23}r_{34}}
J_0(h|\vec{x}_1+\vec{x}_3-\vec{x}_2-\vec{x}_4|)\\
&+\frac{r_{14}r_{23}}{r_{12}r_{13}r_{24}r_{34}}
J_0(h|\vec{x}_1+\vec{x}_4-\vec{x}_2-\vec{x}_3|)\,.
\end{split}
\end{equation}

The logarithmic enhancement in the integral arises from the regions in which two points are taken far away from the other two. To see this, consider the quadratic term in the Taylor expansion of the integrand
\begin{equation}
\overline{\langle\mO(x_1)\mO(x_2)\mO(x_3)\mO(x_4)\rangle^{(h)}_c}-\langle\mO(x_1)\mO(x_2)\mO(x_3)\mO(x_4)\rangle^{(0)}_c=h^2 D_4+\mO\left(h^4\right)\,,
\end{equation}
\begin{equation}
\begin{split}
D_4
&=-\frac{H_4}{128}\sum_{(ij|kl)}
\frac{r_{ij}r_{kl}}{r_{ik}r_{il}r_{jk}r_{jl}}
|\vec{x}_i+\vec{x}_j-\vec{x}_k-\vec{x}_l|^2+\frac{1}{16}\sum_{(ij|kl)}
\frac{\sinh K_0(r_{ij})\sinh K_0(r_{kl})}{r_{ij}r_{kl}}
(r_{ij}^2+r_{kl}^2)\,.
\end{split}
\label{eq:D-full}
\end{equation}
Setting, for instance,
\begin{equation}\label{eq_app_R_u_v}
\vec{x}_1=0\,,\qquad
\vec{x}_2=\vec{v}\,,\qquad
\vec{x}_3=\frac{\vec{R}+\vec{u}}{2}\,,\qquad
\vec{x}_4=\frac{\vec{R}-\vec{u}}{2}\,,
\end{equation}
the region in which the pairs $(x_1,x_2)$ and $(x_3,x_4)$ are widely separated
corresponds to $R\gg u,\, v,\sim O(1)$, where $u=|\vec{u}|$, $v=|\vec{v}|$ and $R=|\vec{R}|$. Expanding $D_4$ at large $R$, we obtain
\begin{equation}\label{eq_app_D4}
D_4=-\frac{uv\sinh K_0(u)\sinh K_0(v)}{2R^2}
\left(1+\frac12\cos 2\theta+\frac12\cos 2\phi\right)
+O(R^{-3})\,,
\end{equation}
where $\vec{u}\cdot \vec{R}=u R\cos\theta$ and $\vec{v}\cdot \vec{R}=v R\cos\phi$. Integrating~\eqref{eq_app_D4} over $d^2R$ the terms proportional over $\cos 2\theta$ and $\cos 2\phi$ vanish, while the remaining contribution produces a logarithmic divergence, signalling that the expansion in $h$ breaks down when $hR\sim 1$. We can estimate then the coefficient of the $h^2\log h$ term expanding the integrand at small $h$ and integrating up to an infrared cutoff $\sim 1/h$. Using that the integration measure becomes
\begin{equation}
d^2x_2\,d^2x_3\,d^2x_4
=\frac14\,d^2R\,d^2u\,d^2v\,,
\end{equation}
and accounting for the three possible pairings, we obtain 
\begin{equation}
\frac{3}{4}\int_{\mO(1)\lesssim R\lesssim h^{-1}} d^2R\,d^2u\,d^2v \,D_4=
-\frac{3I^2_s(\pi)}{8}\int_{\mO(1)\lesssim R\lesssim h^{-1}} \frac{d^2R}{R^2}=-\frac{3\pi I^2_s(\pi)}{4}\log\left(\frac{1}{h}\right)+\mO(1)\,.
\end{equation}
Comparing with~\eqref{eq_app_I4}, we conclude
\begin{equation}
A=\frac{3\pi I^2_s(\pi)}{8}\,,
\end{equation}
in agreement with~\eqref{eq_app_A_consistency}.

These considerations also suggest a strategy to compute $B$. We first restrict the integration domain so that only one of the three possible configurations of two widely separated pairs remains. We then subtract its large-distance contribution before expanding in $h$, and add back the integral of the subtracted expression. The finite part of this add-back must be retained, since it contributes to $B$.

Since the connected correlator is symmetric under permutations of
$x,y,z$, we can impose the ordering
\begin{equation}\label{eq_app_region_Omega}
|x|\leq |y|\leq |z|\,.
\end{equation}
Within this domain, the region responsible for the logarithmic enhancement is $|x|\sim |y-z|\sim O(1)$, with $|y|\sim |z|\gg 1$. It is further convenient to change coordinates as in~\eqref{eq_app_R_u_v} 
\begin{equation}
\vec{x}=\vec{v}\,,\qquad
\vec{y}=\frac{\vec{R}+\vec{u}}{2}\,,\qquad
\vec{z}=\frac{\vec{R}-\vec{u}}{2}\,.
\end{equation}
Then, including the Jacobian and accounting for the six possible permutations, we see that we can replace
\begin{equation}
\int d^2x\,d^2y\,d^2z
\longrightarrow \frac32 \int_\Omega d^2R\,d^2u\,d^2v
\end{equation}
in~\eqref{eq_app_I4}, where $\Omega$ is the region specified by~\eqref{eq_app_region_Omega}:
\begin{equation}\label{eq_app_ordered_domain}
\Omega:\qquad
2v\leq|\vec{R}+\vec{u}|
\leq|\vec{R}-\vec{u}|\,.
\end{equation}

To construct the subtraction, it is useful to identify a region of $\Omega$ where the condition imposed by the inequalities~\eqref{eq_app_ordered_domain} on the relative angle $\theta$ between $\vec{R}$ and $\vec{u}$ simplifies. It is simple to realize that for
\begin{equation}\label{eq_app_subtraction_radius}
R>1+u+2v\,,
\end{equation}
the angle $\theta$ ranges over the half-circle $\pi/2\leq\theta\leq3\pi/2$. Note also that the relative angle $\phi$ between $\vec{R}$ and $\vec{v}$ always spans a full circle. We are then led to consider the following subtraction
\begin{equation}\label{eq_app_Fsub}
\begin{split}
F_{\rm sub}(h;\vec{R},\vec{u},\vec{v})={}&\Theta(R-1-u-2v)\,\sinh K_0(u)\sinh K_0(v)\\&\times\left\{
\frac{2uv}{R^4}\big[J_0(hR)-1\big]+\frac{\cos(\theta+\phi)}{2R^2}
\left[J_0(h|\vec{u}+\vec{v}|)-J_0(h|\vec{u}-\vec{v}|)\right]\right\}\,,
\end{split}
\end{equation}
where $\Theta$ is the Heaviside step function. We add and subtract
this expression from the integrand.

The expansion of~\eqref{eq_app_Fsub} to order $h^2$ reproduces
the leading $R^{-2}$ term in~\eqref{eq_app_D4}, including its
angular dependence. The angular term integrates to zero on the
support of the subtraction. Subtracting this term nevertheless improves the numerical integration. 

We first evaluate the contribution that must be added back.
Only the angle-independent term contributes. This gives
\begin{equation}
\begin{split}
\frac32\int_\Omega d^2R\,d^2u\,d^2v\,F_{\rm sub}(h)
={}&12\pi^3\int_0^\infty du\int_0^\infty dv\,
u^2v^2\sinh K_0(u)\sinh K_0(v)
\\
&\times\int_{1+u+2v}^\infty
\frac{dR}{R^3}\big[J_0(hR)-1\big]\,.
\end{split}
\end{equation}
Integrating by parts, we find that the integral over $R$ gives
\begin{equation}\label{eq_app_radial_addback}
\int_L^\infty\frac{dR}{R^3}\big[J_0(hR)-1\big]
=
\frac{h^2}{4}
\left[
\log\left(\frac{hL}{2}\right)+\gamma_E-1
\right]+\ldots\,,
\end{equation}
where we neglected terms which are higher order in $h$. Using~\eqref{eq_app_radial_addback} with $L=1+u+2v$, we obtain
\begin{equation}\label{eq_app_addback}
\begin{split}
&\frac32\int_\Omega d^2R\,d^2u\,d^2v\,F_{\rm sub}(h)
=\frac{3\pi I_s(\pi)^2}{8}h^2\log h^2
\\
&+3\pi^3h^2\int_0^\infty du\int_0^\infty dv\,
u^2v^2\sinh K_0(u)\sinh K_0(v)
\left[
\log\left(\frac{1+u+2v}{2}\right)+\gamma_E-1
\right]+\ldots\,.
\end{split}
\end{equation}

Expanding~\eqref{eq_app_Fsub}, we finally obtain the following finite expression for $B$:
\begin{equation}\label{eq_app_B_ordered}
\begin{aligned}
B={}&3\pi^3\int_0^\infty du\int_0^\infty dv\,
u^2v^2\sinh K_0(u)\sinh K_0(v)
\left[
\log\left(\frac{1+u+2v}{2}\right)+\gamma_E-1
\right]
\\
&+\frac32\int_\Omega d^2R\,d^2u\,d^2v\,
\Bigg[
D_4+\Theta(R-1-u-2v)\,
\frac{uv\sinh K_0(u)\sinh K_0(v)}{2R^2}
\left(1+\frac{\cos2\theta+\cos2\phi}{2}\right)
\Bigg]\,.
\end{aligned}
\end{equation}
Here $D_4$ is given in~\eqref{eq:D-full}, and the integration domain is specified in~\eqref{eq_app_ordered_domain}. Both integrals in~\eqref{eq_app_B_ordered} are finite and can be evaluated numerically. Using randomized quasi-Monte Carlo integration, we obtain\footnote{The first, two-dimensional integral in \eqref{eq_app_B_ordered} was evaluated by adaptive quadrature. For the second integral, we performed the integration over the overall orientation analytically and split the domain into the regions $R<1+u+2v$ and $R>1+u+2v$. Each resulting five-dimensional integral was mapped to the unit hypercube and evaluated using independently scrambled Sobol sequences. The quoted uncertainty was estimated from the standard error among the independent scramblings.}
\begin{equation}
B=96.15(4)\,.
\end{equation}
Using this result in~\eqref{eq_app_M_sol}, we find
\begin{equation}
M_{\rm sol}
=
0.9945(2)\,
m\,e^{-A_s\frac{g^2}{m^2}}\,,
\end{equation}
corresponding to a prefactor $K=0.9945(2)$ in~\eqref{eq_kink_mass_integrability}.

\bibliography{Biblio}
\bibliographystyle{JHEP.bst}

\end{document}